\documentclass[
  aip,%
  cha,%
  amsmath,amssymb,
  svgnames,dvipsnames,floatfix,
  twocolumn,
  reprint, 
]{revtex4-2}

\usepackage{graphicx}%
\usepackage{dcolumn}%
\usepackage{bm}%
\usepackage[utf8]{inputenc}
\usepackage[T1]{fontenc}
\usepackage{mathptmx}
\usepackage{etoolbox}

\usepackage[version=4]{mhchem}
\usepackage{siunitx}
\usepackage{algorithm}
\usepackage[noEnd=true]{algpseudocodex}
\usepackage{mathtools}
\usepackage{tabularray}
\usepackage{empheq}
\usepackage{soul}
\usepackage{hyperref}
\usepackage[capitalize]{cleveref}
\usepackage{overpic}

\labelformat{section}{\S\thesection}
\labelformat{subsection}{\S\thesection.\thesubsection}
\renewcommand{\eqref}[1]{Eq.~(\ref{#1})}
\newcommand{\lineref}[1]{line~(\ref{#1})}
\newcommand{\pref}[1]{(\ref{#1})}
\makeatletter
\newcommand{\citeb}[2][\@nil]{%
  \def\tmp{#1}%
  \ifx\tmp\@nnil
    [\onlinecite{#2}]%
  \else
    [\onlinecite{#2}, #1]%
  \fi}
\makeatother

\newcommand{\half}{\frac{1}{2}}
\newcommand{\grad}{\nabla}

\DeclarePairedDelimiter\parensImpl{\lparen}{\rparen}
\DeclarePairedDelimiter\bracketsImpl{\lbrack}{\rbrack}
\DeclarePairedDelimiter\bracesImpl{\lbrace}{\rbrace}

\DeclarePairedDelimiter\absImpl{\lvert}{\rvert}
\DeclarePairedDelimiter\normImpl{\lVert}{\rVert}

\newcommand{\parens}{\parensImpl*}
\newcommand{\brackets}{\bracketsImpl*}
\newcommand{\braces}{\bracesImpl*}
\newcommand{\abs}{\absImpl*}
\newcommand{\norm}{\normImpl*}

\newcommand{\pvec}{\mathbf{p}}
\newcommand{\uvec}{\mathbf{u}}
\newcommand{\cO}{C_{\ce{O}}}
\newcommand{\cR}{C_{\ce{R}}}
\newcommand{\cH}{C_{\ce{H^+}}}
\newcommand{\cHi}{\cH^{(0)}}
\newcommand{\cT}{C_{\textrm{T}}}
\newcommand{\cTi}{c_{0}}
\newcommand{\nox}{n_{\text{O}}}
\newcommand{\nre}{n_{\text{R}}}
\newcommand{\nel}{n_{\text{e}}}
\newcommand{\npr}{n_{\text{p}}}
\newcommand{\phiL}{\phi_{\scriptscriptstyle L}}
\newcommand{\phiS}{\phi_{\scriptscriptstyle S}}
\newcommand{\phiLt}{\tilde{\phi}_{\scriptscriptstyle L}}
\newcommand{\Eeq}{E_\text{eq}}
\newcommand{\Estd}{E^{\circ}}
\newcommand{\Eadj}{{E^{\circ}}'}
\newcommand{\etaact}{\eta_\text{act}}
\newcommand{\etat}{\tilde{\eta}}
\newcommand{\etamax}{\etat_\text{max}}
\newcommand{\Var}{V_\text{ar}}
\newcommand{\Vart}{\tilde{V}_\text{ar}}
\newcommand{\St}{\tilde{S}}
\newcommand{\Rt}{\tilde{R}}

\newcommand{\seq}{s_\text{eq}}
\newcommand{\socin}{s_{\text{in}}}

\newcommand{\smt}{s_{\text{mt}}}
\newcommand{\selec}{s_\text{elec}}
\newcommand{\socmin}{s_{\text{min}}}
\newcommand{\socmax}{s_{\text{max}}}
\newcommand{\Itot}{I_\text{tot}}
\newcommand{\Fin}{\mathcal{F}_{\text{in}}}
\newcommand{\Fout}{\mathcal{F}_{\text{out}}}
\newcommand{\Fnet}{\mathcal{F}_{\text{net}}}
\newcommand{\thetamrc}{\theta_{\text{mrc}}}
\newcommand{\thetamin}{\theta_{\text{min}}}
\newcommand{\thetamax}{\theta_{\text{max}}}
\newcommand{\umt}{U_{\text{mt}}}
\newcommand{\amat}{\mathsf{A}}
\newcommand{\dmat}{\mathsf{D}}
\newcommand{\fmat}{\mathsf{F}}
\newcommand{\smat}{\mathsf{s}}
\newcommand{\etamat}{\mathsf{\eta}}
\newcommand{\logit}{\text{logit}}
\newcommand{\ccfl}{C_\textrm{CFL}}
\newcommand{\rn}{\textrm{Re}}

\newcommand{\kg}{\kilogram}
\newcommand{\m}{\meter}
\newcommand{\cm}{\centi\meter}
\newcommand{\um}{\micro\meter}
\renewcommand{\sec}{\second}
\newcommand{\mv}{\milli\volt}
\newcommand{\amp}{\ampere}
\newcommand{\ma}{\milli\ampere}
\newcommand{\coul}{\coulomb}
\newcommand{\pa}{\pascal}

\makeatletter
\def\@email#1#2{%
 \endgroup
 \patchcmd{\titleblock@produce}
  {\frontmatter@RRAPformat}
  {\frontmatter@RRAPformat{\produce@RRAP{*#1\href{mailto:#2}{#2}}}\frontmatter@RRAPformat}
  {}{}
}%
\makeatother

\makeatletter
\@addtoreset{ALG@line}{algorithm}
\makeatother

\begin{document}
\title[Digital Twin for Porous Electrodes in Redox Flow Batteries]
{Digital Twin for Porous Electrodes in Redox Flow Batteries}
\author{Michael S. Emanuel}
\affiliation{ 
School of Engineering and Applied Sciences, 
Harvard University, 29 Oxford Street, Cambridge MA, United States
}%
\author{Chris H. Rycroft}%
\affiliation{Department of Mathematics, University of Wisconsin--Madison, Madison, WI 53706, USA}
\affiliation{Mathematics Group, Lawrence Berkeley Laboratory, Berkeley, CA 94720, USA}
\email{rycroft@wisc.edu.}
\date{11 February, 2025}
\graphicspath{{figs/}}

\begin{abstract}
Porous electrodes are a vital component of redox flow batteries, fuel cells and electrolyzers.
We present a 3D digital twin for a porous electrode by direct numerical solution of the governing Navier-Stokes and Nernst-Planck equations for incompressible flow and electrochemical mass transport with Butler-Volmer reaction kinetics. 
We demonstrate our method by simulating a laboratory sized system at a sub-fiber scale of \qty{1.25}{\um} using a single workstation. 
We establish convergence and the consistency of mesh refined lattices, and review an experimental validation of our model. 
We show the efficacy of two novel techniques to speed convergence to steady state, iterative upsampling and model refinement.
We prove that the state of charge ($s$) at steady state can be formulated as the solution to a single PDE in $s$ with a Dirichlet boundary condition, and introduce a novel figure of merit $\umt$, the utilization of an electrode geometry at the mass transport limit, both of which can be rapidly simulated with explicit methods from a given fluid velocity field.
All of our code is highly performant, parallelizable, open source C++ that is published with this work and compatible with the largest modern scientific supercomputers available today.
\end{abstract}
\maketitle

\section{Introduction}
\label{sec:intro}
Anthropogenic climate change poses a generational challenge to human well-being and prosperity \cite{RTPS22}.
The clean energy transition will require vast amounts of renewable energy from intermittent sources
such as solar and wind, which will in turn drive demand for long duration energy storage \cite{KEG17, GSC15}.
Redox flow batteries (RFBs) are a promising energy storage technology with the unusual feature 
that their storage capacity can be scaled independently of their power, 
making them ideally suited to long duration grid scale energy storage applications \cite{Gur18, APW19}.

Porous electrodes are a vital component of flow battery systems and strongly influence 
mass transfer, charge transfer, ohmic resistance and hydraulic resistance \cite{FCB19}.
They are porous media that admit hydraulic flow and also provide the active surface for electrochemical reactions.
This dual role creates an inherent design tension between greater porosity to reduce hydraulic resistance
and greater surface area to improve mass and charge transfer.

The oldest approach to model porous electrodes dating to Newman et al. in the 1960s
 is to use volume averaging over sufficiently large regions of the electrode 
to treat it as homogeneous and isotropic \cite{NT62, NT75}.
This approach has been highly successful and is still leading to new refinements, e.g.
Smith and Bazant's extension of Newman's porous electrode theory 
to phase-separating materials \cite{SB17}.

A popular approach to modeling porous electrodes in recent literature is to use the COMSOL Multiphysics 
package \cite{COMSOL} to solve the Nernst-Planck equation,
typically using some form of volume averaging above pore scales, and often in 2D.
This has the advantages of avoiding time consuming effort in software development 
and ready access to other COMSOL features such as optimization engines,
against the disadvantages of costly software licenses 
and opaque, unmodifiable source code for the underlying model \cite{SB17}.
Lee et al. used COMSOL to simulate the cathode of a vanadium flow battery consisting 
of a ``sandwich'' of three planar layers with varying fiber sizes and porosity and optimize its design \cite{LG24},
while Lin et al. performed topology optimization on 3D flow fields for RFBs \cite{LB22}
and Beck et al. computationally designed porous electrodes with spatially varying porosity \cite{BW21}.

Pore Network Models (PNMs) have also been used to simulate and optimize porous electrodes.
As suggested by their name, PNMs treat the electrode as a collection of spherical pores (``throats'') 
connected by a network of cylindrical ``pipes'' via hydraulic flow
in a modeling approach dating back to the 1980s \cite{CD85, LSD81, DP86}.
Gostick et al. introduced the OpenPNM package \cite{OpenPNM16} to solve PNMs.
Later van der Heijden et al. experimentally assessed their validity with two commercial carbon electrodes \cite{HFC22},
and went on to use PNMs as the fitness function for an optimization using a genetic algorithm \cite{HFC24}.

The lattice Boltzmann method (LBM) is a second technique that approximates 
the governing equations for the sake of computational speed.
The key idea in LBM is to describe the motion of each species in a fluid using discrete particle density distribution functions.
Zhang et al. used LBM to simulate the electrochemical behavior of three commercially available carbon electrodes,
using a geometry obtained from X-ray computed tomography and LBM equations from Dawson et al. 
to obtain good agreement with experimental characterization of the electrodes \cite{ZFC19, DCD93}.

In contrast to PNMs and LBMs, which make compromises on the governing equations,
and COMSOL approaches, which accept volume averages above pore scale and closed source software,
we present here a constructive, direct numerical simulation of the Nernst-Planck equation
using only open source C++ code which is published along with this work.
The spatial resolution of \qty{1.25}{\um} is substantially below the pore scale, 
and is in fact a fraction of a typical fiber diameter in carbon electrodes, e.g. \qty{8}{\um} for SGL39AA \cite{LG24}, 
hence we refer to this as a \textit{fiber-scale} simulation.
We employ the framework introduced by Dussi and Rycroft \cite{DR22},
stepping up from two to three dimensions and making some technical adjustments that were necessary.
We also introduce several new ideas, including reformulating the governing PDE to have a Dirichlet boundary condition, 
which collectively make it feasible to calculate a converged steady state.
This model was used to simulate a microfluidic flow battery system described by Barber et al.~\cite{BEE24}
and achieved good qualitative agreement with experimental images,
faithfully reproducing convective tails and large scale concentration gradients.

A redox flow battery includes two half cells separated by an ion exchange membrane.
Each half cell includes an electrolyte tank, a porous electrode, and a current collector.
Electrolyte is pumped through the two electrodes and undergoes a reaction
with electrons moving between the current collectors 
while charge is balanced by an offsetting ionic current through the membrane.
A voltage source is applied during charging, and a load can be powered during discharge.

\begin{figure}[ht]
\centering
\includegraphics[width=1.00\linewidth]{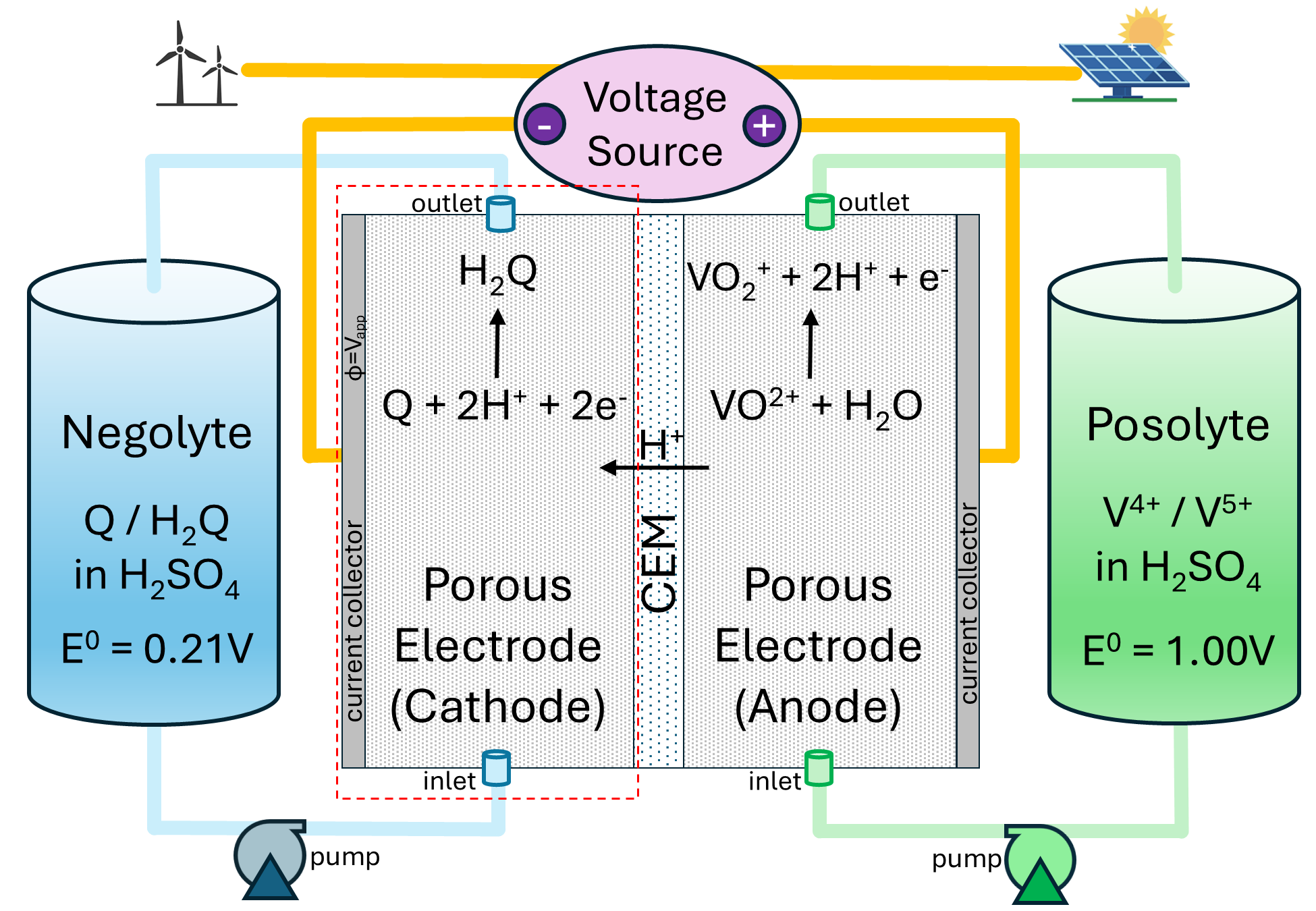}
\caption{
\textbf{A redox flow battery being charged.}
A quinone \ce{Q} is reduced to a hydroquinone \ce{H_{2}Q} in the cathode by an applied reducing voltage. 
\ce{V^{4+}} is oxidized to \ce{V^{5+}} in the anode. 
Protons flow across a cation exchange membrane (CEM) to balance charge. 
Only the negolyte electrode (in red dashed rectangle) is simulated.}
\label{fig:flow-battery}
\end{figure}

This work presents simulation results and modeling approaches that are broadly applicable to porous electrodes in RFBs.
Notwithstanding this general approach, the simulation was designed with a particular model system in mind,
a quinone based negolyte that is being reduced (charged) as shown in Figure \ref{fig:flow-battery}.
This system was chosen because our collaborators fabricated and experimentally characterized it \cite{BEE24}.
The active surface for the reaction is a 3D printed electrode with precisely 
engineered features matching a design geometry of cylindrical conducting 
surfaces and a relatively low specific area compared to commercial electrodes.
The counterelectrode is a commercial carbon paper with much larger surface area 
that is assumed to be faster due to its larger area. 
The posolyte is a \ce{V^{+4} / V^{+5}} redox pair with $\Estd = \qty[retain-explicit-plus]{+1.00}{\volt}$.
The quinone negolyte consists of a dilute 20 mM solution of anthraquinone-2,7-disulfonate (AQDS)
in a supporting electrolyte of 1M sulfuric acid and $\Estd = \qty[retain-explicit-plus]{+0.21}{\volt}$ \cite{HMS14}.
\begin{figure}[ht]
\centering
\includegraphics[width=1.00\linewidth]{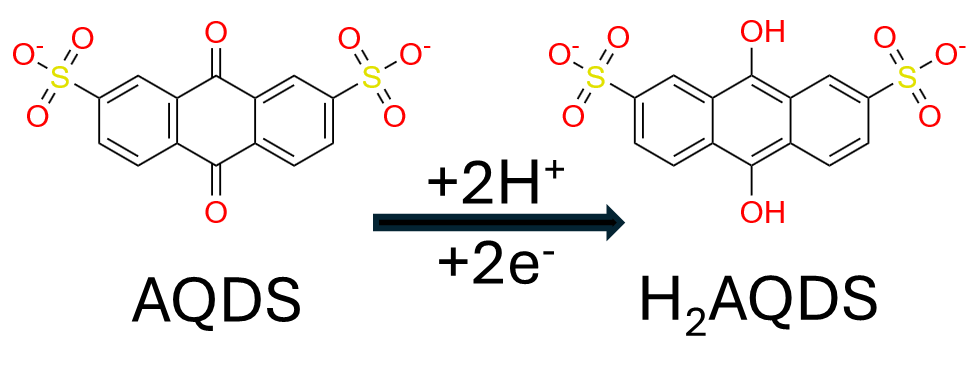}
\caption{
\textbf{Negolyte redox pair.}
The oxidized species, the anthraquinone \ce{AQDS},
is reduced to its hydroquinone \ce{H_{2}AQDS} with the addition of two protons and two electrons.
Both species have a \ce{-2} charge in solution, and 
$\Estd = \qty[retain-explicit-plus]{0.21}{\volt}$ \cite{HMS14}.
}
\label{fig:aqds-structure}
\end{figure}
One molecule of \ce{AQDS} is reduced to one molecule of \ce{H_{2}AQDS} in a 
proton coupled electron transfer where two electrons are transferred (Figure \ref{fig:aqds-structure}). 
Both species have a $-2$ charge in solution.

All of the simulations presented in this work pertain to this model system.
These are half cell simulations for just the AQDS side of the system rather than full cell simulations.
The simulated applied electric potential is the potential difference between the electrode and the electrolyte at the membrane, i.e. $\phiS - \phi_{\scriptscriptstyle M}$.
This voltage difference is \textit{not} the same as the full cell potential, and omits effects including the standard potential of the counterelectrode and ohmic losses through the membrane.
The voltage difference between the electrode and electrolyte can be measured experimentally using a reference electrode that is placed in the spent electrolyte as it exits the electrode.
We emphasize here that to obtain a steady state simulation of a full cell, 
two half cell simulations at the same current can be combined along 
with a simple model for ohmic losses through the membrane.
This procedure is significantly simpler and computationally cheaper than a coupled full cell simulation 
that would need to resolve transient effects on both half cells and either 
resolve the electric field through the membrane or solve for the 
current balancing applied potential on the two half cells at every time step.

\begin{figure*}[ht]
\centering
\includegraphics[width=1.00\linewidth]{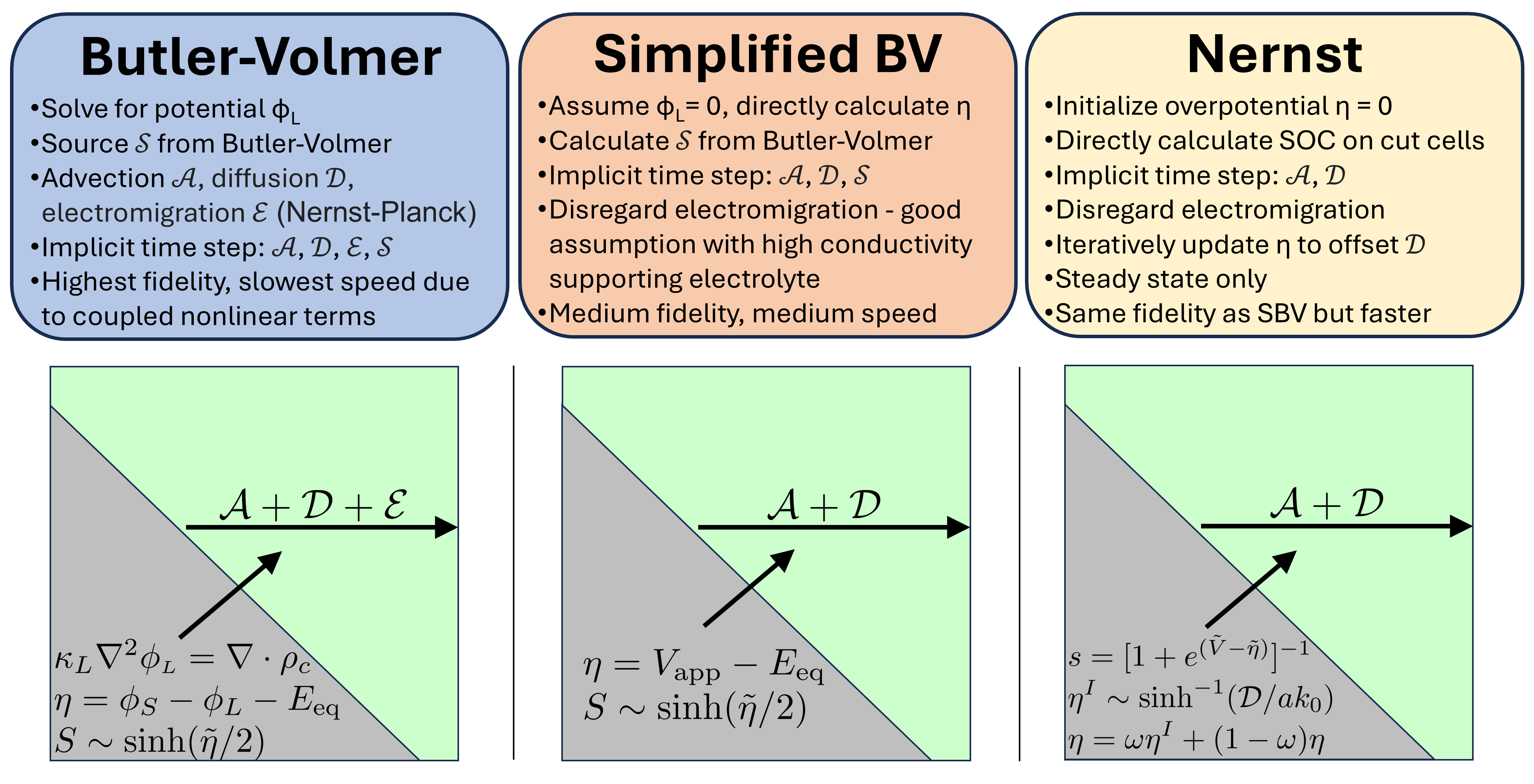}
\caption{
\textbf{Overview of three reaction models.} 
The Butler-Volmer model solves the Nernst-Planck equation including advection, diffusion and electromigration. 
It has the highest fidelity and slowest speed.
The simplified Butler-Volmer model disregards electromigration to gain significant speed with a small compromise in accuracy. The Nernst model makes the same assumption, but only solves for steady state.
\textcolor{gray}{Gray} regions represent solids (electrode) and 
\textcolor{Green}{green} regions represent liquids (electrolyte).
See \ref{sec:theory} for definitions of all algebraic symbols used.
}
\label{fig:model-overview}
\end{figure*}

We develop three models for the electrochemical reaction in a half cell with varying levels of fidelity and speed
(Figure \ref{fig:model-overview}).
The Butler-Volmer (BV) model is the most faithful and computationally expensive.
It solves the Nernst-Planck equation including advection, diffusion and electromigration.
The chemical source term is obtained using Butler-Volmer reaction kinetics.
The simplified Butler-Volmer (SBV) model makes the simplifying assumption that 
the potential in the electrolyte $\phiL$ is uniformly zero, and therefore drops the electromigration term.
This approximation is good in the presence of a strong supporting electrolyte such as 1M sulfuric acid,
where the high conductivity leads to small potential differences in the electrolyte;
and its validity is verified numerically for the specific systems we simulated here.
The SBV model is significantly faster than the BV model because it eliminates 
coupled nonlinear terms arising from electromigration and solving a Poisson equation for $\phiL$.
Finally, the Nernst model makes the same physical assumption as the BV model, 
but directly calculates the steady state. This is done by iteratively updating estimates
of the activation overpotential and state of charge on the reactive surface
until mass balance between diffusion and the reaction is achieved.

One goal of this work is to introduce a numerical solution to the Nernst-Planck equation
of sufficient fidelity that it is accepted as a digital twin of the true solution to the PDE.
This would allow it to be used to validate other models that trade fidelity for greater speed, 
including PNMs, LBMs, and PDE solutions at lower spatial resolution.
A second goal is to build useful intuition for designers of porous electrodes and batteries
by developing a theory to compute the mass transport utilization limit of a given velocity field.

\section{Theoretical Background}
\label{sec:theory}
We model the behavior of a porous electrode at steady state by direct numerical simulation of the governing equations. 
While our simulation apparatus can resolve transient phenomena, we limit our investigation to the steady state
because it is of predominant importance for applications of porous electrodes in areas including flow batteries, electrolyzers, and fuel cells.
We treat this system as having two separable parts, an incompressible Newtonian fluid flow and a mass transport system including an electrochemical source term with Butler-Volmer reaction kinetics.
We first solve for the incompressible fluid flow at a steady state, making the assumption that the chemical reactions do not affect the fluid rheology or the flow. Once the fluid velocity field $\uvec$ is obtained, we simulate the mass transport phenomena of advection, diffusion, electromigration, and electrochemical reaction on the electrode surface.
In this section, we state the governing equations and outline the mathematical theory used to solve them, 
beginning with the incompressible fluid flow.
We close by introducing various approximations that simplify the computation 
without sacrificing the essential accuracy in the results.

\begin{figure}[ht]
\centering
\includegraphics[width=1.00\linewidth]{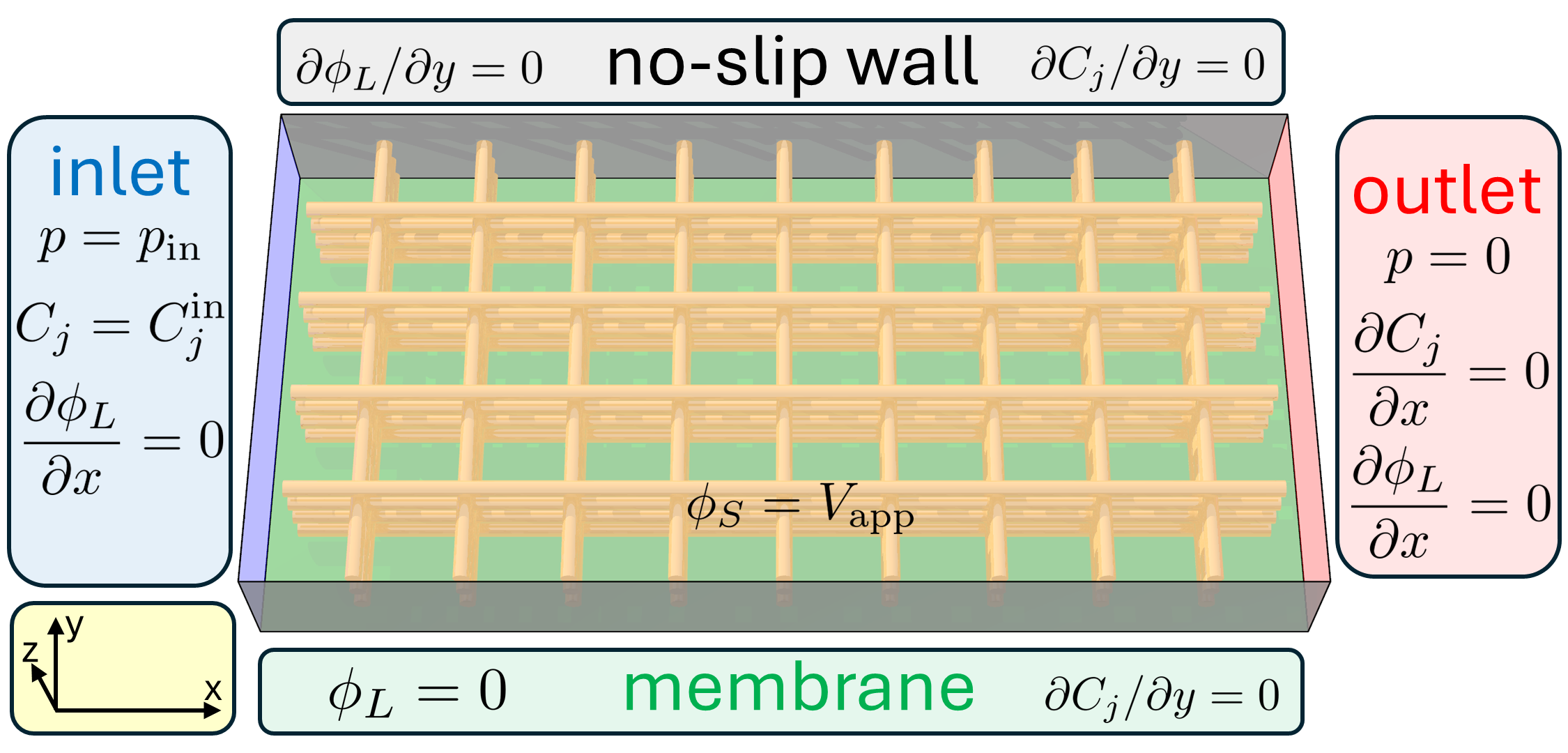}
\caption{
\textbf{Boundary conditions for negolyte half cell simulation.}
Electrolyte at pressure $p_\text{in}$ and known concentration enters the \textcolor{blue}{inlet} ($x=0$) 
and flows along the $x$-axis, exiting through the \textcolor{red}{outlet} at zero pressure. 
The \textcolor{ForestGreen}{membrane} ($z=0$) is a no-slip wall with fixed potential $\phiL=0$ 
and the electrode surface is held at a reducing potential $\phiS =V_\text{app}$. 
The top, front and back faces are all \textcolor{gray}{no slip walls} with zero 
concentration gradient and zero electric field normal to the wall.}
\label{fig:boundary-cond}
\end{figure}

Figure \ref{fig:boundary-cond} shows the boundary conditions for the simulation.
The domain $\Omega$ is a rectangular channel, 
though more complex geometries can be accommodated using embedded boundaries.
Flow moves along the $x$-axis, and at the inlet $(x=0)$ there is an applied pressure $p_\text{in}$,
known concentrations $C_j^\text{in}$ for each species, and zero electric field component $E_x$ normal to the wall.
The outlet conditions are zero pressure and no electric field or concentration gradient along $x$.
The top, left and right walls (in gray) have a no-slip condition with zero fluid velocity 
and no concentration gradient or electric field normal to the wall.
The bottom wall corresponds to the membrane, and has the same no-slip conditions
as the top, except it has a fixed potential $\phiL = 0$ rather than zero electric field $E_y$.
The surface of the electrodes is held at a fixed potential $\phiS$, 
with a lower potential $\phiS < \phiL$ required to drive a reduction reaction.

\subsection{Incompressible Fluid Flow}
\label{sec:flow-theory}
We begin with the Navier-Stokes equation for an incompressible Newtonian fluid 
with constant density $\rho$, a standard result \citeb[Eq. 7.1.6]{Ruderman19},
\begin{equation}
\label{eq:navier-stokes}
\rho \parens{\frac{\partial \uvec}{\partial t} + (\uvec \cdot \grad) \uvec}
 = -\grad p + \mu \grad^2 \uvec,
\end{equation}
where $\uvec$ is the the fluid velocity, $p$ is the pressure, and $\mu$ is the dynamic viscosity 
(in $\unit{\pa.\sec} = \unit{\kg.\m^{-1}.\sec^{-1}}$). 
We also have the incompressibility condition \citeb[Eq. 5.1.5]{Ruderman19},
\begin{equation}
\label{eq:incompressibility}
\grad \cdot \uvec = 0.
\end{equation}
The incompressible flow problem of Eqs.~(\ref{eq:navier-stokes}--\ref{eq:incompressibility})
is solved using the Chorin projection method \cite{Chorin68} as detailed in \ref{sec:flow-methods}.

The Reynolds number $\rn$ is the dimensionless ratio $\rn = \rho L V / \mu$,
where $L$ is the characteristic length scale and $V$ is the characteristic speed.
For this problem, we propose that $L = L_z$ is the height of the channel (the smallest dimension) \cite{AML11}
and $V = q$ is the volumetric flux (superficial velocity) as discussed below in \eqref{eq:darcys-law}.

A classical result of fluid dynamics \citeb[Eq. 7.4.3]{Ruderman19} 
states that in the limiting case of a small Reynolds number $\rn \ll 1$,
the steady state fluid flow simplifies from \eqref{eq:navier-stokes} to Stokes's approximation,
\begin{equation}
\label{eq:stokes-flow}
\grad p = \mu \grad^2 \uvec.
\end{equation}
We refer to flows in this regime as Stokes flows, and note the physical intuition that 
in a Stokes flow, an applied pressure gradient is dissipated by viscous forces,
while momentum transport is negligible.
One immediate consequence of \eqref{eq:stokes-flow} is that the steady state fluid velocity is linear in the applied pressure.
This enables us to approximate a whole series of flows in the Stokes regime by running a single 
simulation at one applied pressure and then scaling the results for varying pressures.

The presence of gravitational forces is a complication that can be avoided by incompressible flow theory.
The dynamic pressure $\mathcal{P}$ is defined by
\begin{equation}
\label{eq:dynamic-pressure-def}
\grad \mathcal{P} = \grad p - \rho g,
\end{equation}
where $g$ is the gravitational field.
As discussed by Deen \citeb[\S 6-3]{Deen12}, it is often advantageous to treat incompressible
flow problems in the presence of gravity in terms of the dynamic pressure $\mathcal{P}$.
For the case of a steady state flow of an incompressible fluid with fixed solid boundaries as we have here,
the solution of a problem in terms of the true pressure $p$ including gravitational body forces
will match the solution in terms of the dynamic pressure.
In the remainder of this work, we will only consider the dynamic pressure and disregard gravitational body forces.
We will denote the dynamic pressure by $p$ moving forward for notational ease.
There is no risk of confusion of pressure gradients between applied pumping pressure and gravity,
because we always have the applied pressure and flow in the $x$ direction.

Darcy's law describes the flow of fluid through a porous medium \cite{Darcy1856} and can be written 
\begin{equation}
\label{eq:darcys-law}
q = - \parens{\frac{\kappa}{\mu}} \grad p,
\end{equation}
where $q$ is the volumetric flux in $\unit{\m.\sec^{-1}}$ (sometimes called the superficial velocity), 
and $\kappa$ is the permeability of the porous medium in \unit{\m^{2}}.
Darcy's law can be derived directly from the Navier-Stokes equations 
by volume averaging Stokes flow through a porous medium \cite{Hubbert56, Whitaker86}.
Even though the porous electrodes considered here are far from being homogeneous porous media,
we can nevertheless empirically describe the flow through a porous electrode using Darcy's law.
Let $Q$ denote the volumetric flow rate in \unit{\m^{3}.\sec^{-1}},
$A$ the cross sectional area of the inlet ($yz$ plane) in \unit{\m^{2}},
and $\Delta p = p_\text{in} - p_\text{out}$ the pressure drop.
Volumetric flux is flow per unit area, so $Q = Aq$, and the mean pressure gradient is $\grad p = \Delta p / L$.
This gives us the integral form of Darcy's law for a homogeneous permeable medium,
\begin{equation}
\label{eq:darcys-law-integral}
Q = \parens{\frac{\kappa A}{\mu L}} \Delta p .
\end{equation}
We can define the hydraulic resistance $R_H$ so it appears in the denominator 
on the right hand side of \eqref{eq:darcys-law-integral}, giving us
\begin{equation}
\label{eq:hydraulic-resistance}
R_H = \frac{\mu L}{\kappa A} = \frac{\Delta p}{Q} .
\end{equation}
As long as a fluid flow is in the Stokes regime, the hydraulic resistance will be approximately constant.
$R_H$ will increase as the flow becomes turbulent, a result seen in \ref{sec:results-flow-stokes}.

\subsection{The Nernst Equation and Butler-Volmer Reaction Kinetics}
\label{sec:bv-theory}
In this section we review the Butler-Volmer theory of reaction kinetics.
Consider a reversible proton-coupled redox reaction between two electroactive species, denoted \ce{O} and \ce{R}
for the oxidized and reduced species respectively, which exchange $\nel$ electrons and $\npr$ protons in the reaction,
\begin{equation}
\label{eq:chem-rxn-gen}
\ce{O + $\npr$ H^{+} + $\nel$ e^{-} <=> R}
\end{equation}
The equilibrium potential $\Eeq$ for this reaction is given by the Nernst Equation 
\citeb[Eq. 8]{DR22} or \citeb[Eq. 3.7.11]{BardFaulkner22},
\begin{equation}
\label{eq:nernst-full}
\Eeq = \Estd + \frac{RT}{\nel F} \log \parens{\frac{\cO \; \left(\cH\right)^{\npr}}{\cR}},
\end{equation}
were $E^0$ is the voltage under standard conditions, $R$ is the ideal gas constant, 
$T$ is the temperature, and $\log$ denotes the natural logarithm.
The activation overpotential, denoted $\etaact$ with units of volts,
 is the difference between the interfacial voltage 
 across the electrode $\phiS - \phiL$ and its equilibrium value $\Eeq$, and is given by
\begin{equation}
\label{eq:etaact-def}
\etaact = \phiS - \phiL - \Eeq.
\end{equation}

Denote by $\cT$ the total concentration of electroactive species, $\cT = \cO + \cR$.
We will show in \ref{sec:rxn-sbv-model} that under the conditions in the model AQDS system, 
$\cT$ is mathematically conserved.
We will further demonstrate that even in systems where $\cT$ is not exactly conserved, 
it is very close to being a spatially uniform constant in practice.
We can therefore think of $\cT$ as simplifying to a constant $\cTi$ for practical purposes,
and in the results presented here $\cTi = 20 \; \unit{\mole.\m^{-3}}$.
(Note that $1 \; \unit{\mole.\m^{-3}}$ corresponds to a one millimolar (mM) concentration.)
Define the local state of charge (SOC) $s$ by
\begin{equation}
\label{eq:soc-def}
s = \frac{\cR}{\cO + \cR} = \frac{\cR}{\cT} \approx \frac{\cR}{\cTi}.
\end{equation}
The state of charge is a dimensionless parameter that ranges between $0$ and $1$, 
which correspond respectively to completely discharged and completely charged internal states for a battery.
While SOC is often treated as a single scalar to describe an entire battery system, 
here it is a scalar field defined throughout the electrolyte, i.e. over the same region as $\phiL$.
An immediate consequence of \eqref{eq:soc-def} is that we can make the substitutions 
\begin{equation}
\label{eq:conc-from-soc}
\cR = s \, \cT \approx s \, \cTi, \quad \cO = (1-s) \, \cT \approx (1-s) \, \cTi.
\end{equation}

The Butler-Volmer theory predicts the current density $j$ (in $\unit{\ampere.\m^{-2}}$) 
of an electrochemical reaction at the surface of an electrode.
Following Bard and Faulkner \citeb[\S 3.4]{BardFaulkner22}, 
define the exchange current density $j_0$ 
($\unit{\ampere.\m^{-2}}$) as
\begin{equation}
\label{eq:exchange-current-def}
j_0 = \nel \, F \, a \, k_0 \, \cO^{(1-\alpha)} \, \cR^{\alpha},
\end{equation}
where $F$ is Faraday's constant, $k_0$ is the rate constant in $\unit{\m.\sec^{-1}}$
 and $\alpha$ is a dimensionless charge transfer coefficient.
The specific area $a$ is the ratio of surface area to volume, $a = A / V$ integrated over a test region.
$a$ has units of \unit{\m^{-1}} and scales as one over the fiber diameter 
in real world electrodes, e.g. carbon cloth, paper or felt.

Define the thermal voltage $V_T = RT/F$, which is approximately 25.7 mV at standard conditions.
The Butler-Volmer equation \citeb[Eq. 3.4.11]{BardFaulkner22} states that the current density $j$ is given by
\begin{equation}
\label{eq:butler-volmer}
j = j_0 \brackets{e^{-\alpha \nel \etaact / V_T} - e^{(1-\alpha) \nel \etaact / V_T}}.
\end{equation}
Let $S$ be the chemical source term, the rate at which the entire reaction 
in \eqref{eq:chem-rxn-gen} moves forward in units of $\unit{\mole.\m^{-3}.\sec^{-1}}$,
and let $n_k$ denote the signed stoichiometric number of species $k$.
For the example reaction shown, $\nre = 1$ and $\nox = -1$, since the reaction 
produces one molecule of $\ce{R}$ and consumes one molecule of $\ce{O}$.
The source term $S_k$ for species $k$ is the the change in concentration over time for that species, 
and therefore $S_k = n_k \, S$.

We can relate the description of the reaction rate in terms of current in \eqref{eq:butler-volmer}
 to the source term $S$ by conservation of charge in a test region.
If $A$ is the area of the electrode surface and $V$ is the liquid volume,
then the current $A j$ must match the rate of consumption of electrons $\nel F V S$.
Setting these two equal and solving for $S$ we find $S = (A / \nel F V) j = (a / \nel F) j$.
Substituting for the Butler-Volmer equation \eqref{eq:butler-volmer}, the terms $\nel F$ cancel and we 
recover \citeb[Eq. (6)]{DR22}, obtaining
\begin{equation}
\label{eq:source-term-bv1}
S = a \, k_0 \, \cO^{(1-\alpha)} \, \cR^{\alpha} \brackets{e^{-\alpha \nel \etaact / V_T} - e^{(1-\alpha) \nel \etaact / V_T}}.
\end{equation}

The Butler-Volmer equation appears in a number of variations, some of which 
introduce a mass transfer coefficient $k_m$ linking the bulk and surface concentrations
\cite{DW20, DGM16, MvS12, ZGL24}. 
In this work, we do \textit{not} use these surface / bulk formulations of the Butler-Volmer equation 
because we are explicitly modeling the concentration in cells on a sub-fiber scale. 
The role of the mass transfer coefficient is therefore supplanted by 
a direct simulation of mass transfer due to diffusion and advection near the reactive surface.
Eqs.~(\ref{eq:exchange-current-def}) and (\ref{eq:butler-volmer}) follow 
what Dickinson and Wain call the electroanalytical formulation of the Butler-Volmer equation \cite{DW20}.

We make the common assumption that the charge transfer coefficient $\alpha = 1/2$ \cite{DR22}.
This allows us to simplify \eqref{eq:source-term-bv1} with the relation $\sinh(x) = (e^x - e^{-x})/2$.
Define the dimensionless reducing overpotential $\etat$ by
\begin{empheq}[box=\fbox]{equation}
\label{eq:etat-def}
\etat = -\frac{\nel \etaact}{V_T}.
\end{empheq}
Note that the sign of $\etat$ is reversed from the definition of the overpotential $\etaact$;
$\etat$ is a nondimensionalized overpotential in the reducing direction, 
where a lower value of $\etaact$ drives the reducing reaction forward at a faster rate.
This definition of $\etat$ is motivated by the model system, 
in which AQDS that starts at state of charge near zero is reduced. 
In an electrode that is being charged for energy storage applications, a positive value of $\etat$ 
would be selected with the goal of achieving a state of charge $s$ near one for the solution at the outlet.
The intuition is that $\etat$ is a control variable- 
the operator selects a reducing potential to apply to the flowing electrolyte -
and the result is the state of charge at the outlet.
A higher reducing voltage leads to a higher SOC at the outlet.
Substitute for the two concentrations in \eqref{eq:source-term-bv1} using \eqref{eq:conc-from-soc}
and for the overpotential terms using \eqref{eq:etat-def} and we obtain
\begin{empheq}[box=\fbox]{equation}
\label{eq:source-term-bv}
S = 2a \, k_0 \, \cT \sqrt{s (1-s)} \sinh(\etat / 2).
\end{empheq}

We now rewrite the Nernst equation (\ref{eq:nernst-full}) in terms of the state of charge $s$ 
and separate the small effect of changes in the proton concentration $\cH$ from its initial value $\cHi$. 
Define the adjusted standard potential $\Eadj$ by \cite{DR22}
\begin{equation}
\label{eq:E0-prime-def}
\Eadj = \Estd + \frac{\npr RT}{\nel F} \log \parens{\cHi}.
\end{equation}
Substituting for $\cR$ and $\cO$ in terms of the state of charge with \eqref{eq:conc-from-soc}, 
and shifting the reference from $\Estd$ to $\Eadj$, we obtain
\begin{equation}
\label{eq:nernst-adj}
\begin{split}
\Eeq = \underbrace{\Eadj
\vphantom{\frac{\cH - \cHi}{\cHi}}
}_{\text{Standard}} 
- \underbrace{\parens{\frac{V_T}{\nel}\right) \log \left( \frac{s}{1-s}}
\vphantom{\frac{\cH - \cHi}{\cHi}}
}_{\text{SOC dependence}} + \\
\underbrace{\parens{\frac{\npr V_T}{\nel}\right)  \log \left( 1 + \frac{\cH - \cHi}{\cHi}}
}_{\text{pH Dependence}}.
\end{split}
\end{equation}
The leading terms are the adjusted standard potential $\Eadj$ and the dependence on the state of charge.
The last term reflects the change in proton concentration (pH) from its starting values.
For a strong supporting electrolyte as is typically the case in practice, 
this last term is dominated by the state of charge dependence.
In the model system, $\cHi$ is larger than $\cTi$ by a factor of 50, and numerical simulations show only very small 
relative changes in $\cH$ as the simulation runs, i.e.  $\abs{\cH - \cHi} \ll \cHi$.

In the ensuing discussion, we will disregard the pH dependence and rely on the simplified
Nernst equation for this system,
\begin{equation}
\label{eq:nernst-simp}
\Eeq \approx \Eadj - \parens{\frac{V_T}{\nel}} \log \parens{\frac{s}{1-s}}.
\end{equation}

\subsection{Electrochemical Reaction with Variable Potential in Electrolyte - ``Butler-Volmer Model''}
\label{sec:rxn-bv-model}
The Nernst-Planck equation describes mass transport phenomena due to advection, diffusion, and electromigration,
\begin{empheq}[box=\fbox]{flalign=left}
\label{eq:nernst-planck}
\scalebox{0.94}
{$\displaystyle
\frac{\partial C_j}{\partial t} = 
-\underbrace{\vphantom{\frac{z_j D_j F}{RT} \grad \cdot \parens{C_j \grad \phi_L}}
\uvec \cdot \grad C_j}_{\text{Advection}}
+ \underbrace{\vphantom{\frac{z_j D_j F}{RT} \grad \cdot \parens{C_j \grad \phi_L}}
D_j \grad^2 C_j}_{\text{Diffusion}} 
+ \underbrace{
\frac{z_j D_j F}{RT} \grad \cdot \parens{C_j \grad \phi_L}}_{\text{Electromigration}}
\;+ \underbrace{\vphantom{\frac{z_j D_j F}{RT} \grad \cdot \left( C_j \grad \phi_L \right)}
S_j }_{\text{Source}},
$}
\end{empheq}
where $C_j$ is the concentration of species $j$ in $\unit{\mole.\m^{-3}}$, 
$D_j$ is the diffusivity of species $j$ in $\unit{\m^{2}.\sec^{-1}}$, 
$z_j$ is the dimensionless charge number as an integer, 
$\phi_L$ is the electric potential in the liquid electrolyte in volts, 
and $S_j$ is the chemical source term, also in $\unit{\mole.\m^{-3}.\sec^{-1}}$
\citeb[Eq. 11.30]{NewmanBalsara21}.
The form of the equation shown in \eqref{eq:nernst-planck} can be derived from the more
common form with fluxes, e.g.  \citeb[Eq. 4.1.10]{BardFaulkner22},
by noting that ${\partial C_j}/{\partial t} = - \grad \cdot \mathbf{N}_j$,
where $\mathbf{N}_j$ is the flux of species $j$ in $\unit{\mole.\m^{-2}.\sec^{-1}}$.
The boundary conditions at the inlet include known concentrations for all the species.

We obtain the potential in the electrolyte $\phiL$ by solving a Poisson equation for $\phiL$
with a suitable adjustment for a solution where electroneutrality holds, \citeb[Eq. 9]{DR22},
\begin{empheq}[box=\fbox]{equation}
\label{eq:poisson-phi}
\kappa_L \grad^2 \phiL  = S_\phi - F \sum_{k} z_k D_k \grad^2 C_k,
\end{empheq}
where $S_\phi = \nel F S$ is the source term for electrons generated in the reaction.
\eqref{eq:poisson-phi} is analogous to Gauss's law in electrostatics $\grad \cdot E = \rho_c / (\epsilon_0 \epsilon_r)$
where $E = -\grad \phiL$ is the electric field; $\rho_c$ is the charge density; 
$\epsilon_0$ is the permittivity of a vacuum; and $\epsilon_r$ is the relative permittivity.
The ionic conductivity of the electrolyte $\kappa_L$ can be obtained as the sum of the conductivities
of each charged species in the electrolyte,
\begin{equation}
\label{eq:kappa}
\kappa_L = \frac{F^2}{RT} \sum_{k} z_k^2 D_k C_k.
\end{equation}
This is a standard result \citeb[Eq. (10)]{DR22}, and can be derived by substituting
for mobility $u_i$ in Newman and Balsara's discussion \citeb[Eq. 11.7]{NewmanBalsara21} 
with the formula $u_i= D_i / RT$ \citeb[Eq. 11.41]{NewmanBalsara21}.
In the model system, we explicitly simulate the concentration of three species, \ce{AQDS}, \ce{H_{2}AQDS}, and \ce{H^+}.
The fourth species in the simulation is \ce{HSO_4^{-}}, which is treated implicitly by imposing an electroneutrality condition,
\begin{equation}
\label{eq:electroneutrality}
\sum_{j}  z_j C_j = 0.
\end{equation}
We are making a modeling approximation by ignoring the second dissociation 
of \ce{HSO_4^-} into \ce{SO_4^{2-}} and \ce{H^+},
but in principle the approach here could be extended to include one more species explicitly.
The topics of electroneutrality and electric fields in dilute solutions are discussed extensively 
by Newman and Balsara \citeb[\S 11]{NewmanBalsara21}, and this treatment is consistent.
For the common case where a flow battery is operated with a supporting electrolyte, 
$\kappa_L$ will be dominated by the supporting electrolyte 
and can be treated as a constant $\kappa_0$ with minimal error.

The potential at the membrane is also a boundary condition, and in this work it is set to zero, 
i.e. $\phi_L = 0 $ on the bottom wall where $z=0$ \cite{DR22}.
The governing equations are invariant to a constant offset to the potential, 
so the only assumption made here is that the membrane potential is uniform.
We further assume a high electrical conductivity in the electrode,
and set the potential $\phiS$ to a constant that is specified as one of the operating conditions in the simulation.
We define the applied reducing potential $\Var$ so that a higher value of $\Var$ drives a higher state of charge at equilibrium,
and a setting of $\Var = 0$ corresponds to a state of charge of $1/2$ for a solution 
in contact with an electrode at this potential and a liquid electrolyte at zero potential.
These choices lead to the definition
\begin{empheq}[box=\fbox]{equation}
\label{eq:Var-def}
\phiS = \Eadj - \Var.
\end{empheq}
At the walls other than the membrane, the boundary condition on $\phiL$ is $\grad \phiL \cdot \hat{n} = 0$
where $\hat{n}$ is a normal vector pointing outward through the exterior wall.
This is justified because there cannot be any net charge flux moving through the walls other than the membrane.

\begin{figure}[ht]
\centering
\includegraphics[width=1.00\linewidth]{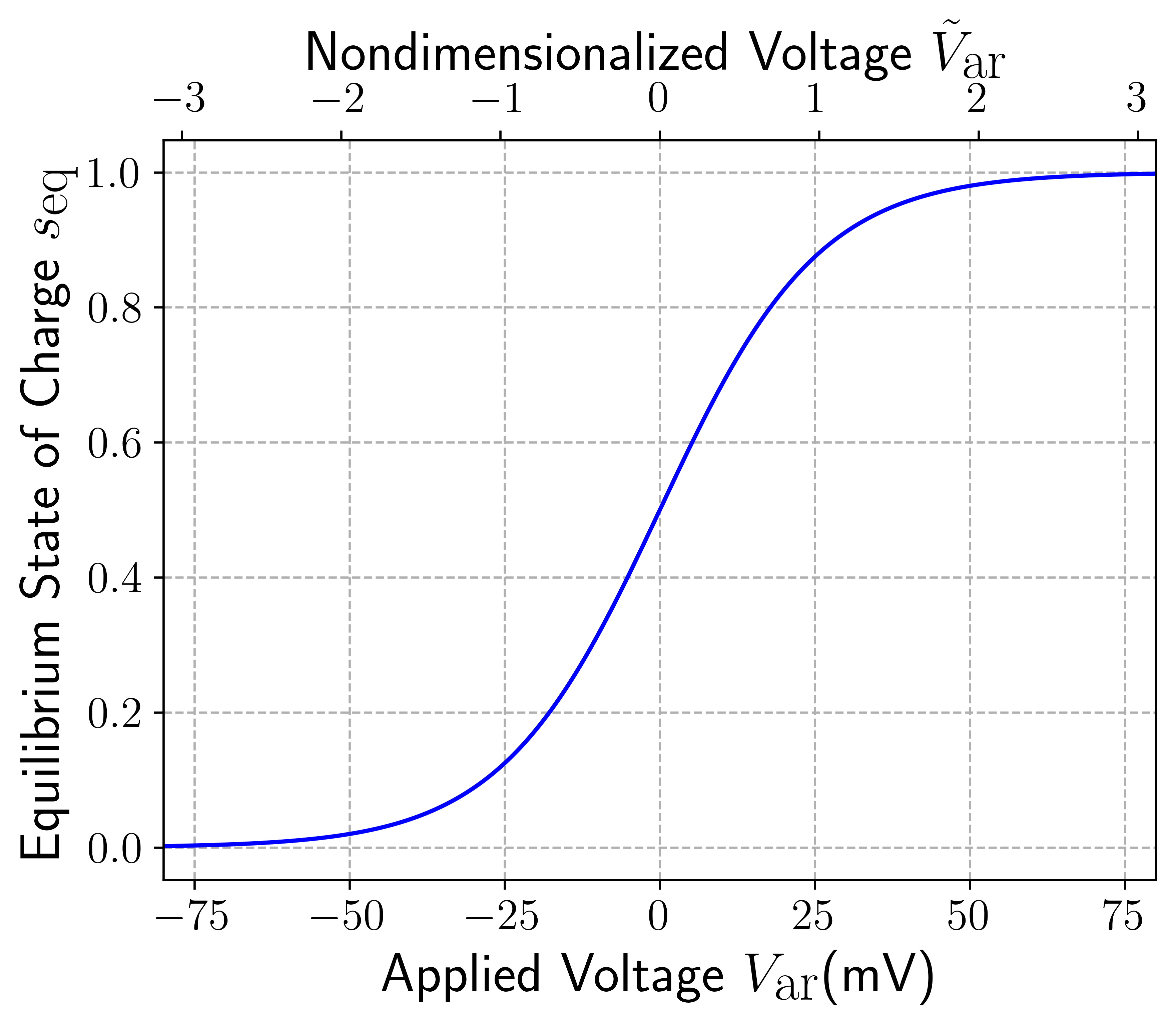}
\caption{
\textbf{Dependence of equilibrium SOC ($\seq$) on applied reducing voltage ($\Var$).}
Calculated from \eqref{eq:seq-from-Var} with $\phiL = 0$.
$\seq$ has a sigmoid shape like $\sigma(x)$ and is monotonically increasing in $\Var$.
}
\label{fig:seq-from-Var}
\end{figure}

We can verify that this definition of $\Var$ has the desired properties by solving for the equilibrium 
state of charge $\seq$ in \eqref{eq:nernst-simp} when $\Eeq$ is set to $\phiS$, 
in which case we obtain
$(\Eadj - \Var) - \phiL = \Eadj - \frac{V_T}{\nel} \log \left( \frac{\seq}{1-\seq}\right) $.
The adjusted standard potential $\Eadj$ cancels out, leaving
\begin{equation}
\label{eq:Var-from-seq}
\Var + \phiL =\frac{V_T}{\nel} \log \left( \frac{\seq}{1-\seq}\right).
\end{equation}
Define the dimensionless applied reducing voltage $\Vart$ 
and the dimensionless potential in the electrolyte $\phiLt$ by
\begin{equation}
\label{eq:Vart-def}
\Vart = \frac{\nel \Var}{V_T},
\quad
\phiLt = -\frac{\nel \phiL}{V_T}.
\end{equation}
The negative sign on $\phiLt$ is introduced for consistency with $\Vart$.
A lower value of $\phiL$ boundary cells corresponds to greater losses due to charge transport in the electrolyte.
A more conductive electrolyte (larger $\kappa_0$) will minimize this loss term.
The sigmoid function $\sigma(x) = 1 / (1 + e^{-x})$ used in machine learning
is the inverse of the logit function, $\logit(x) = \log(x / (1-x))$.
A straightforward simplification shows that the equilibrium state of charge in contact with the electrodes is given by
\begin{equation}
\label{eq:seq-from-Var}
\seq = \sigma(\Vart - \phiLt) = \left( 1 + e^{-\nel (\Var + \phiL) / V_T}\right)^{-1}.
\end{equation}
In physical terms, this definition of $\Var$ implies that when $\Var = 0$ and $\phiL = 0$, 
the state of charge for an electrolyte solution in equilibrium with an electrode is $1/2$, 
and the equilibrium state of charge $\seq$ increases monotonically with $\Var$ 
according to \eqref{eq:seq-from-Var} as shown in Figure~\ref{fig:seq-from-Var}.\\

We can extend this idea to see the dependence of the nondimensionalized reducing overpotential $\etat$
on the nondimensionalized applied reducing potential $\Vart$.
Start with \eqref{eq:etaact-def} and substitute for $\phiS$ using \eqref{eq:Var-def} 
and for $\Eeq$ with \eqref{eq:nernst-simp}. After cancelling out $\Eadj$ we find
$\etaact = \frac{V_T}{\nel} \log\left( \frac{s}{1-s}\right) - \Var - \phiL$.
Now nondimensionalize $\eta$, $\Var$ and $\phiL$ by multiplying the whole equation by $-\nel / V_T$,
substitute for $\etaact$ with \eqref{eq:etat-def} 
and finally substitute for $\Var$ and $\phiL$ with \eqref{eq:Vart-def} to obtain
\begin{empheq}[box=\fbox]{equation}
\label{eq:etat-from-s-bv}
\etat = \Vart - \phiLt - \log\left( \frac{s}{1-s}\right).
\end{empheq}
This expression for $\etat$ immediately shows the separate contributions from the applied reducing voltage 
$\Vart$, the electric potential $\phiLt$ in the electrolyte touching the electrode surface, and the local state of charge $s$.
When there is a strong supporting electrolyte, $\kappa_L$ will be large and the electric field $\phiL$ will be small
and dominated by $\Vart$. The dependence on $s$ is a form of mass transport overpotential.
As $s$ becomes higher on the reactive surface, a greater reducing voltage is necessary to overcome 
the unfavorable concentration ratio $s/(1-s)$ in the Nernst equation.
This effect can become quite important at high states of charge; 
to achieve $s=0.99$ for example, the Nernst term leads to 59 mV of mass transport overpotential.
\eqref{eq:seq-from-Var} for the equilibrium state of charge $\seq$ is immediately seen to be consistent with 
setting the overpotential to zero in \eqref{eq:etat-from-s-bv}.

The Butler-Volmer model in this work jointly solves the Nernst-Planck equation \pref{eq:nernst-planck} 
and the Poisson equation for potential \pref{eq:poisson-phi} with implicit time steps.
The chemical source term in \eqref{eq:nernst-planck} is evaluated using 
Butler-Volmer kinetics with \eqref{eq:source-term-bv},
and the overpotential $\etat$ is calculated from \eqref{eq:etat-from-s-bv}.
To obtain the steady state of a known fluid flow with a specified applied potential $\Var$,
we run time steps until convergence criteria are met for small changes in both the state of charge and potential.
The Butler-Volmer model uses adaptive time stepping because it has a nonlinear source term
that can vary significantly as the simulation evolves.
For that reason, we introduce a dimensionless measure of the relative change over a time step.
Let $\tau$ denote the nominal time for fluid to flow through the electrode; $\tau = V / Q$
where $V$ is the volume of the electrode in \unit{\m^3} and 
$Q$ is the volumetric flow rate in $\unit{\m^{3}.\sec^{-1}}$.
The convergence criteria for the Butler-Volmer model are
\begin{equation}
\label{eq:convergence-bv-change}
\begin{aligned}
\left(\tau / \Delta t \right) \norm{s^{(n+1)} - s^{(n)}} &< \varepsilon_{s} \\
\left(\tau / \Delta t \right) \norm{\phiL^{(n+1)} - \phiL^{(n)}} &< \varepsilon_{\phi} \norm{\phiL^{(n)}}.
\end{aligned}
\end{equation}
The norm used to calculate the size of the change for convergence criteria
is the root mean square, i.e. $\norm{x} = \norm{x}_{2} / \sqrt{N}$.

An additional convergence criterion may be introduced to ensure that electrochemical conversion 
and mass transport between the inlet and outlet are in balance.
Denote the flow of electrons carried by electroactive species at the inlet and outlet by
\begin{equation}
\label{eq:electron-flow}
\Fin = \nel \cTi \sum_{\text{inlet}} A_i u_i s_i ,\quad
\Fout = \nel \cTi \sum_{\text{outlet}} A_i u_i s_i,
\end{equation}
where $A_i$ is cross sectional area in the inlet or outlet plane of cell $i$ ($h^2$ for each cell), 
$u_i$ is the $x$ component of the flow velocity, and $s_i$ is the state of charge of cell $i$.
$\Fin$ and $\Fout$ are current-like quantities with units of \unit{\amp}.
The net current flow
\begin{equation}
\label{eq:electron-flow-net}
\Fnet = \Fout - \Fin
\end{equation}
is therefore the number of moles of electrons that are being added to the electrolyte per second.
At steady state, $\Fnet$ must be equal to the total current generated electrochemically on the electrode surface,
\begin{equation}
\label{eq:I-tot-def}
\Itot = \nel F \sum_{\text{elec}} V_i S_i.
\end{equation}
A tolerance $\varepsilon_{\text{I}}$ is set on the maximum relative difference 
between these currents which must be equal at steady state,
\begin{equation}
\label{eq:convergence-bv-flow}
\abs{I_\text{tot} - \mathcal{F}_\text{net}} < \varepsilon_{\text{I}} \Itot.
\end{equation}

The comparison of the total current and the flow of charge through the electrode
motivates the definition of an important figure of merit.
Define the utilization $U$ as the ratio of the current at steady state $\Itot$ 
to the maximum current that would be possible
if the electrolyte at the outlet were completely reduced \cite{WA20}, i.e.
\begin{equation}
\label{eq:utilization-def}
U = \frac{\Itot}{\nel \cTi F Q (1 - \socin)} 
= \frac{\displaystyle\sum_{\text{outlet}} A_i u_i s_i}{(1 - \socin)\displaystyle\sum_{\text{outlet}} A_i u_i}.
\end{equation}
The second expression for $U$ is shown as ratio of two sums over cells comprising the outlet in a discretized geometry.
It cancels $\nel F \cTi$ and uses the fact that the volumetric flow rate $Q$ 
is the sum of the flows out of each cell at the outlet, i.e.
$Q = \sum_{\text{out}}{A_i u_i}$.

\subsection{Electrochemical Reaction with Uniform Potential in Electrolyte - ``Simplified Butler-Volmer Model''}
\label{sec:rxn-sbv-model}
The Butler-Volmer model (\ref{sec:rxn-bv-model}) provides a rich description of an operating porous electrode,
but it requires solving coupled mass transport and charge transport equations, making it computationally demanding.
A number of simplifications can be made to this model without sacrificing its essential accuracy.
The biggest simplification is to disregard the electromigration term 
in the Nernst-Planck equation \pref{eq:nernst-planck} to obtain
\begin{equation}
\label{eq:nernst-planck-sbv}
\frac{\partial C_j}{\partial t} = - \uvec \cdot \grad C_j + D_j \grad^2 C_j + S_j, 
\end{equation}
thereby sidestepping the entire charge problem.
This turns out to be an excellent approximation in the model system because the supporting electrolyte is 
highly effective at moving charge to neutralize any standing electric field.
This fact does not come as a surprise, since that is precisely the engineering function served by the supporting electrolyte.

The nondimensionalized reducing overpotential $\etat$ is obtained 
from \eqref{eq:etat-from-s-bv} by ignoring $\phiLt$,
which is implicitly set to zero in this model, yielding
\begin{empheq}[box=\fbox]{equation}
\label{eq:etat-from-s}
\etat = \Vart - \log\left( \frac{s}{1-s}\right).
\end{empheq}
The source term $S$ does not change and is still given by \eqref{eq:source-term-bv}.

We now demonstrate that under relatively mild assumptions that are true in the model system,
this system can be further simplified to a mass transport problem in only the state of charge $s$.
Suppose that the diffusion coefficients $D_{\ce{O}}$ and $D_{\ce{R}}$ 
for the oxidized and reduced species are identical, and denote their common value by $D$.
This assumption holds for the model system, where the experimental values for \ce{AQDS} and \ce{H_{2}AQDS}
both match to the one significant figure available with $D = 4 \cdot 10^{-10} \; \unit{\m^2.\sec^{-1}}$,
and is close to holding for most redox pairs, which typically have two species of similar size and shape.
With this one assumption, the total concentration of redox active species $\cT$ will remain a uniform constant.
Since \eqref{eq:nernst-planck-sbv} is linear, we can sum the diffusion and advection terms 
for the oxidized and reduced species to get corresponding terms for $\cT$. 
The source term for $\cT$ is zero, because the reduction reaction 
converts \ce{O} to \ce{R} without changing the total concentration $\cT$; 
this assumption would break down if we were modeling degradation side reactions.
The resulting transport equation for $\cT$ is thus
\begin{equation}
\label{eq:nernst-planck-cT}
\frac{\partial \cT}{\partial t} = - \uvec \cdot \grad \cT + D \grad^2 \cT.
\end{equation}
The fluid at the inlet is uniformly mixed with $\cT = \cTi$, a known constant set at 20 mM in these simulations.
Evaluating \eqref{eq:nernst-planck-cT} when $\cT(x) = \cTi$ is spatially uniform, 
we find $\frac{\partial \cT}{\partial t} = 0$ at $t=0$, and $\cT$ does not change over time.
In fact, the diffusion term in \eqref{eq:nernst-planck-cT} also shows that any initial inbalance would even out 
over time as diffusion moved material from regions of high concentration to lower concentration. 
The advection term also tends to make $C_T$ spatially uniform as the uniformly mixed inlet propagates through the channel.
This argument can thus be extended to explain why $\cT$ is a uniform constant
and justifies the further simplification $\cT \approx \cTi$ in the simplified Butler-Volmer model.

Since $\cT$ is a constant, we can take the transport equation for the reduced species, divide it by $\cT$, 
and obtain the Nernst-Planck transport equation for the state of charge,
\begin{empheq}[box=\fbox]{equation}
\label{eq:nernst-planck-soc}
\frac{\partial s}{\partial t} = - \uvec \cdot \grad s + D \grad^2 s + 2 a k_0 \sqrt{s(1-s)} \sinh( \etat / 2) .
\end{empheq}
The source term for SOC on the right hand side is found by taking the source 
term $S$ in \eqref{eq:source-term-bv} and dividing by $\cT$.
An immediate consequence of \eqref{eq:nernst-planck-soc} is that for the dilute theory developed
here where chemical activities are assumed to match concentrations, 
not only is the total concentration $\cTi$ of redox species invariant, 
it does not have any effect on the state of charge distribution or utilization of an electrode at steady state. 
$\cTi$ is simply a scalar that factors out of the PDE to be solved for the state of charge at steady state.
The richer Butler-Volmer model will show a slight scaling up in $\phiL$ as $\cTi$ increases.
The convergence criteria for steady state are the same as those given in 
\eqref{eq:convergence-bv-change} and \eqref{eq:convergence-bv-flow},
except the potential $\phiL$ and its change are no longer applicable.

The right hand side of \eqref{eq:nernst-planck-soc} includes a source term
for the state of charge with units of \unit{\sec^{-1}}. Denote this by $\St = S / \cT$.
We can completely nondimensionalize the reaction rate by factoring out 
the specific area $a$ and the rate constant $k_0$ to obtain
\begin{equation}
\label{eq:rate-soc-eta}
\Rt = \frac{S}{a k_0 \cT}= 2 \sqrt{s(1-s)} \sinh(\etat / 2).
\end{equation}
We can simplify \eqref{eq:rate-soc-eta} to eliminate $\etat$ and express the reaction rate $\Rt$
using only the state of charge $s$ and the applied reducing voltage $\Vart$.
Substitute for $\etat$ using \eqref{eq:etat-from-s} and a straightforward simplification shows that
\begin{empheq}[box=\fbox]{equation}
\label{eq:rate-from-soc}
\Rt = \exp(\Vart/2) -  \braces{2 \cosh(\Vart/2)} s.
\end{empheq}
We can verify that \eqref{eq:rate-from-soc} is consistent with 
the equilibrium state of charge calculation in \eqref{eq:seq-from-Var} by solving $\St(\seq)=0$.

We can generalize \eqref{eq:rate-from-soc} to the case 
where the charge transfer coefficient $\alpha$ is not necessarily one half.
The source term in \eqref{eq:source-term-bv} generalizes to
\begin{equation}
\label{eq:source-term-bv-gen}
S = a k_0 \cT  s^{\alpha} (1-s)^{(1-\alpha)} \braces{e^{\alpha \etat} - e^{(\alpha - 1) \etat}}
\end{equation}
and the reaction rate in \eqref{eq:rate-from-soc} generalizes to 
\begin{empheq}[box=\fbox]{equation}
\label{eq:rate-from-soc-gen}
\Rt = e^{\alpha \Vart} \braces{1 - \parens{1+ e^{-\Vart}} s}.
\end{empheq}
We can verify the consistency of \eqref{eq:rate-from-soc-gen} with the 
special case $\alpha = 1/2$ to recover \eqref{eq:rate-from-soc}.
We can also see on inspection that $\Rt = 0$ when $s = \sigma(\Var) = \seq$.
Equation (\ref{eq:rate-from-soc-gen}) is a striking result.
It states that for any charge transfer coefficient $\alpha$, 
the reaction rate is an affine function of the state of charge, 
with a maximum source term $S(0) = a k_0 e^{\alpha \Vart}$.
The reaction rate $\Rt$ decreases linearly in $s$ with slope $e^{\alpha \Vart} + e^{(1-\alpha) \Vart}$.
This matches the physical intuition that a higher local state of charge leads to a lower reaction rate,
with a high enough SOC leading to a negative sign.
While this is a direct consequence of a simple calculation on the Butler-Volmer rate law,
it provides a different perspective on the relationship between the rate and the local state of charge
that abstracts away a complex nonlinear dependence on the activation overpotential $\etat$.
To our knowledge this is a novel result relating local reaction rates to local state of charge
that will apply whenever a battery system follows dilute solution thermodynamics and Butler-Volmer rate kinetics
and is operated at a controlled voltage.
\begin{figure}[ht]
\centering
\includegraphics[width=1.00\linewidth]{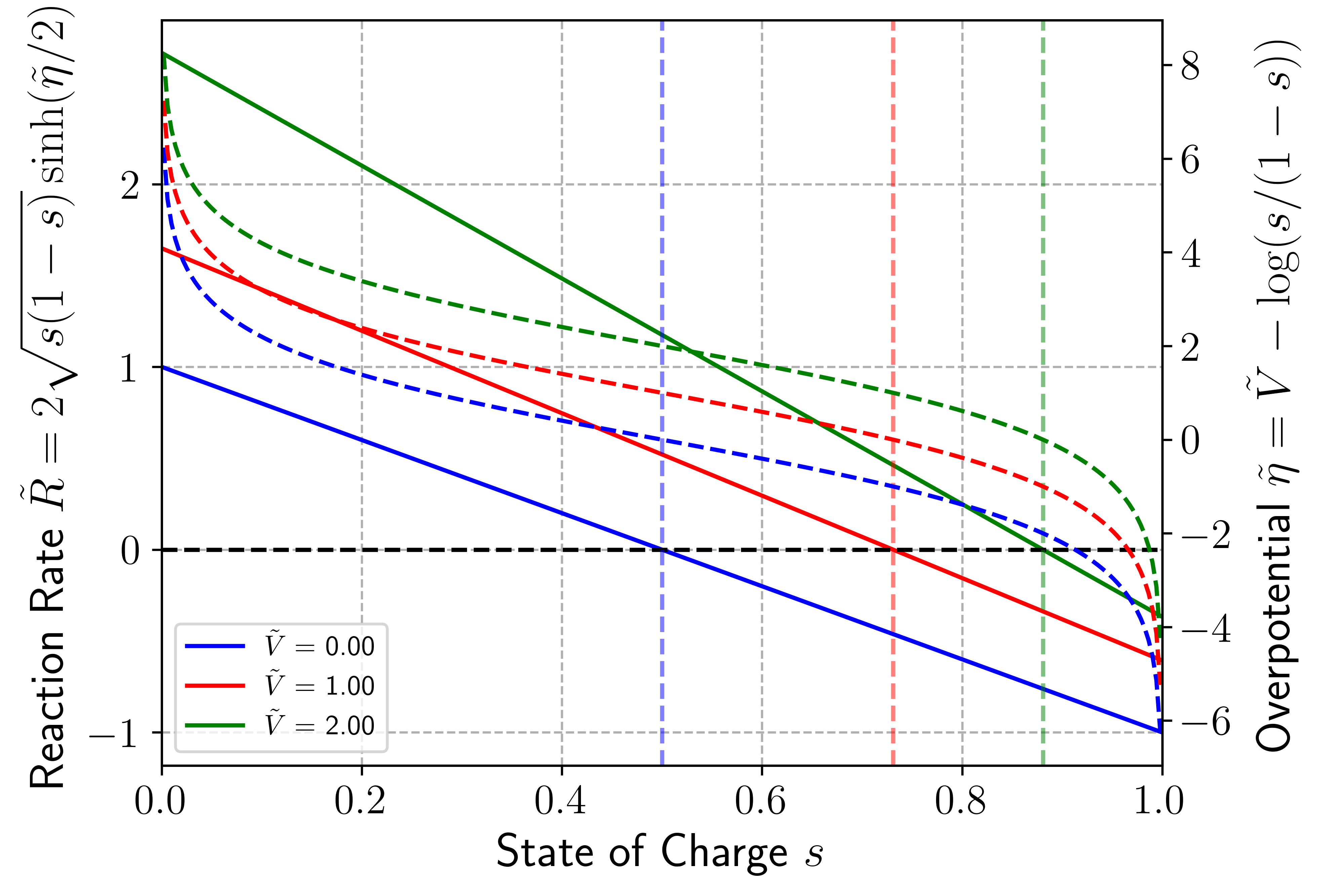}
\caption{
\textbf{Nondimensionalized reaction rate $\bm{\Rt}$ and overpotential $\bm{\etat}$ vs. the state of charge $\bm{s}$.}
The reaction rate (solid lines) is an affine function of $s$. 
The overpotential (dashed lines) has a sigmoid shape.
The horizontal line at $\Rt =0$ corresponds to the equilibrium condition.
The vertical lines indicate the equilibrium SOC at each applied voltage.
Increasing voltage drives faster rates and higher equilibrium SOC.
}
\label{fig:rate-vs-soc}
\end{figure}

\subsection{Reaction Steady State from Nernst Equilibrium Boundary Condition - ``Nernst Model''}
\label{sec:rxn-nernst-model}
In this section we introduce a second model to obtain the steady state of the 
simplified Butler Volmer model shown in \ref{sec:rxn-sbv-model}
with less computational work.
The obvious way to use the PDE \eqref{eq:nernst-planck-soc} to find the steady state 
is to time step the system until steady state is achieved.
This can require many time steps, and the presence of a nonlinear source term implies that the 
time steps may be significantly shorter than the CFL criterion alone would imply.
To motivate this model, consider the limiting case where the kinetic rate $k_0$ became infinitely fast.
In this limit, the state of charge on the electrode surface would equal its equilibrium value $\seq$ 
implied by \eqref{eq:seq-from-Var}, since any departure would lead to an infinitely fast source term to neutralize it.
In the fast kinetic limit, the Nernst-Planck equation for the state of charge at steady state therefore simplifies 
to advection and diffusion terms on the interior with a Dirichlet boundary condition on the electrode surface:
\begin{empheq}[left=\empheqlbrace]{align}
\label{eq:np-fast-interior}
\uvec \cdot \grad s  &= D \grad^2 s & \text{(electrolyte interior)} \\
\label{eq:np-fast-boundary}
s &= \sigma(\Vart) & \text{(electrode surface).}
\end{empheq}
\eqref{eq:np-fast-interior} on the interior is obtained from \eqref{eq:nernst-planck-soc} by omitting the reaction term, which is zero away from the electrode surface, and using the fact that $\partial s / \partial t = 0$ at steady state.
This system can be solved more straightforwardly than the original Nernst-Planck equation for the state of charge 
\eqref{eq:nernst-planck-soc} because there is no longer a nonlinear source term.
It is a linear PDE and in can be solved efficiently by either explicit time stepping or with a direct linear solver.
A further simplification is that all of the boundary cells are controlled 
by the Dirichlet boundary condition in \eqref{eq:np-fast-boundary},
so the PDE to be solved in \eqref{eq:np-fast-interior} is confined to regular cubic cells,
allowing faster and simpler explicit methods to be used.

With this idea in mind, consider the same PDE in \eqref{eq:nernst-planck-soc}, 
but this time at steady state with a finite rate constant $k_0$.
The SOC at steady state $s$ is an unknown scalar field, and there is an overpotential $\etat$ associated with it.
At this steady state, the three terms on the right hand side of \eqref{eq:nernst-planck-soc}
must sum to zero on the boundary cells.
The advection term vanishes because of the no-slip condition.
The diffusion and source terms must cancel, with the chemical source producing reduced species 
that is transported away from the electrode surface by diffusion, so
\begin{equation}
\label{eq:nernst-model-elec}
2 a k_0 \sqrt{s(1-s)} \sinh(\etat / 2) = - D \grad^2 s.
\end{equation}

Observe that because \eqref{eq:nernst-model-elec} includes the specific area $a$, 
it is implicitly defined only for a small test region, e.g. one cell in a finite volume method.
We can also write a local version of \eqref{eq:nernst-model-elec} 
that applies on the boundary itself without any averaging over cells,
by setting the source flux equal to the diffusive flux,
\begin{equation}
\label{eq:nernst-model-flux}
2 k_0 \sqrt{s(1-s)} \sinh(\etat / 2) = -D \grad s \cdot \hat{n},
\end{equation}
where $\hat{n}$ is a normal pointing out of the surface of the electrode.
In \eqref{eq:nernst-model-flux}, the specific area term is dropped from \eqref{eq:nernst-model-elec}
because this an equality of fluxes, and the diffusive flux on the right hand side 
is now given by Fick's first law, $J = - D \grad s$.

We can invert \eqref{eq:nernst-model-elec} to solve for $\etat$ in terms of $\grad^2 s$.
This amounts to an instantaneous estimate of $\etat$ that is consistent 
with the current rate of mass transport away from the reactive surface.
Denote this new estimate $\etat_I$, with
\begin{empheq}[box=\fbox]{equation}
\label{eq:nernst-eta-inst-def}
\etat_I = 2 \sinh^{-1} \left( \frac{- D \grad^2 s}{2 a k_0 \sqrt{s(1-s)}} \right).
\end{empheq}
On the other hand, $\etat$ can be mapped monotonically to $s$ by inverting \eqref{eq:etat-from-s}:
\begin{equation}
\label{eq:s-from-etat}
s(\etat) = \sigma( \Vart - \etat).
\end{equation}
Putting all of these ideas together, the following coupled PDE reformulates the steady state problem as a Dirichlet boundary condition with an auxiliary variable for the overpotential,
\begin{empheq}[box=\fbox]{align}
\label{eq:np-dirichlet-interior}
\uvec \cdot \grad s  &= D \grad^2 s & \text{(interior)}\\
\label{eq:np-dirichlet-boundary}
s &= \sigma(\Vart - \etat) & \text{(electrode)}\\
\label{eq:np-dirichlet-overpot}
\sinh({\etat}/{2}) &= - \parens{\frac{D/2 ak_0}{\sqrt{s(1-s)}}} \grad^2 s & \text{(electrode).}
\end{empheq}

This problem formulation leads to the Nernst model, which jointly builds consistent estimates of $s$ and $\etat$
by updating an iterative estimate of the overpotential $\etat$ as a moving average of $\etat_I$.
This is done by replacing $\etat$ with a convex combination of its previous value and the new estimate $\etat_I$,
with the weighting of the moving average controlled by a relaxation parameter 
$\omega$ via \eqref{eq:nernst-model-eta-update}.
The value of $s$ on the electrode surface is a Dirichlet boundary condition from \eqref{eq:s-from-etat} 
evaluated with the overpotential estimate $\etat$ as in \eqref{eq:nernst-model-s-update}.
The entire update for the Nernst model is therefore given by
\begin{empheq}[box=\fbox]{align}
\label{eq:nernst-model-eta-update}
\etat^{(n+1)} &= (1-\omega) \etat^{(n)} + \omega \etat_I(s^{(n)}), \\
\label{eq:nernst-model-s-update}
s^{(n+1)} &= \sigma\left(\Vart - \etat^{(n+1)}\right).
\end{empheq}
The Nernst model can be initialized with $s^{(0)}$ matching the value at the inlet and $\etat^{(0)} = 0$.

We can combine the ideas of the fast kinetic limit and and the utilization \eqref{eq:utilization-def}
to define a figure of merit for an electrode's flow velocity field: the utilization in the mass transport limit $\umt$.
Let $\smt$ denote the state of charge field that solves the Nernst-Planck equation in the fast
kinetic limit (Eqs.~\ref{eq:np-fast-interior}-\ref{eq:np-fast-boundary})
where the boundary conditions are $\smt = 0$ at the inlet and $\smt = 1$ on the electrode surface.
Then we define the mass transport utilization by substituting $\smt$ into \eqref{eq:utilization-def} to obtain
\begin{empheq}[box=\fbox]{equation}
\label{eq:umt-def}
\umt = \left.\left({\sum_{\text{outlet}} A_i u_i \brackets{\smt}_i}\right) \right/ \left({\sum_{\text{outlet}} A_i u_i}\right).
\end{empheq}
$\umt$ abstracts out any dependence on the reaction kinetics or the applied voltage.
It establishes an upper bound on the utilization that could be achieved in an operating electrode
run at a high reducing potential and measures how effectively an electrode geometry 
and its velocity field accomplish mass transport.
Algorithm \ref{alg:umt} shows how to efficiently compute $\umt$.

\section{Computational Methods}
\label{sec:methods}
This work is built using the rincflo code of Dussi and Rycroft \cite{DR22} 
and extends from two dimensions to three.
Rincflo, in turn, employed the incflo code developed by Sverdrup et al.~\cite{SAN19} for incompressible flow problems.
Both of these codes are built using the AMReX framework of Zhang et al.~\cite{AMReX19}.
AMReX uses a finite volume approach for solving PDEs and includes facilities for constructing
problem geometries using an embedded boundary (EB) description of solid / liquid boundaries.
AMReX refers to a cell containing both solid and liquid as a \textit{cut cell}.
AMReX can do adaptive mesh refinement, allowing a geometric discretization 
to focus computing resources on the areas of greatest importance.
We use a static refined mesh, refining cells that are within one cell of a cut cell.

The top (coarsest) level of the mesh contains the whole domain with no refinement and is indexed by $0$.
Cells are always cubes, with aspect ratios equal to one.
At the top level, denote by $h$ the cell size and the number of cells along each axis by $(N_x, N_y, N_z)$,
so the dimensions of the box containing the problem domain are $(L_x, L_y, L_z) = (N_x h, N_y h, N_z h)$.
Each level of refinement models a subset of the previous level, with twice the resolution,
so the cell size on level $k$ is $h_k = h / 2^k$.

\begin{figure}[ht]
\centering
\includegraphics[width=0.80\linewidth]{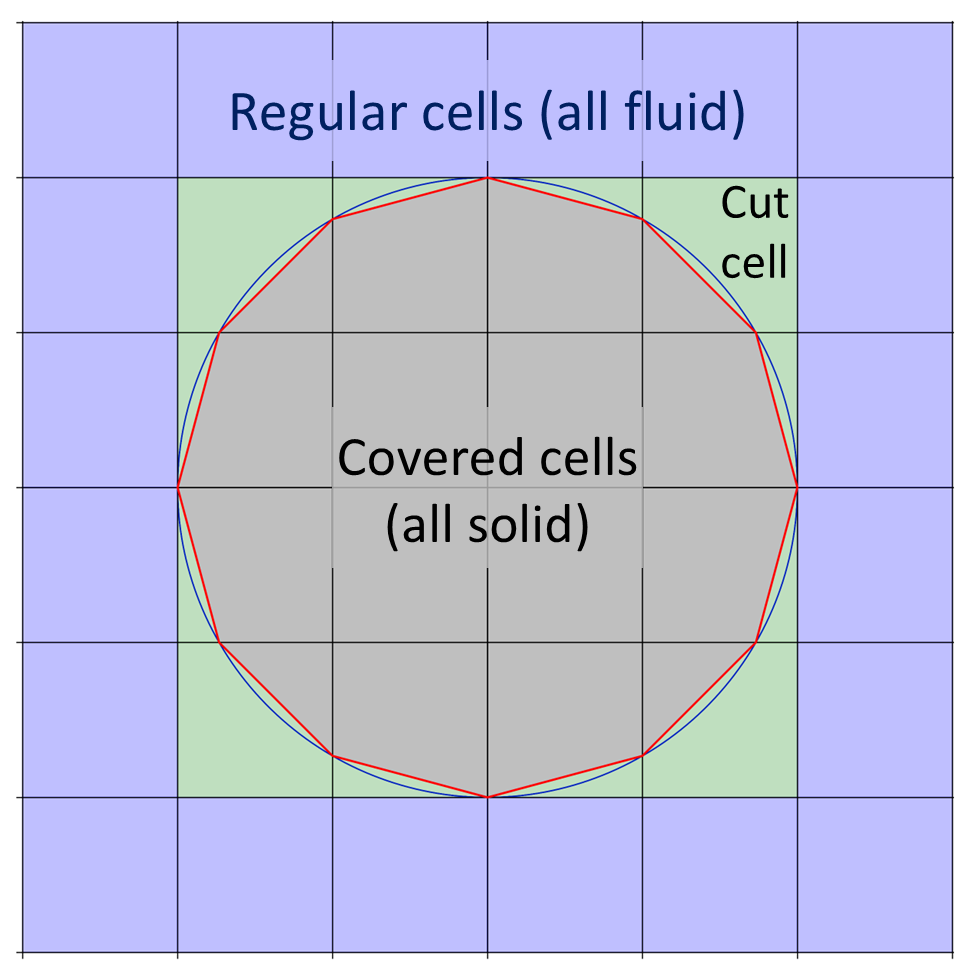}
\caption{
\textbf{Embedded boundary approach in AMReX.}
The domain is split into a grid of cubic cells, shown in black lines. 
Regular cells contain only fluid and covered cells (all solid) are disregarded. 
The electrode surface is composed of cut cells, 
which are modeled using an implicit function and its normal vector.
The exact solid/fluid boundary is shown as a \textcolor{Blue}{blue} circle,
while its approximation from normal vectors is shown as set of \textcolor{Red}{red} chords.
}
\label{fig:cut-cells}
\end{figure}

Figure \ref{fig:cut-cells} illustrates the AMReX approach to describing embedded boundaries with cut cells.
The key abstraction is to treat each cut cell as having a plane that separates it into solid and liquid phases.
This plane is built by calculating the normal vector from a real valued implicit function 
which follows the convention that positive values represent points outside the solid,
negative values are inside the solid, and zero values represent the surface.
The implicit function must be differentiable on the boundary so a normal vector can be constructed.
The canonical choice for the implicit function is a signed distance, e.g. $f(\pvec) = \norm{\pvec - \mathbf{c_0}}_2 - R$ for a sphere with center $\mathbf{c_0}$ and radius $R$. The implicit function for cylinders, the only primitive shape used in this work,
follows a similar idea and is the distance from an arbitrary point to the nearest point on the cylinder's symmetry axis.
Figure \ref{fig:cut-cells} depicts the embedded boundaries in red as matching up on cell faces.
In general, this will \textit{not} be exactly true when the implicit function is approximated to first order about the cell center.
However, when the grid resolution is adequate, the boundaries will be close,
and in any case the discretized linear operators do not assume overlapping boundaries.
Depending on the signs of the implicit function at the corners of a cell, 
it is not always possible to cleanly cut the cell with a single plane.
AMReX calls such cells \textit{multiply cut}, and because they are inconsistent with the conceptual geometry model,
they pose significant challenges.

AMReX provides facilities for constructing linear operators and for solving linear equations of the general form
$(\alpha A - \beta \grad \cdot B \grad) \psi = f$, where $\alpha$ and $\beta$ are known scalar constants,
$A$ and $B$ are known scalar fields, $\psi$ is the unknown scalar field to be solved for, 
and $f$ is a known scalar field on the right hand side.
AMReX includes built in solvers for equations of this type 
as well as an interface to delegate their solution to the hypre library \cite{hypre}.
Following Dussi and Rycroft, we use the built-in geometric multigrid solver 
for diagonally dominated linear systems where $\alpha \ne 0$.
We solve equations where $\alpha = 0$ using the generalized minimal residual method (GMRES)
\cite{TrefethenBau97, Demmel97}
and an algebraic multigrid preconditioner from the hypre library called BoomerAMG.
AMReX is built on an MPI back end and can be used with modern supercomputing facilities for massively parallel processing.

\subsection{Incompressible Fluid Flow}
\label{sec:flow-methods}
Fluid flows along the $x$-axis as shown in Figure \ref{fig:boundary-cond}
with an applied pressure drop from the inlet to the outlet,
and no-slip conditions on the exterior walls and on cut cells comprising the electrode surface.
Experience modeling diverse geometries %
suggests that a simulation is more likely to run successfully 
when the constituent simple shapes do not intersect on more than a tangent line of contact.
Complex geometries with three dimensional overlapping volumes between primitive shapes 
can be processed by AMReX, but they lead to large numbers of multiply cut cells 
and often resulted in convergence problems in the multi- level multi-grid linear solver.
We calculate the steady state fluid flow by taking time steps until a convergence criterion is met,
\begin{equation}
\label{eq:flow-cnvg}
\norm{\uvec^{(n+1)} - \uvec^{(n)}} < \varepsilon_\text{flow} \norm{\uvec^{(n)}},
\end{equation}
where $\norm{\uvec}$ is the $\ell_2$ norm and 
$\varepsilon_\text{flow}$ is a dimensionless tolerance for the flow to converge
that was set to $10^{-9}$ in the flow simulations presented in this work.

Eqs.~(\ref{eq:navier-stokes}) and (\ref{eq:incompressibility})
are jointly solved using the Chorin projection method \cite{Chorin68}. 
The key idea is to advance the discretized velocity $\uvec^{(n)}$ one time step $\Delta t$ 
to $\uvec^{(n+1)}$ by an application of the incompressibility condition to construct a Poisson equation for the new discretized pressure $p^{(n+1)}$. First compute an intermediate velocity $\uvec^{*}$ that accounts for the momentum advection and viscosity terms,
\begin{equation}
\label{eq:cp1}
\frac{\uvec^{*} - \uvec^{(n)}}{\Delta t} = 
-\left( \uvec^{(n)}  \cdot \grad \right) \uvec^{(n)} + \left(\frac{\mu}{\rho}\right) \grad^2 \uvec^{(n)}.
\end{equation}
$\uvec^{*}$ can be obtained either explicitly or implicitly using the discretized 
gradient and Laplacian operators for the problem geometry.
To obtain the velocity $\uvec^{(n+1)}$ at the next time step, we solve 
\begin{equation}
\label{eq:cp2}
\frac{\uvec^{(n+1)} - \uvec^{*}}{\Delta t} = - \frac{\grad p^{(n+1)}}{\rho}
\end{equation}
for the new pressure $p^{(n+1)}$, which is not yet known.
Observe that if we add Eqs.~(\ref{eq:cp1}) and (\ref{eq:cp2}), 
the terms with $\uvec^{*}$ cancel and we recover the discretized analog of 
\eqref{eq:navier-stokes} once it is rearranged so the term 
$\frac{\partial \uvec}{\partial t}$ is isolated on the left hand side. 
Next take the divergence of both sides of \eqref{eq:cp2} to obtain
\begin{equation}
\label{eq:cp3}
\grad \cdot \uvec^{(n+1)} - \grad \cdot \uvec^{*} = - \left(\frac{\Delta t}{\rho}\right) \grad^2 p^{(n+1)}.
\end{equation}
The velocity field at the end of the time step $\uvec^{(n+1)}$ is known to be divergence free by the incompressibility
condition \eqref{eq:incompressibility}, leaving us to solve Poisson's equation for the new pressure,
\begin{equation}
\label{eq:cp4}
 \grad^2 p^{(n+1)} = \left(\frac{\rho}{\Delta t}\right) \grad \cdot \uvec^{*}.
\end{equation}
Once $p^{(n+1)}$ is known, \eqref{eq:cp2} can be easily inverted to directly solve for $\uvec^{(n+1)}$.
The time step $\Delta t$ is derived from the Courant-Friedrichs-Lewy (CFL) condition \cite{CFL67}
and is updated each flow step. A dimensionless constant $\ccfl$ 
is selected as one of the tuned simulation parameters, 
typically $\ccfl = 1/2$. 
The time step $\Delta t$ is then computed following the approach of Sverdrup et al.~\citeb[Eqs. 7-9]{SAN19}.
$\Delta t$ includes terms for advection $C_A$, viscous forces $C_V$ and external forces $C_F$,
\begin{equation}
\label{eq:time-step-flow}
\Delta t = \half \ccfl \left( C_A + C_V + \sqrt{(C_A + C_V)^2 + 4C_F^2}\right)^{-1}.
\end{equation}
The dimensions of $C_A$, $C_F$ and $C_V$ are all in \unit{\sec^{-1}}.
The advective term is
\begin{equation}
\label{eq:time-step-flow-adv}
C_A = \max_{\Omega} \left( \frac{\abs{u}}{\Delta x} + \frac{\abs{v}}{\Delta y} + \frac{\abs{w}}{\Delta z} \right),
\end{equation}
where $u_x$, $u_y$ and $u_z$ are the three spatial components of the flow velocity $\uvec$
and $\Delta x$, $\Delta y$ and $\Delta z$ are the grid spacing of each cell, 
and $\max_{\Omega}$ is a maximum over all the cells in the geometry.
The viscous term for Newtonian flows is given by
\begin{equation}
\label{eq:time-step-flow-viscous}
C_V = \left(\frac{2 \mu}{\rho}\right)\max_{\Omega}
{\left( \frac{1}{\Delta x^2} + \frac{1}{\Delta y^2} + \frac{1}{\Delta z^2} \right)}
\end{equation} 
and the forcing term is
\begin{equation}
\label{eq:time-step-flow-force}
C_F = \max_{\Omega} \left(\sqrt{\frac{\abs{f_x}}{\Delta x} + \frac{\abs{f_y}}{\Delta y} + \frac{\abs{f_z}}{\Delta z}} \right),
\end{equation}
where $\mathbf{f} = (f_x, f_y, f_z)$ is the acceleration due to external forces acting on each cell,
which in this work is limited to the pressure gradient term, $\mathbf{f} = - \grad p / \rho$.

Algorithm \ref{alg:flow-sim} summarizes how a flow simulation is run to steady state.
\begin{algorithm}[H]
\caption{Flow Simulation}
\label{alg:flow-sim}
\begin{algorithmic}[1]

\Function{CalcFlowTimeStep}{$\uvec^{(n)}, p^{(n)}, C_{CFL}$}
\State Calculate $C_C$, $C_V$, $C_F$ from 
Eqs.~(\ref{eq:time-step-flow-adv}), (\ref{eq:time-step-flow-viscous}), and (\ref{eq:time-step-flow-force}).
\State \Return $\half \ccfl \left( C_A + C_V + \sqrt{(C_A + C_V)^2 + 4C_F^2}\right)^{-1}$
\label{alg:flow-sim:time-step}
\EndFunction

\Procedure{FlowStep}{$\uvec^{(n)}, p^{(n)}, \text{ExplicitDiffusion}$}
\State ApplyBoundaryCondition($\partial \Omega$)
\State $\Delta t := \text{CalcTimeStep}(\uvec^{(n)}, p^{(n)}, C_{CFL})$
\If{ExplicitDiffusion}
	\State $\uvec^{*} := 
	\uvec^{n} - \Delta t \left( \uvec^{(n)} \cdot \grad \uvec^{(n)} 
	+ (\mu  / \rho)\grad^2 \uvec^{(n)} \right)$
	\label{alg:flow-sim:u-star-explicit}
\Else 
	\State \textbf{Solve} $(1- (\mu \Delta t / \rho)\grad^2) \textcolor{blue}{\uvec^{*}} = \\
	\uvec^{n} - \Delta t \left( \uvec^{(n)} \cdot \grad \uvec^{(n)} + \rho^{-1} \grad p^{(n)} \right)$
	\label{alg:flow-sim:u-star-linear-system}
	\State $\uvec^{*} := \uvec^{*} + (\Delta t / \rho) \grad p^{(n)}$
	\label{alg:flow-sim:u-star-implicit}
\EndIf
\State \textbf{Solve} $\grad^2 \textcolor{blue}{p^{(n+1)}} = (\rho / \Delta t) \grad \cdot \uvec^{*}$
\label{alg:flow-sim:p-linear-system}
\State $\uvec^{(n+1)} := \uvec^{*} - (\rho^{-1} \Delta t) \grad p^{(n+1)}$
\EndProcedure

\Procedure{FlowSim}{$\uvec^{(0)}, p^{(0)}, \text{ExplicitDiffusion}$}
\State $n := 0$
\Repeat
\State $\text{FlowStep}(\uvec^{(n)}, p^{(n)}, \text{ExplicitDiffusion})$
\State $\Delta \uvec := \uvec^{(n+1)} - \uvec^{(n)}$
\State $n := n + 1$
\Until{$\norm{\Delta \uvec} < \varepsilon_\textrm{flow} \norm{\uvec^{(n)}}$}
\EndProcedure

\end{algorithmic}
\end{algorithm}

The expression for the time step in \lineref{alg:flow-sim:time-step} comes from \eqref{eq:time-step-flow}.
The explicit calculation of $\uvec^{*}$ in \lineref{alg:flow-sim:u-star-explicit} matches \eqref{eq:cp1} 
after it is rearranged so $\uvec^{*}$ appears on the right.
The implicit calculation of $\uvec^{*}$ in \lineref{alg:flow-sim:u-star-implicit}
is obtained by similar reasoning, except the viscous term is treated implicitly.
This corresponds to replacing the momentum diffusion term 
$\grad^2 \uvec^{(n)}$ with $\grad^2 \uvec^{*}$ and then solving the resulting linear equation.
The linear system in \lineref{alg:flow-sim:u-star-linear-system} is solved using the built-in multigrid solver.
The lagged pressure $p^{(n)}$ is used in this step and then added back;
since $p^{(n)}$ solves the Laplace equation on the previous time, 
its gradient is also divergence free and \lineref{alg:flow-sim:u-star-implicit} 
retains the incompressibility condition on $\uvec^*$.
The linear system for the new pressure in \lineref{alg:flow-sim:p-linear-system} 
matches \eqref{eq:cp4} and is solved using hypre.
The unknown in these linear systems is shown in \textcolor{blue}{blue} typeface.
The density $\rho$ and the viscosity $\mu$ are treated as constant scalars in this work.
Note that this algorithm is modeling advection explicitly and 
solving for the viscous term (momentum diffusion) either explicitly or implicitly as a parameter.
Sverdrup et al. advise against using explicit momentum diffusion with cut cells in incflo \cite{SAN19}.
The elliptic equation for the pressure must always be solved implicitly with this approach.

The electrode is modeled as a solid occupying the union of a collection 
of simple convex shapes such as cylinders and spheres.
Multiply cut cells pose a significant problem for Algorithm \ref{alg:flow-sim}.
The default AMReX geometric multigrid solver relies on building coarse grids by coarsening,
and that procedure breaks down when the geometry contains multiply cut cells.
The simulations presented by Dussi and Rycroft \cite{DR22} 
were done in two dimensions with well separated circular electrode fibers.
Those geometries did not produce multiply cut cells because a circle is a convex shape and
the separation between circles was large enough that two of them never occupied the same grid cell.

When running a simulation in three dimensions, there are many more configurations of the 
sign pattern for the eight corners of a cube leading to a multiply cut cell than in two dimensions,
where only 2 of 16 patterns fail.
Most plausible three dimensional electrode geometries contain contact points 
between convex primitive shapes, such as stacked cylinders in the ``logpile'' geometry simulated here.
The complex geometries simulated with Barber et al.~\cite{BEE24} contained three dimensional overlapping
regions between pairs of filaments whose center lines were less than a diameter apart.
This led to large numbers of multiply cut cells and posed substantial convergence challenges.
One of the significant changes from Dussi and Rycroft's original incflo code
is a revised treatment of the parameters related to the handling of these multiply cut cells.
In particular, we switched the mode used by AMReX to build the coarse levels
in its geometric multigrid solver, changing from the default mode that builds a coarse level 
by geometrically coarsening a finer level.

We introduce a novel technique in this work to obtain the steady state flow for a 
refined mesh which we term \textit{iterative upsampling}.
The idea behind iterative upsampling is to first solve the entire system to steady state 
without any mesh refinement, i.e. using an unrefined domain with side length $h$.
The next step is to solve the the problem with one level of mesh refinement.
Every cell that is either cut, or within one cell of a cut cell (by Manhattan distance) will be refined.
This simulation is initialized with the converged steady state velocity $\uvec^{(0)}$ from the previous step.
It is simple to do the initialization using the built-in facilities in AMReX for interpolating from coarse to fine levels.
This one idea can dramatically speed up convergence to steady state when using even a single level of mesh refinement.
A second advantage is that refined geometries initialized with a warm start in this way 
are less likely to have convergence failures on the first time step than problems initialized with a cold start.

Iterative upsampling can be seen as an extension of the multigrid method.
The multigrid method is more commonly used to solve a single time step in a PDE.
In this case, when we solve the entire incompressible flow problem to its converged steady state at a 
coarse resolution, it is analogous to the smoothing step in the traditional multigrid method.
The upsampling of the coarse solution to the refined grid is likewise analogous to interpolation in multigrid.
Heath gives a brief introduction of multigrid methods for linear systems \citeb[\S 11.5.7]{Heath02}
while classic extended treatments are given by Hackbusch \cite{Hackbusch85} 
and Wesseling \cite{Wesseling95}, the latter being freely available.
Previous development of cascadic multigrid methods 
by Bornemann and Deuflhard \cite{BD96} and Pan et al.~\cite{PHH17} take a one-shot approach from coarse to fine.
Chow and Tsitliklis applied one-way multigrid methods to stochastic control problems \cite{CT91},
proceeding ``one way'' from coarse to fine grids.

\subsection{Direct Reaction Models (Butler-Volmer and Simplified Butler-Volmer)}
\label{sec:rxn-bv-methods}
The main idea of the Butler-Volmer reaction model is to
use a fixed point iteration to find a jointly consistent potential $\phiL$ and concentration $C$ at the end of the time step.
The concentration is initialized from the previous step with explicit advection.
On each step in the fixed point iteration, the latest estimates for the concentration and
potential are used as inputs for all the terms in the Nernst-Planck equation \pref{eq:nernst-planck}.
Diffusion is treated implicitly, and the concentration is updated by 
putting the electromigration and chemical source terms on the right hand side.
The potential is then updated implicitly by solving the Poisson equation \pref{eq:poisson-phi}
where the right hand side uses the latest estimates of the ionic flux and electron source terms.
The fixed point iteration can optionally use a relaxation parameter $\omega$
to control the rate that updates are applied.
The simulations shown here all use $\omega =0$, i.e. no relaxation.

The specific area $a$ of a cut cell is the ratio of surface area to volume, $a = A / V$,
where $A$ denotes the surface area of the embedded boundary 
(i.e. the plane that cuts the cell between the solid and liquid regions) 
and $V$ is the volume of the liquid part of the cell.
$a$ has units of \unit{\m^{-1}} and scales as one over the fiber diameter in physical electrodes,
and as $1/ h$ multiplied by a dimensionless parameter of order 1 
describing the local geometry in a single cut cell of side length $h$.

A small change from \cite{DR22} is clipping the range of $s$ and $\etat$ into admissible limits during time steps.
The equilibrium state of charge $\seq$ defined in \eqref{eq:seq-from-Var} 
is the maximum value that would be possible at steady state 
for a charging flow battery where the inlet concentration $\socin$ is smaller than $\seq$.
The overpotential is also clipped to a range controlled by the parameter $\etamax$.
The maximum overpotential was typically set at 100-200 mV, 
corresponding to $\etamax$ ranging between 3.9 and 7.8.
When $\etamax$ is set properly, it will not be a binding constraint at steady state,
but limiting the overpotential makes the early time steps better behaved.
These constraints are imposed at each step to mitigate problems arising from highly nonlinear 
behavior in the source term when $s$ is near 0 or 1.
Algorithm \ref{alg:rxn-bv-fixed-point} summarizes the 
calculation of the source term and one step of the fixed point iteration.

\begin{algorithm}[H]
\caption{Butler-Volmer Fixed Point Iteration}
\label{alg:rxn-bv-fixed-point}
\begin{algorithmic}[1]

\Procedure{CalcSource}{$C, \phiL$}
\State $s:= \text{Clip}(\cR / (\cO+\cR), \socmin, \socmax)$
\State $\etat := \Vart + (\nel \phiL / V_T)  - \log(s / (1-s))$
\label{alg:bvfp:etat}
\State $\etat := \text{Clip}(\etat, -\etat_\text{max}, \etat_\text{max})$
\State $S := 2a \cT k_0 \sqrt{s(1-s)} \sinh( \etat / 2)$
\label{alg:bvfp:source}
\State $\St := S / \cT$
\EndProcedure

\Procedure{FixedPointStepBV}{$C^*, C^{(k)}, \phiL^{(k)}, \Delta t$}
\State CalcSource($C^{(k)}, \phiL^{(k)}$)
\State $\varepsilon_j := (z_j D_j F / RT) \grad \cdot (C^{(k)}_j \grad \phiL^{(k)})$
\label{alg:bvfp:electromigration}
\State $\kappa_L :=  (F^2 / RT) \sum_{j} z_j^2 D_j C^{(k)}_j$
\label{alg:bvfp:kappa}
\State $F_I := F \sum_{j} Z_j D_j \grad^2 C^{(k)}_j$
\label{alg:bvfp:ionic-flux}
\State $S_\phi := \nel F S $
\State \textbf{Solve} $(1 - \Delta t D_j \grad^2) \textcolor{blue}{C_j^{(k+1)}} = 
C_j^{*} + \Delta t \left( \varepsilon_j + n_j S \right)$
\label{alg:bvfp:linear-system-conc}
\State \textbf{Solve} $ \left( \grad \cdot (\kappa_L \grad) \right) \textcolor{blue}{\phiL^{(k+1)}} = S_\phi - F_I $
\label{alg:bvfp:linear-system-phi}
\State $C_j^{(k+1)} := (1-\omega) C_j^{(k+1)} + \omega C_j^{(k)}$
\State $\phiL^{(k+1)} := (1-\omega) \phiL^{(k+1)} + \omega \phiL^{(k)}$
\EndProcedure

\end{algorithmic}
\end{algorithm}

The dimensionless reducing overpotential $\etat$ on \lineref{alg:bvfp:etat} comes from \eqref{eq:etat-from-s}.
The source term on \lineref{alg:bvfp:source} comes from \eqref{eq:source-term-bv}.
The more general form with a charge transfer coefficient $\alpha \ne 1/2$ 
is also supported in the code; the symmetric version is shown here for clarity.
$S$ and $\etat$ need only be calculated on the cut cells.
The electromigration term on \lineref{alg:bvfp:electromigration} comes from \eqref{eq:nernst-planck}.
The conductivity $\kappa_L$ on \lineref{alg:bvfp:kappa} is from \eqref{eq:kappa}.
There is an option to treat $\kappa_L$ as a constant during the fixed point iteration
by calculating it from the calling routine. This can be done to save time and has minimal effect on the results.
The ionic flux term $F_I$ on \lineref{alg:bvfp:ionic-flux} comes from \eqref{eq:poisson-phi}.
The linear system in \lineref{alg:bvfp:linear-system-conc} 
treats diffusion implicitly and electromigration and the chemical source term explicitly.

The Butler-Volmer model takes a single time step by repeating the fixed point iteration
until convergence criteria for a small change in both the state of charge and potential are met.
It is shown in Algorithm \ref{alg:rxn-step-bv}.

\begin{algorithm}[H]
\caption{Butler-Volmer Time Step}
\label{alg:rxn-step-bv}
\begin{algorithmic}[1]

\Procedure{ReactionStepBV}{$C^{(n)}, \phiL^{(n)}, \thetamrc$}
\State ApplyBoundaryCondition($\partial \Omega, \socin$)
\State CalcSource($C^{(n)}, \phiL^{(n)}$)
\State $\Delta t := \text{CalcReactionTimeStep}(\theta_{mrc} )$
\State $C^* := C^{(n)} - \Delta t \left( \uvec \cdot \grad C \right)$
\State $k := 0$
\Repeat
\State $\text{FixedPointStepBV}(C^*, C^{(n,k)}, \phiL^{(n,k)}, \Delta t)$
\State $\Delta s := s^{(k+1)} - s^{(k)}$, $\Delta \phi := \phi^{(k+1)} - \phi^{(k)}$
\State $k := k+1$
\Until{$\norm{\Delta s} < \varepsilon_s$ and $\norm{\Delta \phi} < \varepsilon_\phi$}
\State $C^{(n+1)} = C^{(n,k)}$, $\phiL^{(n+1)} = \phiL^{(n,k)}$
\EndProcedure

\end{algorithmic}
\end{algorithm}

The Butler-Volmer model runs time steps until convergence criteria 
for small changes in the SOC and potential are met.
It is shown in Algorithm \ref{alg:rxn-sim-bv}.
This time, the changes are over a full reaction step rather than in the fixed point iteration 
as in Algorithm \ref{alg:rxn-step-bv}.
The introduction of adaptive time stepping is a small but important change from 
\cite{DR22} that has allowed simulations to converge without prohibitively short fixed time steps.
These were a problem especially at the early stages of a simulation 
initialized to a uniform and very low state of charge. 

\begin{algorithm}[H]
\caption{Butler-Volmer Simulation}
\label{alg:rxn-sim-bv}
\begin{algorithmic}[1]

\Function{CalcReactionTimeStep}{$s,  \St, \thetamrc, \Delta t_\text{flow}$}
\State $\Delta t_\text{react} := \theta_\text{mrc} \cdot \min_{\Omega} \{\max\left( (1-s) / \St, -s / \St \right)\}$
\State \Return $\min ( \Delta t_\text{react}, \Delta t_\text{flow})$
\EndFunction

\Procedure{ReactionSimBV}{$C^{(0)}, \phiL^{(0)}$}
\State $\Delta t_\text{flow} := \text{CalcFlowTimeStep}(\uvec, p)$
\State $\thetamrc := 1/256$
\State $n := 0$
\Repeat
\State $\text{ReactionStepBV}(C^{(n)}, \phiL^{(n)})$
\If{Successful}
	\State $\thetamrc := \text{Clip}(\thetamrc \cdot 2^{1/1024}, \thetamin, \thetamax)$
	\State $\Delta s := s^{(n+1)} - s^{(n)}$, $\Delta \phi := \phiL^{(n+1)} - \phiL^{(n)}$
\Else
	\State $\thetamrc := \text{Clip}(\thetamrc / 2, \thetamin, \thetamax)$
	\State \text{RollBackReactionStep()}
\EndIf
\Until{$\left(\tau / \Delta t\right) \norm{\Delta s} < \varepsilon_s$ and 
$\left(\tau / \Delta t\right) \Delta \phi <  \varepsilon_\phi \norm{\phiL}$}
\EndProcedure

\end{algorithmic}
\end{algorithm}

The method used to compute $\Delta t_\text{react}$ ensures that the reaction 
doesn't consume too large a large fraction of either the oxidized or reduced species.
$\St$ is the source term for the state of charge in \unit{\sec^{-1}}, 
so $(1-s)/\St$ is the amount of time it would take for the chemical reaction to fully 
deplete the oxidized species when $S$ is positive in the absence of the diffusion term.
Conversely, $-s / \St$ is the amount of time it would take to deplete the reduced species if $S$ is negative.
The prefactor $\theta_\text{mrc}$ controls the largest fraction of either \ce{O} and \ce{R} 
that would be consumed by the reaction term in any one step.
The suffix ``mrc'' stands for maximum reactant consumption.
The reaction time step can never be longer than the flow time step because of the CFL condition,
so this constraint is applied separately.

The parameter $\thetamrc$ is initially set at a low, cautious value, e.g. $\thetamrc = 1/256$.
A failed time step may be due to the problem being too nonlinear over the selected $\Delta t$.
After a successful time step, $\thetamrc$ is slightly increased by a ratio slightly larger than one, $2^{1/1024}$.
After a failed time step, $\thetamrc$ is cut in half and the concentrations and potential are rolled back.
Notwithstanding the above rules to tune $\thetamrc$, it is always kept in a 
band between $\thetamin$ and $\thetamax$, 
which were set at $1/1024$ and $1/16$ respectively for the simulations run in this work.
In practice, we often observed very short time steps at the beginning of the simulation,
but as the reaction slows down due to a buildup of reduced species, 
the time steps become comparable to the flow time step.

The criteria for steady state convergence include a factor $(\tau / \Delta t)$,
where $\tau$ is the mean time for fluid to traverse the electrode as in \eqref{eq:convergence-bv-change}.
This term makes the convergence criterion indifferent to the current time step.
It can be interpreted as testing whether the rate of change in the state of charge 
per unit of time to flow across the channel is small.
This is a more stringent criterion than testing the change over an individual time step,
so suitable thresholds for $\varepsilon_s$ and $\varepsilon_\phi$ are higher.
Different simulations in this work used settings in the range of $10^{-4}$ to $10^{-2}$.

The simplified Butler-Volmer model is analogous to the full Butler-Volmer model,
except it disregards the potential in the electrolyte $\phiL$.
Mathematically, this is equivalent to setting $\phiL = 0$.
Algorithm \ref{alg:rxn-sbv} shows the fixed point iteration 
and a full time step in the simplified Butler-Volmer model.
The full simulation to steady state is essentially identical to Algorithm \ref{alg:rxn-sim-bv}
except for the omission of $\phiL$, and it is thus omitted here.

\begin{algorithm}[H]
\caption{Simplified Butler-Volmer Model}
\label{alg:rxn-sbv}
\begin{algorithmic}[1]

\Procedure{FixedPointStepSBV}{$C^*, C^{(k)}, \Delta t$}
\State CalcSource($C^{(k)}, \phiL = 0$)
\State \textbf{Solve} $(1 - \Delta t D_j \grad^2) \textcolor{blue}{C_j^{(k+1)}} = C_j^{*} + \Delta t \left( n_j S \right)$
\State $C_j^{(k+1)} := (1-\omega) C_j^{(k+1)} + \omega C_j^{(k)}$
\EndProcedure

\Procedure{ReactionStepSBV}{$C^{(n)}, \thetamrc$}
\State ApplyBoundaryCondition($\partial \Omega, \socin$)
\State CalcSource($C^{(n)}, \phiL = 0$)
\State $\Delta t := \text{CalcReactionTimeStep}(\theta_{mrc} )$
\State $C^* := C^{(n)} - \Delta t \left( \uvec \cdot \grad C \right)$
\Repeat
\State $\text{FixedPointStepSBV}(C^{(n,k)})$
\State $\Delta s := s^{(k+1)} - s^{(k)}$
\Until{$\norm{\Delta s} < \varepsilon_s$}
\State $C^{(n+1)} = C^{(n,k)}$
\EndProcedure

\end{algorithmic}
\end{algorithm}

\subsection{Steady State Reaction Model (Nernst)}
\label{sec:rxn-nernst-methods}
Unlike the BV and SBV models shown in the previous section,
the Nernst model does not take time steps per se.
It instead takes iterative steps that move closer to steady state, 
though they are analogous to time steps and have a time-like parameter $\Delta t$.
The assumptions underlying the Nernst model are identical to those of the simplified Butler-Volmer model.
They should both converge to the same steady state solution.
A step in the Nernst model begins by imposing boundary conditions at both the inlet
and on the electrode surface using the latest overpotential estimate from \eqref{eq:nernst-model-s-update}.
Advection and diffusion are then calculated.
The change in the state of charge on the electrode surface due to advection and diffusion
is then used to calculate a new estimate of the instantaneous overpotential according to \eqref{eq:nernst-eta-inst-def}.
The term $D \grad^2 S$ is replaced by $- \Delta \selec / \Delta t$ 
because this is the average rate that the SOC is decreasing on this time step due to mass transport.
The overpotential $\eta$ is calculated as a moving average of $\eta_I$ according to \eqref{eq:nernst-model-eta-update}.
The Nernst model is summarized in Algorithm \ref{alg:rxn-nernst}.
\begin{algorithm}[H]
\caption{Nernst Model}
\label{alg:rxn-nernst}
\begin{algorithmic}[1]

\Procedure{NernstStep}{$s^{(n)}, \etat^{(n)}, \Delta t$}
\State $\socin^{(n)} = \socin$
\State $\selec^{(n)} = 1 / \left(1 + \exp(- \Vart + \etat^{(n)}) \right)$
\State $s^{*} = s^{(n)} - \Delta t (\uvec \cdot \grad s^{(n)})$
\State \textbf{Solve}$\left(1 - \Delta t D \grad^2\right) \textcolor{blue}{s^{(n+1)}}  = s^{*}$
\State $\Delta \selec = \selec^{(n+1)} - \selec^{(n)}$
\State $\etat_I = 2 \sinh^{-1} 
\left[ -\left( \frac{\Delta \selec}{\Delta t}\right)   \left/   \left(2 a k_0 \sqrt{s^{(n)} (1 - s^{(n)})}\right)  \right. \right] $
\State $\etat^{(n+1)} = (1 - \omega) \etat^{(n)} + \omega \etat^{(n+1)}$
\EndProcedure

\Procedure{NernstSim}{$s^{(0)} = \socin, \etat^{(0)} = 0$}
\State $\Delta t := \text{CalcFlowTimeStep}(\uvec, p)$
\State $n := 0$
\Repeat
\State $\text{NernstStep}(s^{(n)}, \etat^{(n)})$
\State $ \Delta s = s^{(n+1)} - s^{(n)}$
\State $n := n+1$
\Until{$\left(\tau / \Delta t\right) \norm{\Delta s} < \varepsilon_s$}
\EndProcedure

\end{algorithmic}
\end{algorithm}

Numerical experiments have verified that the Nernst model converges to consistent 
results with the simplified Butler-Volmer model and that faster convergence is indeed possible.
The simplified models can still be used to advantage even if the desired output is a full Butler-Volmer model at steady state.
We introduce the technique of \textit{model refinement}.
The idea is simple but powerful: run a simpler but faster model to steady state and 
use the resulting state to initialize a richer but slower model.
The comparison showing consistent results and faster convergence with 
both simpler models and model refinement is in \ref{sec:results-rxn-models}.

To close this section, we note that it is possible to reformulate Algorithm \ref{alg:rxn-nernst} 
so that is uses only explicit calculations.
We have done a proof of concept implementation in numpy which is sufficient
to show that this technique works well.
The first step is to construct explicit sparse matrix representations for the advection and diffusion operators.
It is straightforward to define advection on interior cells; the only subtlelty is to use upwinding.
For cells on the boundary of the domain (either walls or the electrode surface) 
the no-slip condition implies that advection vanishes.
The diffusion operator is equally simple on interior cells.
For cells on one of the walls, we use a one-sided second order stencil 
$f''(x) \approx [2 f(x) - 5 f(x+h) + 4f(x+2h) - f(x+3h)] / h^2$.

The mass transport utilization $\umt$ defined in \eqref{eq:umt-def} can be calculated explicitly
using advection and diffusion operators defined only on interior cells 
because the state of charge on the boundary cells is a Dirichlet boundary condition.
Thus we have a simple and efficient method to compute $\umt$ explicitly in Algorithm \ref{alg:umt}.

\begin{algorithm}[H]
\caption{SOC at Mass Transport Limit}
\label{alg:umt}
\begin{algorithmic}[1]
\Procedure{SocMT}{$\uvec$, $\Omega$}
\State $\Delta t$ := CalcFlowTimeStep($\uvec, p=0$)
\State $\amat$ := MakeAdvectionOpInt($\Omega, \Delta t$)
\State $\dmat$ := MakeDiffusionOpInt($\Omega, D, \Delta t$)
\State $\fmat$ := $\amat + \dmat$
\State $\smat_0$ := 0, $\smat_0$[\text{elec}] := 1
\State n:= 0
\Repeat
\State $i:= n \% 2, j:= (n+1) \% 2$
\State $\smat_{j} := \fmat \smat_{i}$ 
\State $\smat_{j}[\text{elec}] := 1$
\State $\Delta \smat := \smat_{j} - \smat_{i}$
\State $n := n+1$
\Until{ $ (\tau / \Delta t) \norm{\Delta s} < \varepsilon_s$ } 
\EndProcedure
\end{algorithmic}
\end{algorithm}

We can also formulate a variation of the Nernst model in Algorithm \ref{alg:rxn-nernst} 
using only explicit operations.
A precise treatment of diffusion on the cut cells 
requires intricate computations as discussed by Sverdrup et al.~\cite[\S III.B.1]{SAN19},
but it is possible to quickly obtain a good approximation 
by imposing the requirement that the diffusion operator 
conserves mass when it operates on interior cells.
Let $i$ and $j$ denote the indices of two neighboring cells in the fluid and boundary, respectively,
and $A$ be a generic operator acting only on interior cells.
Then $A$ can be extended to have a mass-conserving action on boundary cells by the formulas
$A_{ji} = -A_{ij} \cdot V_i / V_j$ and $A_{jj} = - \sum_{i \ne j} A_{ji}$.
Defining advection and diffusion according to this recipe leads to conservative operators
that disregard the interactions between neighboring cut cells but are otherwise correct.

The procedure described above constructs explicit sparse $N \times N$ matrices for advection ($\amat$) and diffusion ($\dmat$).
The flow operator matrix $\fmat = \amat + \dmat$ will have on the order of $7 N$ entries; 
each interior cell will have a diagonal entry plus at most six entries for its neighbors.
We can combine the ideas of Algorithms \ref{alg:rxn-nernst} and \ref{alg:umt}
into Algorithm \ref{alg:rxn-nernst-explicit} for the explicit Nernst model.

\begin{algorithm}[H]
\caption{Explicit Nernst Model}
\label{alg:rxn-nernst-explicit}
\begin{algorithmic}[1]
\Procedure{NernstExplicit}{$\uvec, \Omega, \Vart$}
\State $\Delta t := \text{CalcFlowTimeStep}(\uvec, p=0)$
\State $\amat := \text{MakeAdvectionOpFull}(\Omega, \Delta t)$
\State $\dmat := \text{MakeDiffusionOpFull}(\Omega, $D$, \Delta t)$
\State $\fmat  := \amat + \dmat$
\State $\smat_0 := 0, \smat_0[\text{elec}] := 1$
\State $\etamat := 0, \etamat_I := 0$
\State n:= 0
\Repeat
\State $i:= n \% 2, j:= (n+1) \% 2$
\State $\smat_{j} := \fmat \smat_{i}$ 
\State $\Delta_\text{elec}  := \smat_{j}[\text{elec}] - \smat_{i}[\text{elec}]$
\State $\etamat_I := 2 \sinh^{-1}[ - (\Delta_\text{elec} / \Delta t) / (2 a k_0 \sqrt{\smat_{i} (1-\smat_{i})})]$
\State $\smat_{j}[\text{elec}] := 1$
\State $\Delta := \smat_{j} - \smat_{i}$
\State $n := n+1$
\Until{ $ (\tau / \Delta t) \norm{\Delta s} < \varepsilon_s$ } 
\EndProcedure
\end{algorithmic}
\end{algorithm}

Numerical experiments have demonstrated that the explicit Nernst model
produces consistent results with the implicit Nernst model and runs significantly faster.

\section{Results and Discussion - Fluid Flow}
\label{sec:results-flow}
We begin with a benchmark simulation of a logpile geometry 
that is an idealization of a woven design.
We then demonstrate convergence of the flow simulations without considering mesh refinement.
Once convergence is established, we further demonstrate the consistency of 
flow simulations done with and without mesh refinement.
We demonstrate the efficacy of iterative upsampling and show that it leads 
to consistent results with faster convergence.
We close by examining the effect of varying the applied pressure over five orders of magnitude.
We see a clear transition between a Stokes flow regime at lower pressures 
and the emergence of turbulence with a concomitant increase in both hydraulic resistance and advective mixing.

\subsection{Incompressible Flow Simulation for a Logpile Lattice}
\label{sec:results-flow-logpile}
\begin{figure}[ht]
\centering
\includegraphics[width=1.00\linewidth]{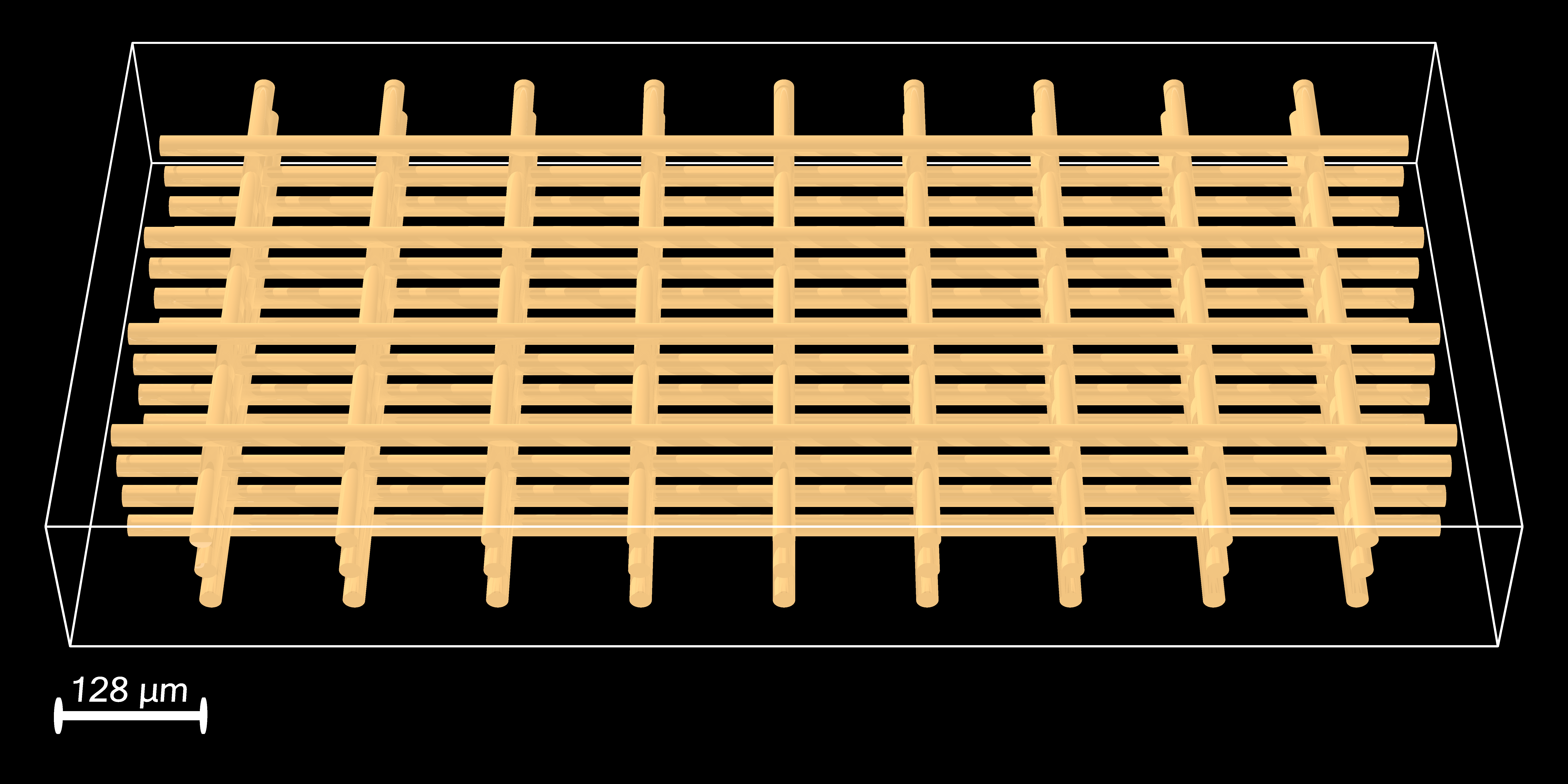}
\caption{\textbf{Logpile lattice geometry}}
\label{fig:logpile}
\end{figure}

Our benchmark geometry is a logpile lattice of dimensions 
\qtyproduct{1280 x 640 x 160}{\um} as shown in Figure \ref{fig:logpile}.
It contains 43 cylindrical rods representing the electrode in seven layers.
The layers are \qty{20}{\um} apart on the $z$-axis, at heights of 20, 40, \ldots 140~\unit{\um}.
The odd layers are oriented along the $x$-axis and the even layers are oriented along the $y$-axis.
The diameter of each rod is \qty{20}{\um}.
The cylinders are \qty{128}{\um} apart in each plane, 
leaving four cylinders in each x-oriented layer and nine cylinders in each y-oriented layer.
We simulated this system with an applied pressure of \qty{100}{\pa} at the inlet.
Our reference simulation was done at a high resolution without any mesh refinement.
We set a spatial resolution $h = \qty{1.25}{\um}$, leading to a discretized geometry 
of \numproduct{1024 x 512 x 128} cells and a total of $N = 2^{26} \approx \num{6.71d7}$ grid cells.
We ran this simulation until it converged with $\varepsilon_\text{flow} = 10^{-9}$ as per \eqref{eq:flow-cnvg}.
The simulation took 2094 time steps to converge. 
The average time step took 170 seconds running on a single server with 192 CPU cores.
The entire job took about 99h to complete.
These steps evolved the system through 9.08 ms.
The predicted volumetric flow rate was 3.45 mL / h.

\begin{figure*}[ht]
\centering
\includegraphics[width=1.00\linewidth]{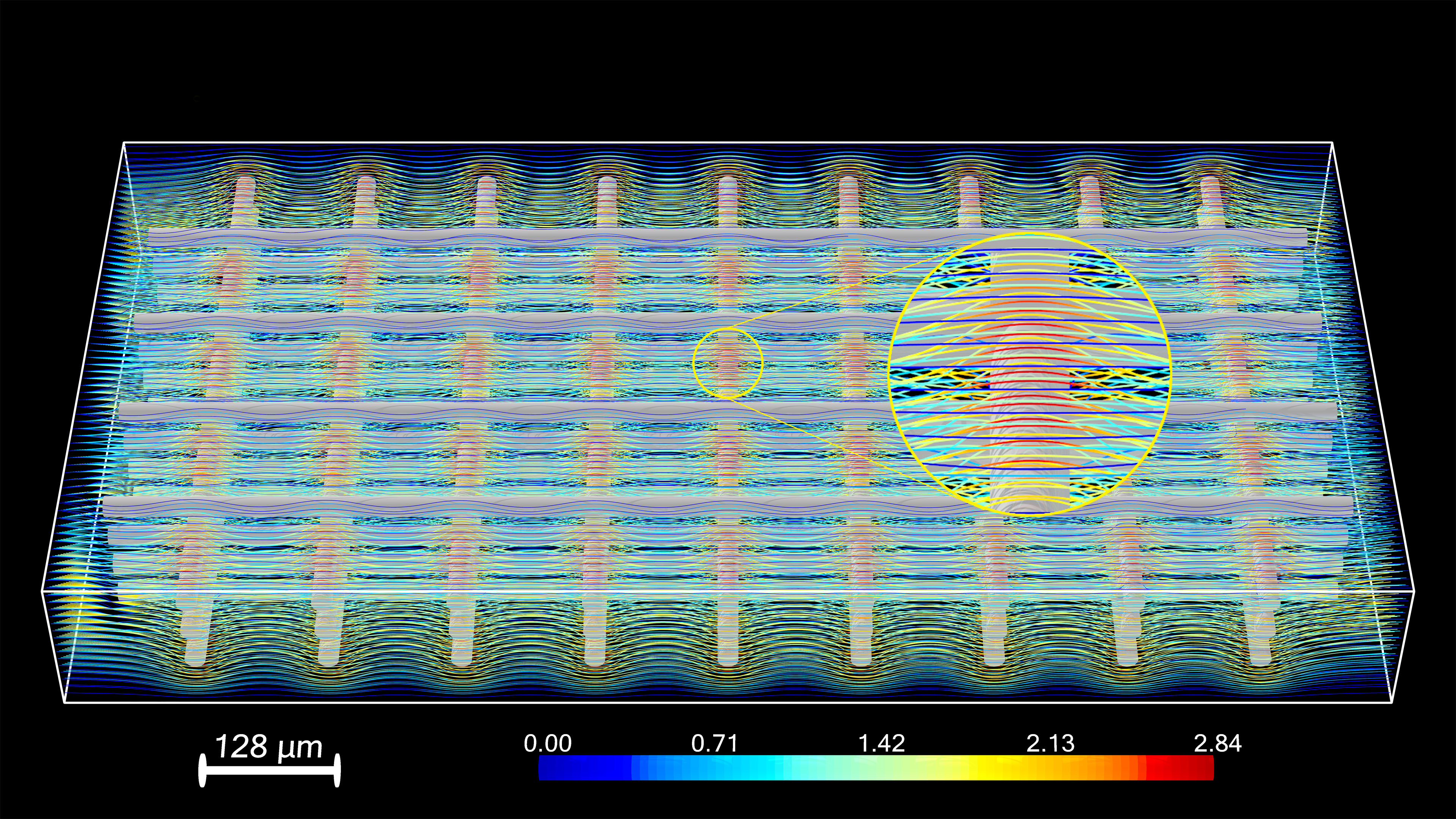}
\caption{
\textbf{Streamlines of steady state flow on logpile geometry with 100 Pa pressure at inlet.}
Inset region is zoomed four times to demonstrate the finest resolution of \qty{1.25}{\um}.
Scalebar indicates speed in \unit{\cm.\sec^{-1}}.}
\label{fig:streamlines}
\end{figure*}

Figure \ref{fig:streamlines} presents a 3D rendering of the integrated streamlines on this flow.
Streamlines were integrated using Ralston's method and evaluating the fluid velocity
at an arbitrary point by linear interpolation against the values at the cell centers from the flow simulation.
As fluid moves from left (inlet) to right (outlet), we can see the streamlines
moving over or under transverse rods and hitting a higher speed as it does.

\subsection{Convergence of Flow Simulations}
\label{sec:results-flow-convergence}
We demonstrate the convergence of our method by simulating three geometries
at a series of different spatial resolutions.
The three geometries are a single rod; two rods; and the logpile in Figure \ref{fig:logpile}.
For each of these geometries, we ran a reference simulation with $n_z = 120$ cells in the $z$-axis,
i.e. spatial resolution of $h = (4/3)~\unit{\um}$ and a geometry with \numproduct{960 x 480 x 120} cells.
When comparing the results of a coarse mesh against a fine mesh,
the only way to avoid interpolation of some kind is for the spacing on the coarse 
mesh to be an integer multiple of the spacing on the fine mesh.
With this in mind, we additionally attempted to simulate each geometry on a series of
coarse meshes with $n_z$ taking on all the proper divisors of $120$ between $5$ and $60$.
In a handful of cases, these simulations either failed to converge or produced anomalous results
due to an unlucky alignment of cut cells with an electrode filament. 
These outlier points were manually excluded from consideration.

For each geometry and coarse resolution with $n_z < 120$, we compute a relative error 
in the velocity field using the $\ell_2$ norm by comparing the coarse resolution 
with the reference simulation done at $n_z = 120$.
It is not immediately obvious how to compare two discretizations of a field on a coarse and a fine mesh. 
Let $\varphi^c$ and $\varphi^f$ denote a generic scalar field on a coarse and fine mesh, respectively,
with refinement ratio $r = n^f_{z} / n^c_{z}$ between the meshes.
We compute the difference by upsampling both fields to the fine resolution, i.e.
$[\varphi^f - \varphi^c]_{i,j,k} = \varphi^f_{i,j,k} - \varphi^c_{i/r, j/r, k/r}$.
A second subtlelty in the convergence analysis is that in the absence of a known analytical solution,
if we treat the reference numerical simulation as the exact solution,
we may introduce a bias to the estimated convergence rate that is artificially flattering.
This bias becomes larger as the coarse resolution gets closer to the reference resolution of $n_z = 120$.
With this in mind, we estimate the convergence rate with the functional form
\begin{equation}
\label{eq:flow-convergence-fit}
\varepsilon(h) = C h^p (1 - \alpha r^p),
\end{equation}
where $h = 1 / n_z$ is the dimensionless step size in the test simulation;
$r = h_f / h_c$ is the ratio of the fine step size to the coarse step size; 
and $\alpha$ is a tunable parameter that is inferred from the data to control the size of the discretization effect.
This power law fit for estimating flow convergence follows the approach of Rycroft et al.
in establishing convergence for the reference map technique for simulating solid / fluid interactions
of elastically deformable solids \cite{RWK20}.
\begin{figure}[ht]
\centering
\includegraphics[width=1.00\linewidth]{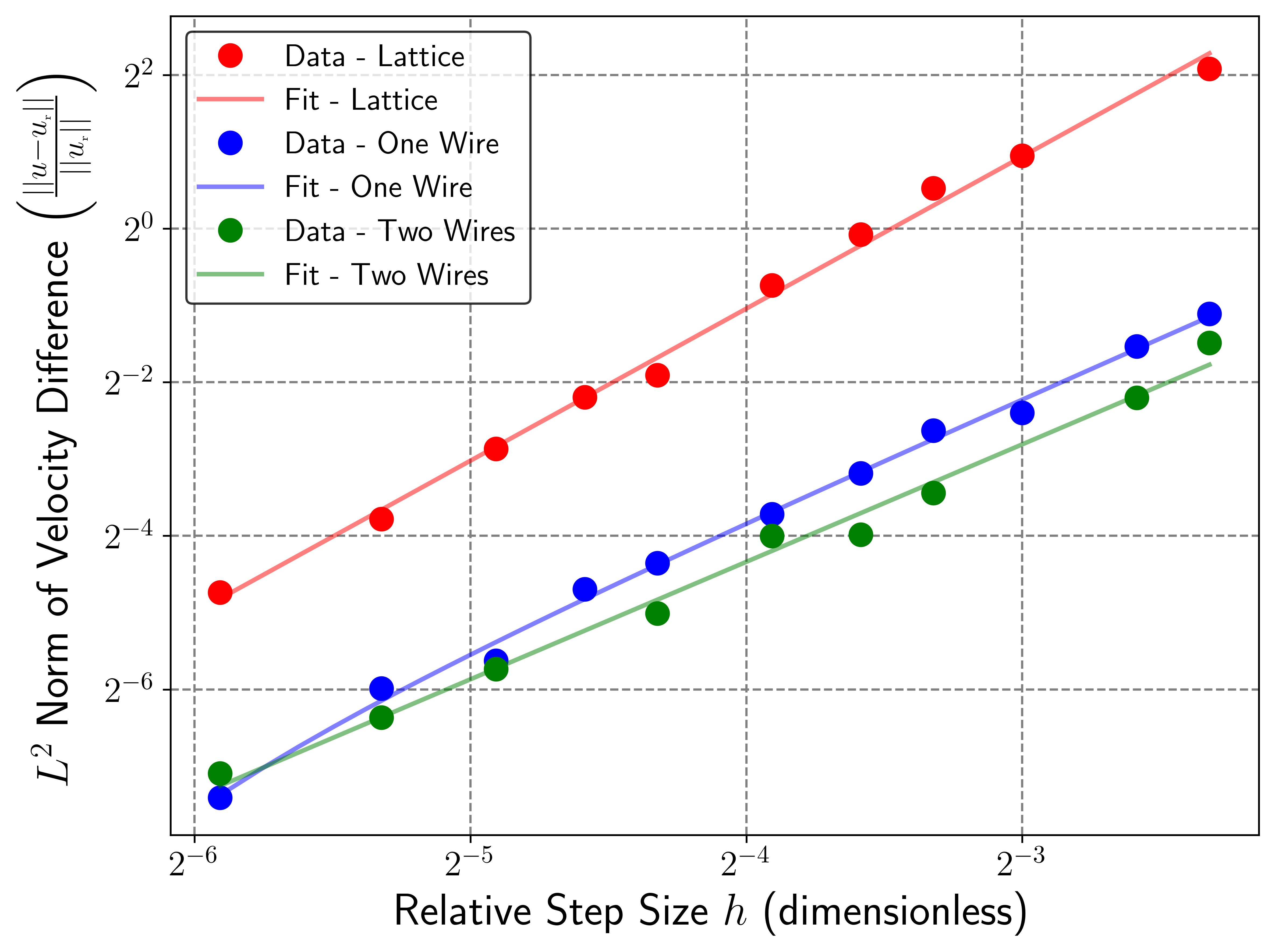}
\caption{
\textbf{Convergence of flow simulations.} 
Three geometries are simulated without mesh refinement (lattice, one filament, two filaments).
The $y$-axis shows the $\ell_2$ norm of the difference between each simulation 
and a reference simulation with $n_z = 120$. 
The $x$-axis is the relative step size $h = 1 / n_z$.
The fit is based on \eqref{eq:flow-convergence-fit}.}
\label{fig:flow-cnvg}
\end{figure}
\begin{table}[ht]
\centering
\begin{tabular}{| l | c | c | c | c |}
\hline
Geometry & $C$ & $p$ & $\alpha$ & RMSE \\
\hline
Lattice & 117.9 & 1.981 & 0.000 & 0.0978 \\
One Filament & 5.7 & 1.575 & 1.000 & 0.0802 \\
Two Filaments & 3.4 & 1.530 & 0.000 & 0.1216 \\
\hline
\end{tabular}
\caption{Flow convergence fit for three simulated geometries.}
\label{tbl:flow-cnvg}
\end{table}
Figure \ref{fig:flow-cnvg} shows a plot of the convergence for all three geometries while
Table \ref{tbl:flow-cnvg} shows the fitted parameter values.
The column RMSE has the root mean squared value of $\log(\varepsilon_j / f_j)$
where $\varepsilon_j$ is the actual $\ell_2$ error of the $j$-th point in the series, 
and $f_j$ is the fitted value based on \eqref{eq:flow-convergence-fit}.
The flow simulations for a logpile geometry are converging with a fitted power $p = 1.981$ 
that is very close to the theoretically expected value of $2$
given that the Chorin projection method is accurate to second order.
The other two geometries appear to be converging at somewhat slower rates, with $p \approx 1.55$.
However, they are also more accurate than the dense geometries in terms of their relative errors to the reference simulation.
These results are in line with those found by Rycroft et al.
for geometries with solid / fluid interactions.\cite{RWK20}\footnote{The boundary treatment in \citeb{RWK20} was significantly different than the one here, limiting the applicability of a direct comparison in convergence rates.}

\subsection{Consistency of Mesh Refined Flow Simulations}
\label{sec:results-flow-mesh}
We evaluate the convergence of mesh refined geometries using a similar approach.
This time, the reference simulation was carried out with $n_z = 128$ and $h = \qty{1.25}{\um}.$
Coarse geometries were also simulated with $n_z \in \{8, 16, 32, 64\}$.
Each coarse geometry was refined successively until the finest resolution matched that of the reference simulation.
The physical condition for each simulation was an applied pressure differential of 100 Pa across the channel.
Errors were fit with the model
\begin{equation}
\label{eq:flow-convergence-mesh}
\varepsilon = \{ C_c h_c^p + C_f h_f^p\} \{1 - \alpha (h_\text{ref} / h_\text{c})^p \},
\end{equation}
where $C_c$ and $C_f$ are fitted coefficients for the error attributed to coarse and fine grained phenomena;
$p$ is the fitted power of convergence; $\alpha$ is a fitted discretization coefficient;
$h_\text{ref}$ is the grid spacing on the reference mesh;
and $h_\text{c}$ and $h_\text{f}$ are the grid spacings on the coarse and fine levels, respectively.
\begin{figure}[ht]
\centering
\includegraphics[width=1.00\linewidth]{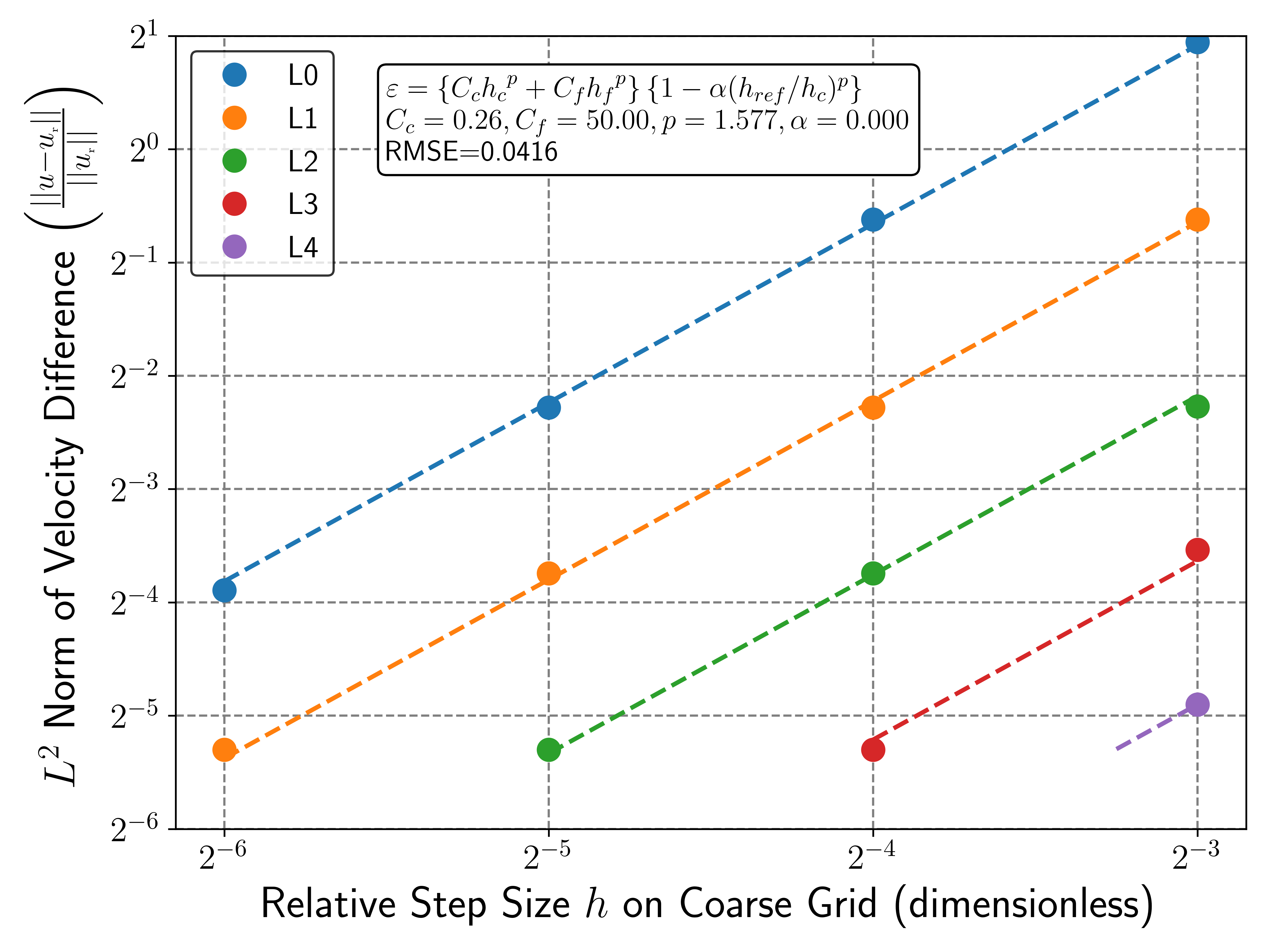}
\caption{
\textbf{Convergence of flow simulations with mesh refinement.}
The $y$-axis shows the $\ell_2$ norm of the difference between a mesh refined
simulation and a reference simulation with $n_z=128$ for the lattice geometry.
The $x$-axis shows $n_z$ at the coarse level of the mesh ($L0$),
with different colors for the level of mesh refinement.
Errors are fit using \eqref{eq:flow-convergence-mesh}.}
\label{fig:flow-cnvg-mesh}
\end{figure}
\begin{table}[ht]
\centering
\begin{tabular}{| l |c | c | c | c | c |}
\hline
Geometry & $C_c$ & $C_f$ & $p$ & $\alpha$ & RMSE \\
\hline
Lattice & 0.26 & 50.00 & 1.577 & 0.000 & 0.0416 \\
One Filament & 0.00 & 1.00 & 1.228 & 0.507 & 0.1840 \\
Two Filaments & 0.00 & 3.85 & 1.410 & 0.542 & 0.4295 \\
\hline
\end{tabular}
\caption{Convergence of mesh refined simulations.}
\label{tbl:flow-cnvg-mesh}
\end{table}

Figure \ref{fig:flow-cnvg-mesh} shows the convergence of the lattice geometry.
The fit captures the data almost perfectly with power of convergence of $p = 1.577$.
The error is dominated by the term with $C_f h_f^p$, 
showing that successive mesh refinements are an effective way to reduce error on this geometry.
The fitted power is somewhat less than two, showing that there is some compromise 
to accuracy in refining only the mesh rather than the entire geometry.
Table \ref{tbl:flow-cnvg-mesh} shows the fitted results for all three geometries.
The results are broadly in line with the trend seen in the lattice geometry,
with high overall accuracy even on quite coarse simulations, but a slower rate of convergence.
Taken together, these results support the conclusion that mesh refinement
is an effective technique in simulating flows of this kind that will produce reliable results
when the coarse and fine resolutions are sufficent to capture the relevant 
phenomena at large and small physical scales, respectively.

\subsection{Efficacy of Iterative Upsampling for Flow Simulations}
\label{sec:results-flow-upsample}
We demonstrate the efficacy of the iterative upsampling technique by comparing 
the reference flow simulation from the previous section to an analogous mesh refined simulation.
The reference simulation was run with a uniform grid and $n_z = 128$.
The mesh refined simulation was run with $n_z = 64$ at the coarse level and 1 level of mesh refinement.
The reference system required 2094 time steps and \qty{99.3}{\hour} to achieve steady state.
The upsampled simulation was first run to steady state without mesh refinement, requiring 1386 time steps and \qty{4.6}{\hour}.
This result was then used to initialized the mesh refined simulation, which required an additional 1575 time steps 
and \qty{31.2}{\hour} to achieve steady state.
The total run time for steady state using upsampling was thus \qty{35.8}{\hour},
and the reference simulation requires 2.77 times more run time on the same system.

\begin{figure}[ht]
\centering
\begin{overpic}[permil,width=1.00\linewidth]{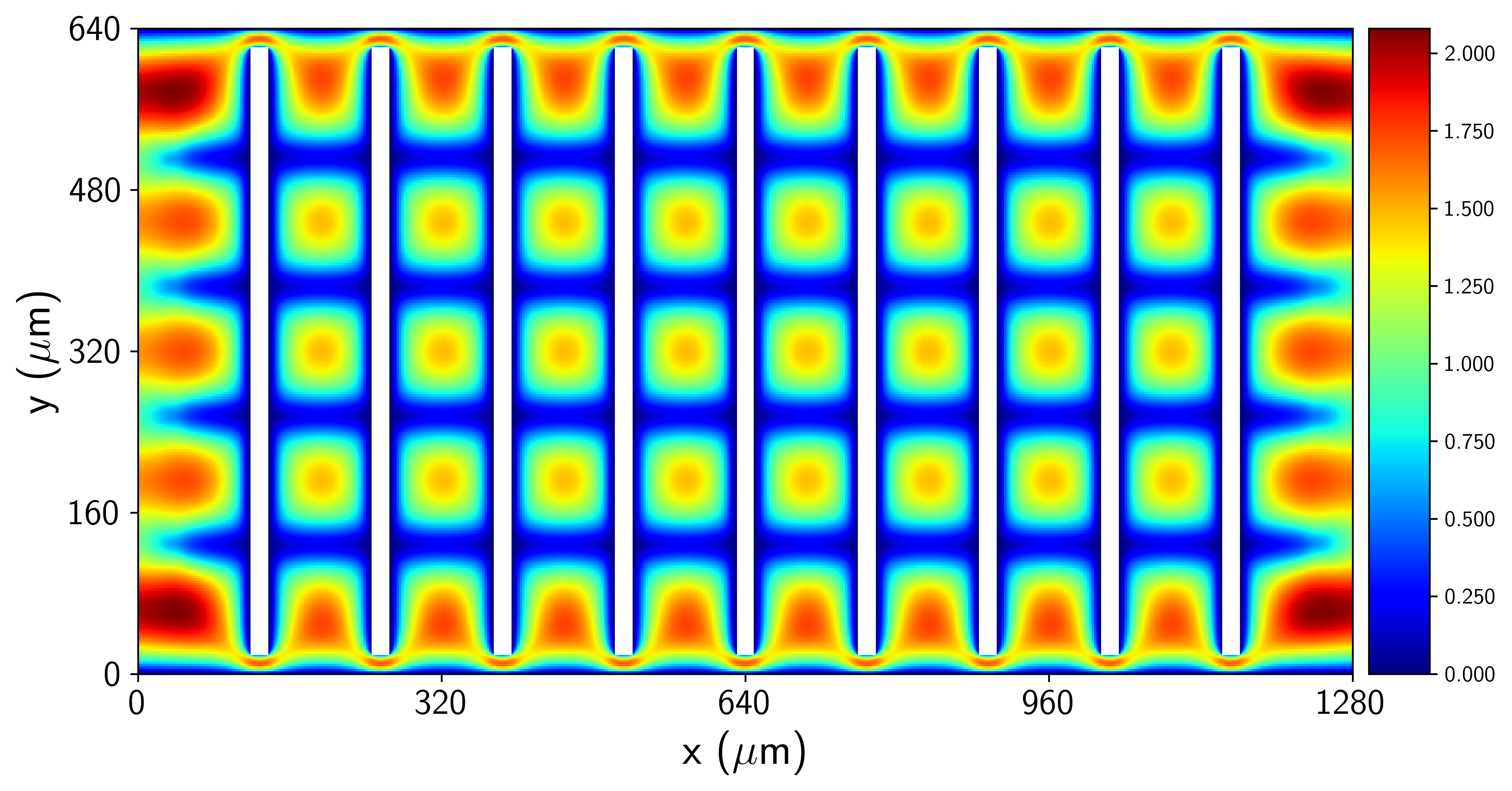}\put(0,450){\color{blue}\large{(a)}}\end{overpic}
\begin{overpic}[permil,width=1.00\linewidth]{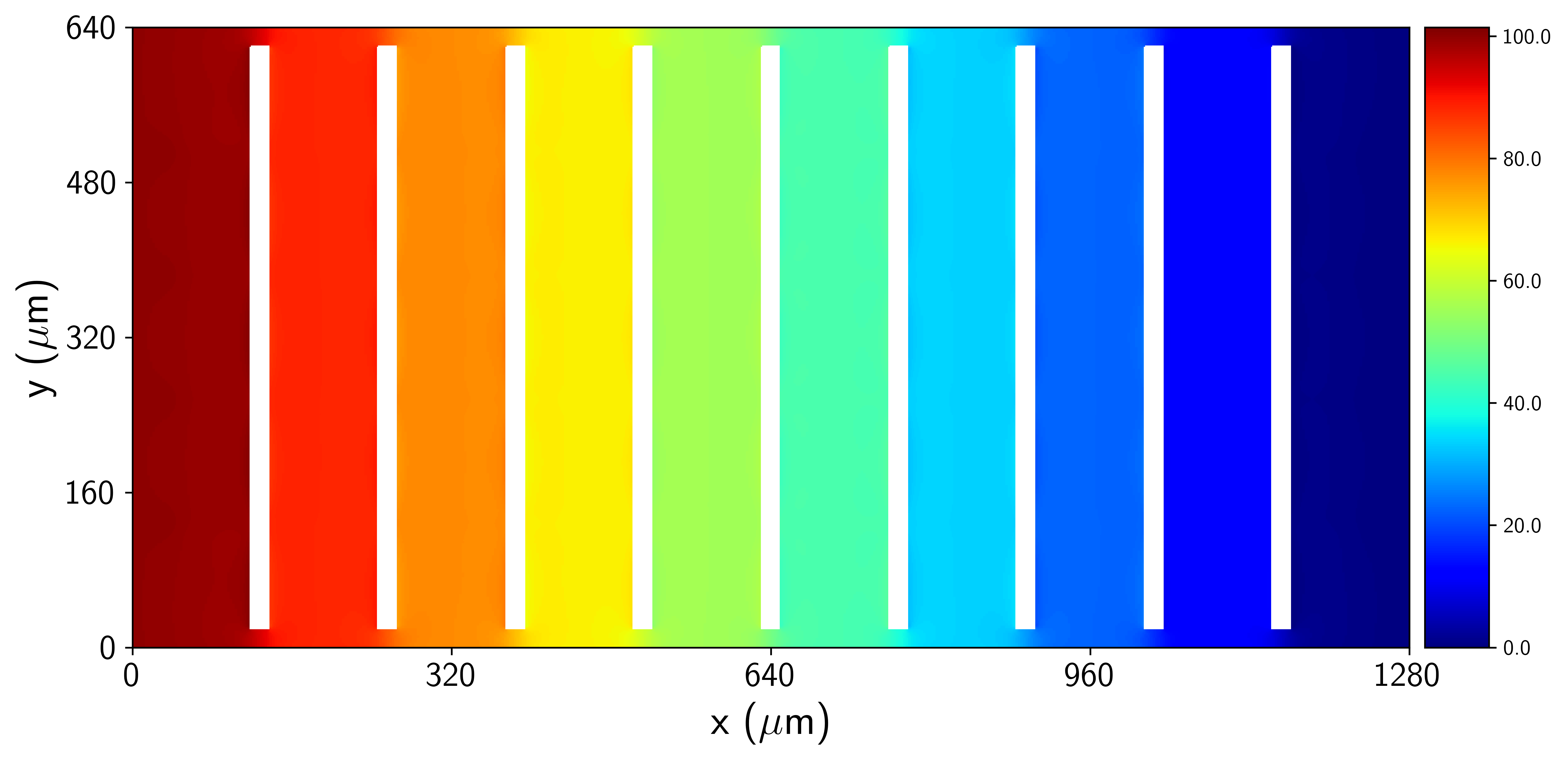}\put(0,400){\color{blue}\large{(b)}}\end{overpic}
\caption{
\textbf{Reference flow simulation at $\bm{p}$ = 100\,Pa on mid-plane $\bm{z = 80}\,\pmb{\unit{\um}}$.}
Panel (a) is speed in \unit{\cm.\sec^{-1}} and (b) is pressure in \unit{\pa}.}
\label{fig:flow-sim-ref}
\end{figure}

\begin{figure}[ht]
\centering
\includegraphics[width=1.00\linewidth]{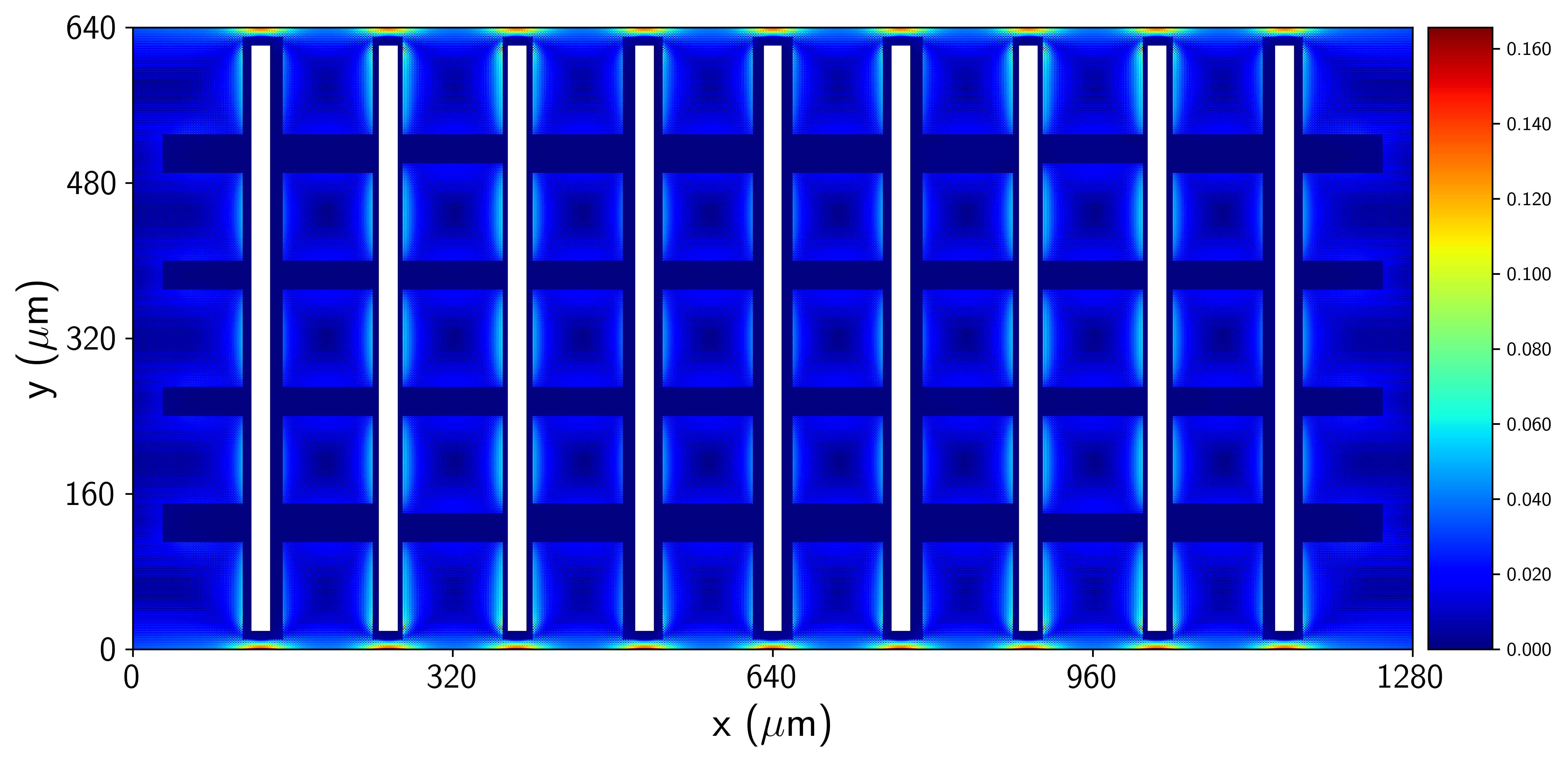}
\caption{
\textbf{Relative error of upsampled simulation vs. reference simulation.}
The reference simulation has $n_z=128$ and no meshing.
The comparison simulation has $n_z = 64$ and one level of mesh refinement.
The error is $\varepsilon_{ij} = \left({\norm{u'_{ijk} - u^{r}_{ijk}}}/{||u^{r}||} \right)$
where $\uvec$ and $\uvec'$ are the velocities on the reference and refined simulations, respectively.}
\label{fig:flow-error-rel}
\end{figure}

The results are substantially similar. The root mean squared error between the two velocity fields is 3.10\%.
The errors are predominantly on unrefined cells that are far away from the reaction.
When we consider only the mesh refined cells, which include the boundary cells and their neighbors,
the root mean squared error drops more than a full order of magnitude to $1.5 \cdot 10^{-3}$.
Figure \ref{fig:flow-sim-ref}(a) plots the speed of the reference flow simulation at the mid-plane ($z = \qty{80}{\um})$.
When this is plotted side by side with the results of the upsampled simulation, the results are almost indistinguishable visually.
Figure \ref{fig:flow-error-rel} shows the relative error, defined as the the norm of the velocity difference 
in a cell over the root mean square velocity.
This plot shows that the differences are small through most of the plane with just a few 
regions where the flow pinches between the end of a rod and the outer boundary wall.
Together with the previous section, these results support the conclusion that
iterative upsampling is an effective technique to resolve a steady state flow field
in significantly less computational work than would be required to simulated
the same flow with a uniform mesh.
Figure \ref{fig:flow-sim-ref}(b) shows the pressure of the reference simulation.
The pressure is almost indistinguishable from a linear gradient.
There are slight variations near the pinch points above and below the transverse filaments.

\subsection{Stokes Flow and Turbulent Regimes as Pressure Varies}
\label{sec:results-flow-stokes}
We simulated the incompressible flow at a broad range of pressures spanning five orders 
of magnitude to explore the Stokes flow regime as well as the onset of turbulent flow.
Simulations were carried out on on the logpile geometry with 16 applied pressures of $1, 2, 5, 10, \ldots 10^5$\,\unit{\pa}.
All simulations were run with the same meshing as the upsampled simulation in the previous section,
i.e. one level of mesh refinement and a fine resolution of \qty{1.25}{\um}.
Simulations were run using the iterative upsampling technique.
We calculated the Reynolds number of each flow as described in \ref{sec:flow-theory}
using the superficial velocity and channel length.

\begin{figure}[ht]
\centering
\begin{overpic}[permil,width=1.00\linewidth]{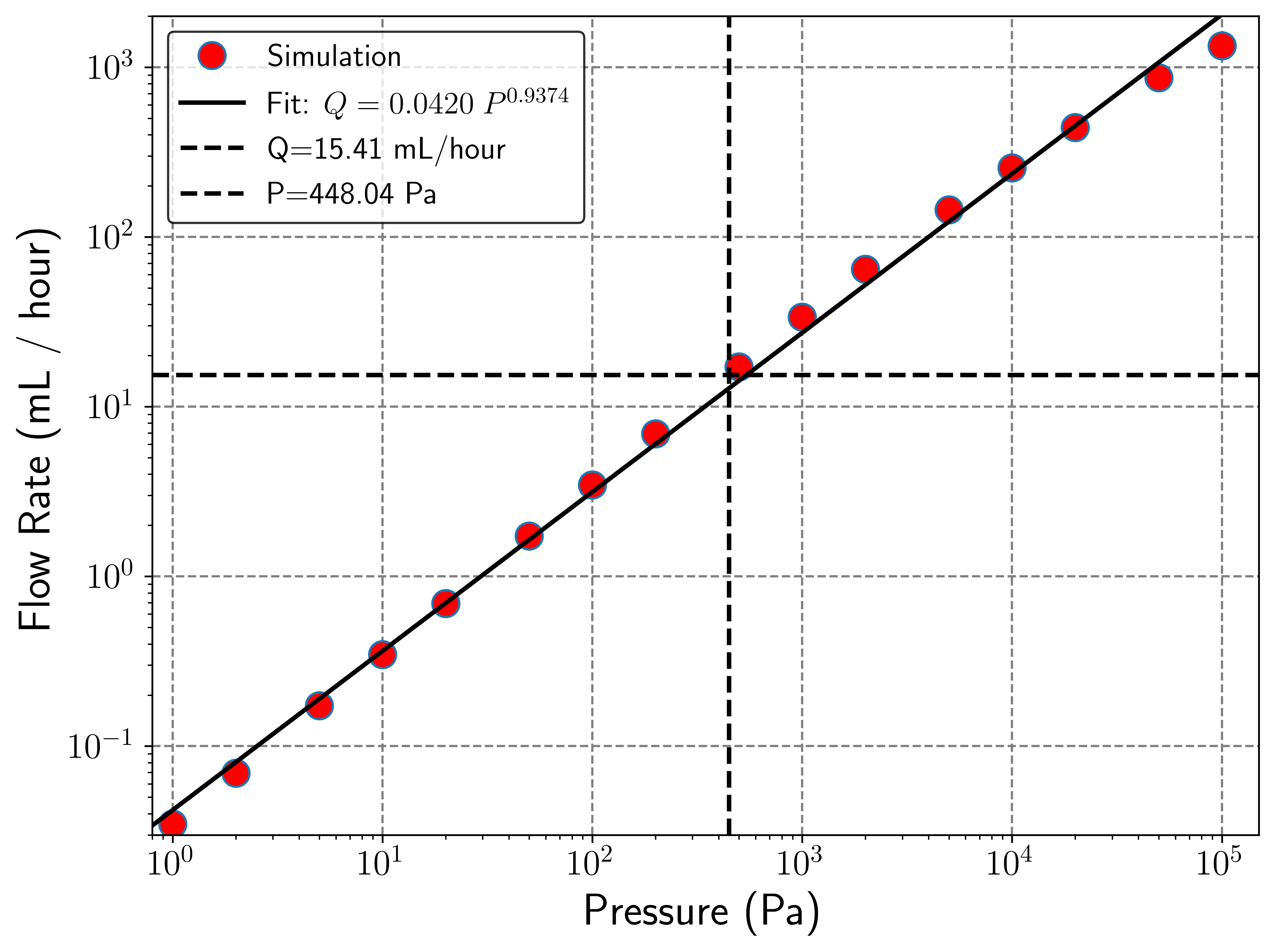}\put(0,740){\color{blue}\large{(a)}}\end{overpic}
\begin{overpic}[permil,width=1.00\linewidth]{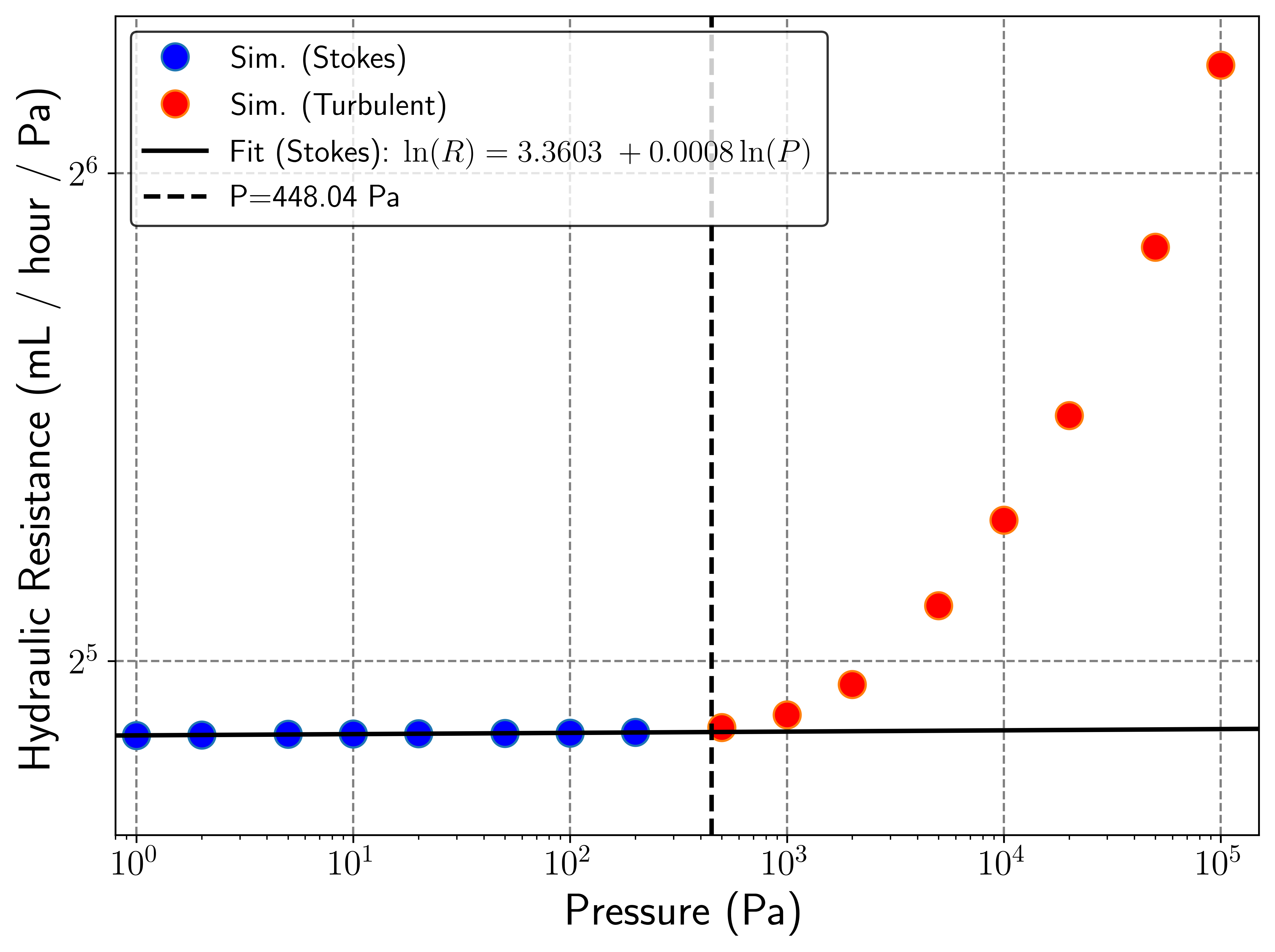}\put(0,740){\color{blue}\large{(b)}}\end{overpic}
\caption{
\textbf{Flow simulations at pressures spanning five orders of magnitude.}
Panel (a) shows volumetric flow rate $Q$ and (b) shows hydraulic resistance $R_H$.
The $x$-axis is applied pressure in \unit{\pa}.}
\label{fig:stokes-flow}
\end{figure}

Figure \ref{fig:stokes-flow}(a) shows the volumetric flow rate plotted against the applied pressure.
A power law fit $Q \sim C P^{r}$ is overlaid along with dashed lines 
corresponding to the predicted onset of turbulence at $\rn = 3000$ \cite{AML11}.
As predicted by theory, the flow vs. pressure relationship is very stable in the Stokes regime.
$Q$ then shows a slight bend lower as the flow starts to become turbulent.
Figure \ref{fig:stokes-flow}(b) shows the hydraulic resistance of each flow simulation 
calculated according to \eqref{eq:hydraulic-resistance}.
The Stokes and turbulent regimes are plotted in different colors.
We can see a rapid nonlinear increase in the hydraulic resistance as the flow exits the Stokes regime.

The entire velocity field is essentially linear in the applied pressure, not just the overall flow rate.
A side by side plot of the flows at the lowest and highest pressure in the 
Stokes regime ($p = 1$ and \qty{200}{\pa}) look almost identical.
The root mean square relative difference between these flows after scaling for pressure is only $2.5\%$,
a remarkably small amount given the factor of 200 spanning the pressures.
Even after the flow starts to become turbulent, there is still a strong similarity to the Stokes flow.
The RMS difference between the flows at $200$ and $\qty{2000}{\pa}$ for instance is $0.164$.
This is close enough that the flows look qualitatively similar and generate similar streamlines.

We conclude from the above plots and calculations that these results are consistent with the prediction made by theory.
Most practical applications of flow batteries are solidly in the Stokes regime because,
aside from the problem of excessive pumping losses with increasing hydraulic resistance,
turbulent flow speeds are far too fast for sufficient electrochemical utilization.
We can therefore further conclude that a single numerical flow simulation 
done at a low Reynolds number broadly representative of battery operating conditions
is sufficient to accurately characterize the whole regime of fluid flows relevant to flow battery operation.
Flows at other pressures can be predicted by scaling them linearly to the one simulated pressure.
If a more accurate flow is required, the scaled flow can be used an a starting point
for a numerical simulation, dramatically speeding up the time to convergence.

\section{Results and Discussion - Chemical Reaction}
\label{sec:results-rxn}
In this section we review the results of our electrochemical simulations on the logpile geometry.
We begin by visualizing the results of a reference simulation of a 
single reaction condition in the simplified Butler-Volmer model.
We then demonstrate spatial convergence and the consistence of mesh refined 
results using analogous techniques to those used for the flow simulations in \ref{sec:results-flow}.
We further demonstrate that the results of various reaction models are all consistent with each other.

We close this section by simulating the logpile electrode at a broad range of voltages and flow rates
using the computationally efficient explicit Nernst model.

\subsection{Reference Reaction Simulation}
\label{sec:results-rxn-logpile}
\begin{figure}[ht]
\centering
\includegraphics[width=1.00\linewidth]{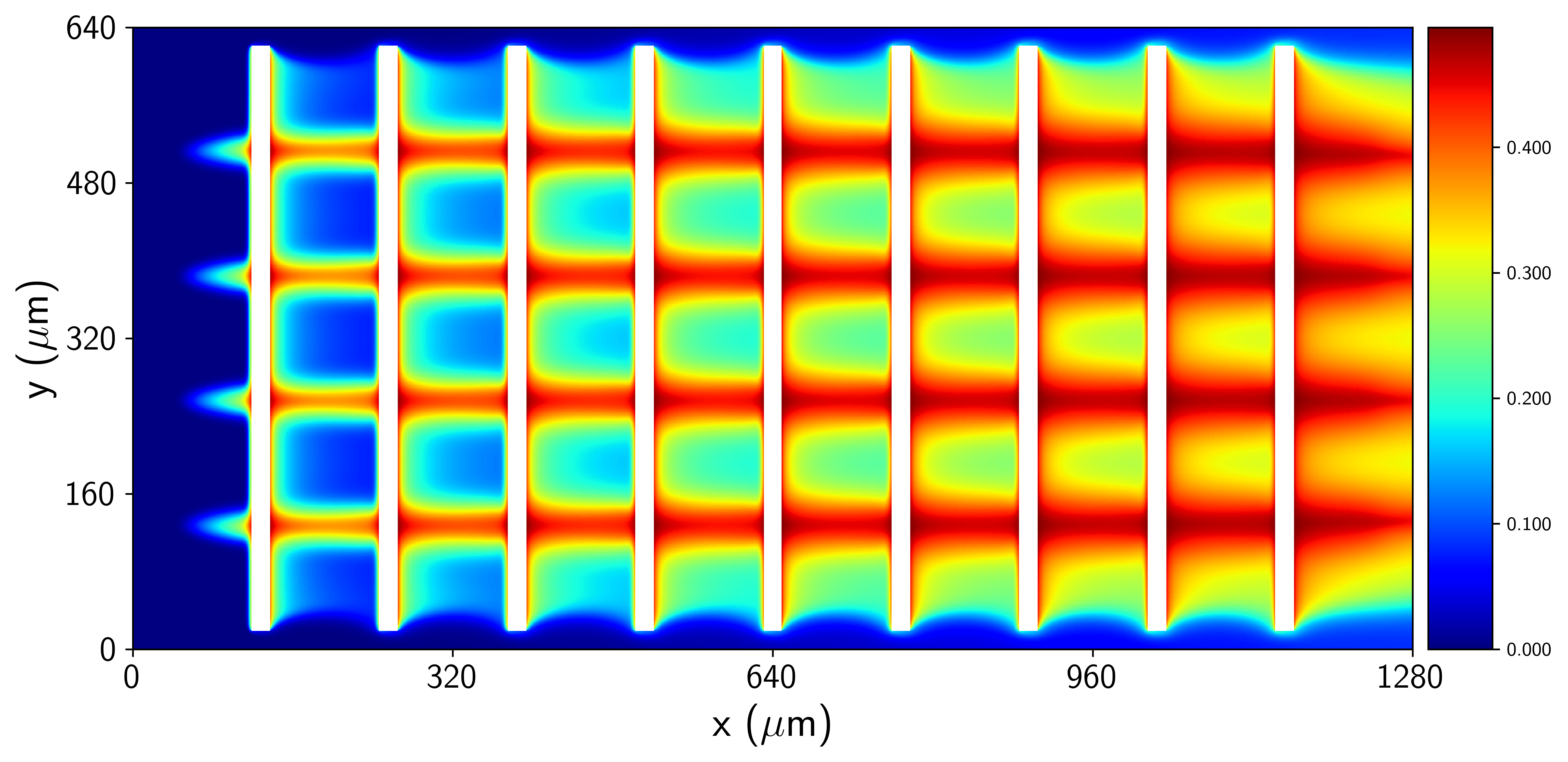}
\caption{
\textbf{Simulated state of charge with $\bm{p}$ =100\,Pa and $\bm{\Var}$ = 0\, mV.}
Plane view at mid-plane $z=\qty{80}{\um}$.}
\label{fig:soc-ref-2d}
\end{figure}

\begin{figure}[ht]
\centering
\begin{overpic}[permil,width=1.00\linewidth]{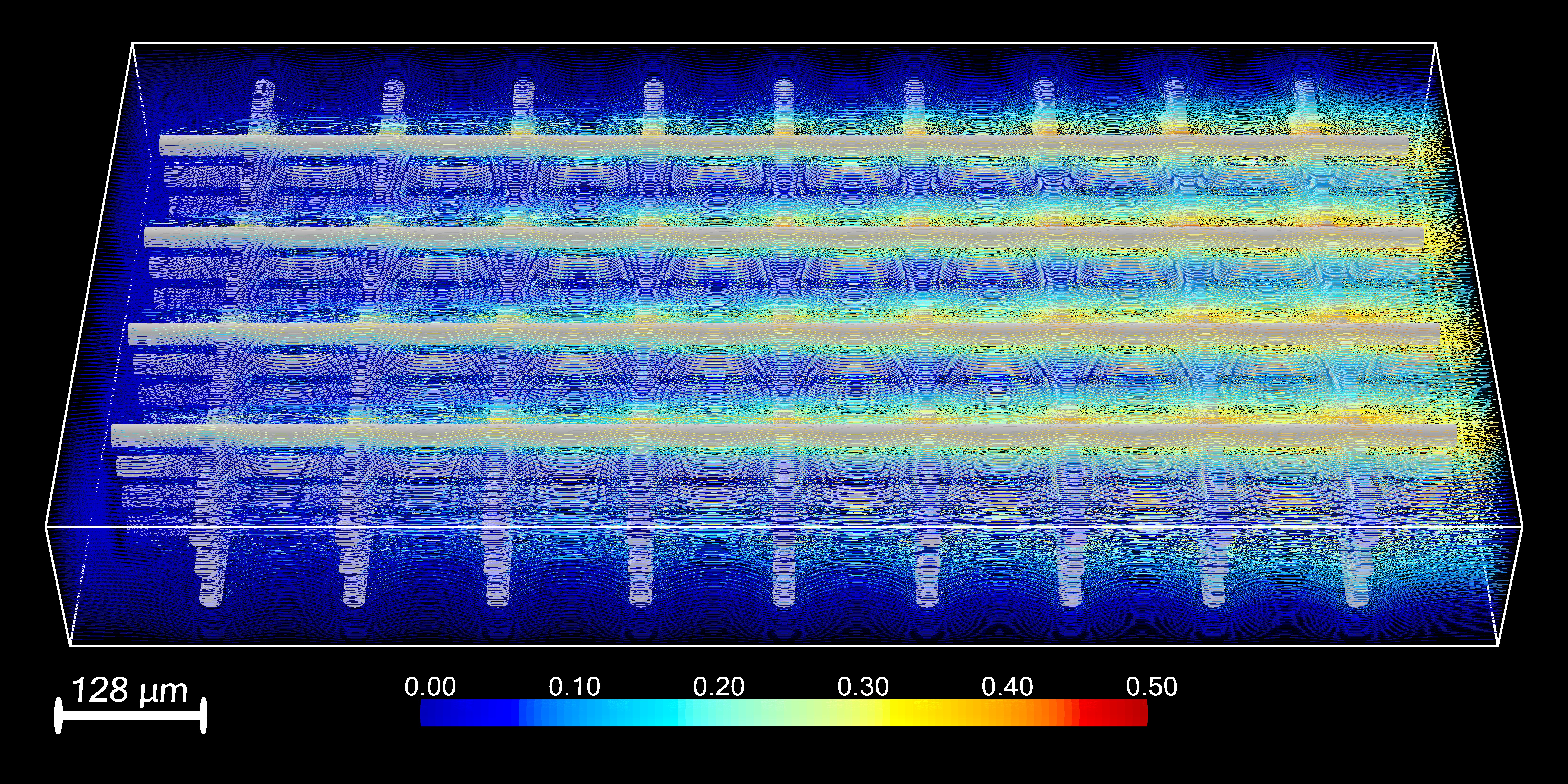}\put(10,460){\color{white}\large{(a)}}\end{overpic}
\begin{overpic}[permil,width=1.00\linewidth]{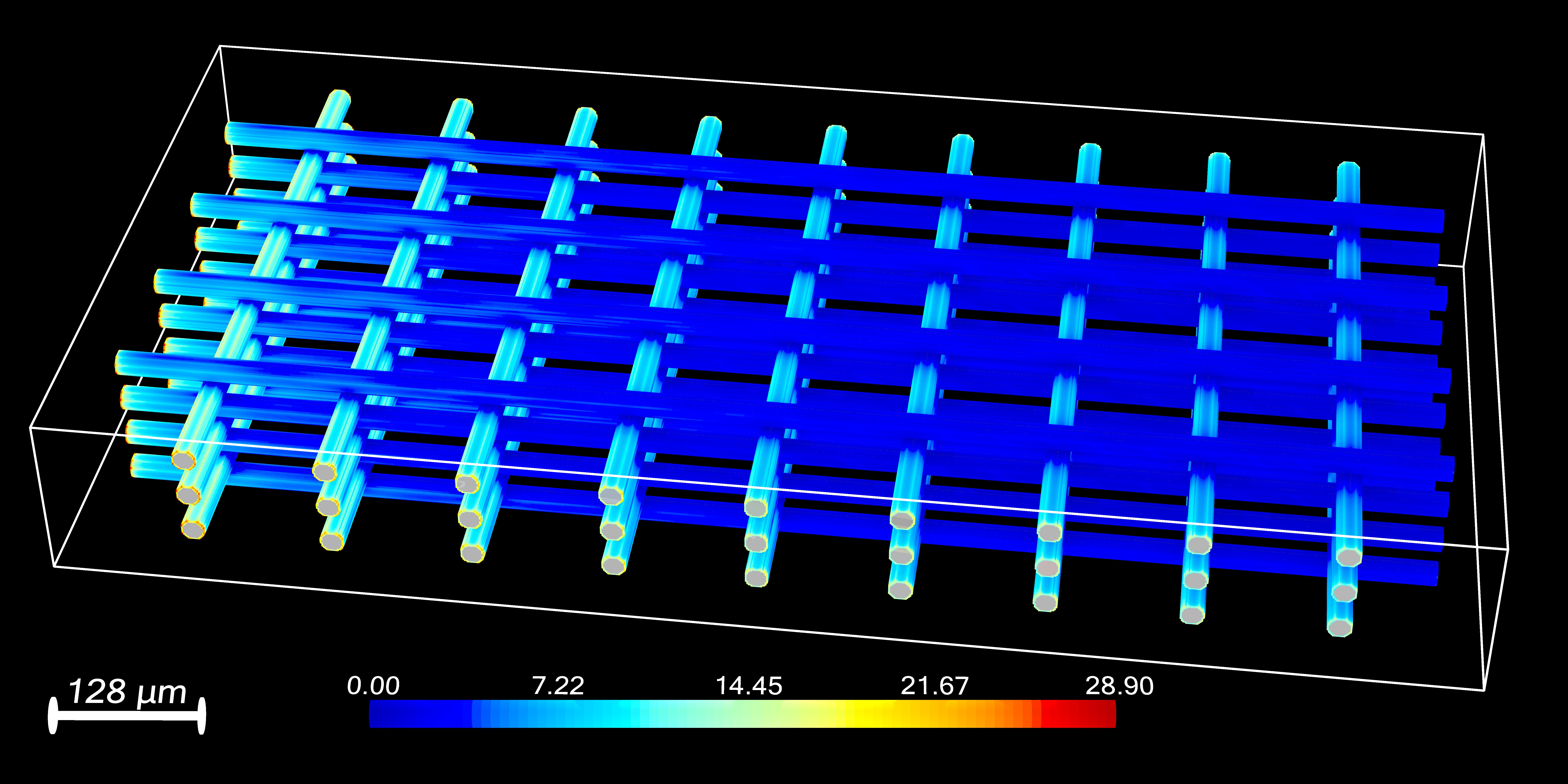}\put(10,460){\color{white}\large{(b)}}\end{overpic} 
\begin{overpic}[permil,width=1.00\linewidth]{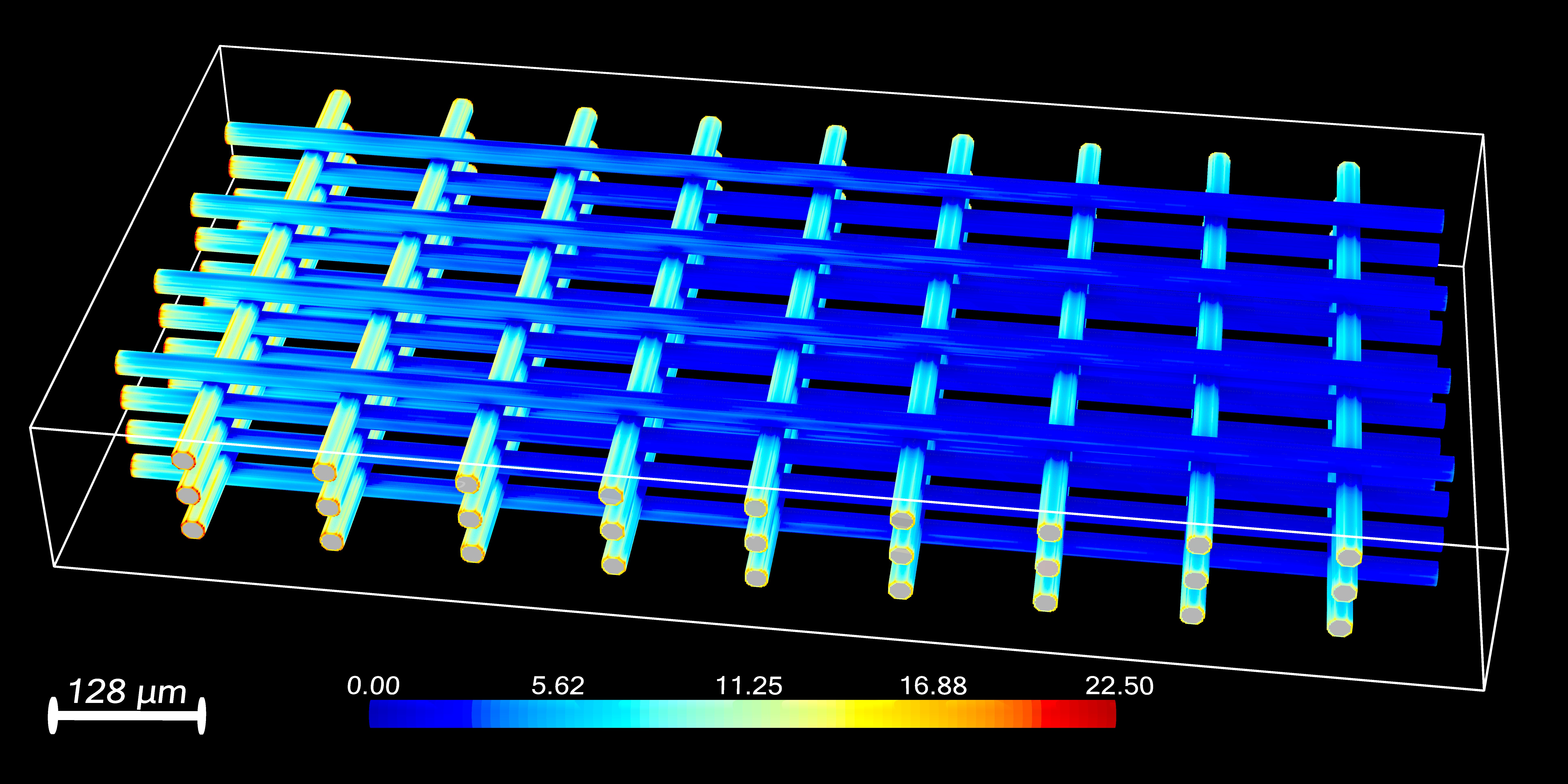}\put(10,460){\color{white}\large{(c)}}\end{overpic}
\caption{
\textbf{Simulated reaction steady state using SBV model with $\bm{p}$ = 10\,Pa and $\bm{\Var}$ = 0\,mV.}
(a) State of charge. (b) Overpotential (\unit{\mv}). (c) Current density (\unit{\ma.\cm^{-2})}.}
\label{fig:rxn-ref-3d}
\end{figure}

Our reference reaction simulation is on the logpile geometry with an applied pressure of 10 Pa
and an applied reducing voltage of 0 mV.
This was run to convergence in the simplified Butler-Volmer model
with a very tight tolerance of $10^{-6}$ on the RMS change in SOC per flow period.
This simulation converged in about 125,000 time steps with 3.35 seconds of simulated reaction time.
The predicted current was 0.09 mA, corresponding to a utilization of 0.243.
The maximum state of charge for electrolyte in equilibrium with the electrode at this voltage 
is 0.50, so this utilization is a bit less than half of that upper bound,
and this is a suitable test case for the simulation apparatus since it is balancing between 
advection, diffusion and reaction without any one dominating the others.
Figure \ref{fig:soc-ref-2d} shows a plane view of the state of charge in the center of the electrode.
We can see characteristic tails of reduced species building up 
along the $x$ oriented fibers above and below this layer
as seen experimentally by Barber et al.~\cite{BEE24}.
The vertically oriented filaments get better mass transport from advection and have 
a smaller buildup slowing down the reaction.
The state of charge becomes more uniform as the flow moves to the right 
and diffusion increasingly mixes the reactants.
Figure \ref{fig:rxn-ref-3d} shows a 3D rendering of key simulation outputs plotted along streamlines.
The three panels show the state of charge, overpotential and current density, respectively.
We note that the current density on the transverse rods is much higher 
than on the axially aligned rods due to their superior mass transport.
Advection moves material away from the reactive surface on the transverse rods,
but along the reactive surface for the axial rods, limiting their reaction rate to diffusion transport.

\subsection{Convergence and Mesh Consistency of Reaction Simulations}
\label{sec:results-rxn-convergence}
We demonstrate convergence with the same approach used for the flow simulations.
We simulated the same operating conditions for the reference reaction simulation 
($p = \qty{10}{\pa}$ and $\Var = \qty{0}{\mv}$) on a series coarse grids with $n_z \in \{ 5, 6, 8, 10, 12, 15, 20, 24, 60\}$.
These were compared with a reference simulation with $n_z = 120$.
The fluid flow for each of these simulations corresponds to the one used earlier in the flow convergence analysis.
Errors were computed by averaging the predicted state of charge in the fine simulation down
to match the coarse geometry and taking the root mean square of the difference in SOC.
\begin{figure}[ht]
\centering
\includegraphics[width=1.00\linewidth]{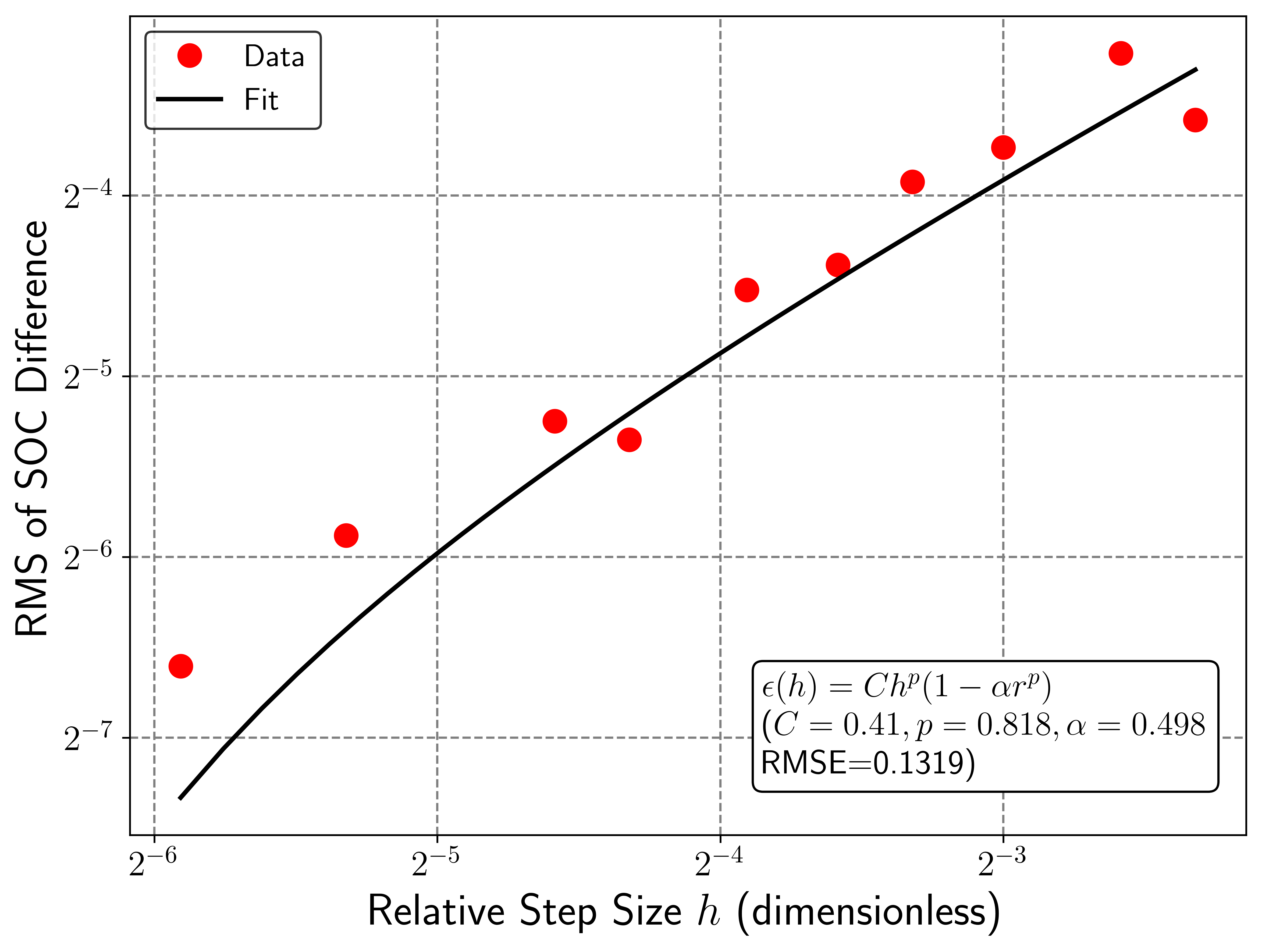}
\caption{
\textbf{Convergence of reaction simulations.}
Simplified Butler-Volmer model, $p = \qty{10}{\pa}$ and $\Var = \qty{0}{\mv}$
on refined grid with $n_z \in \{ 5, 6, 8, 10, 12, 15, 20, 24, 40, 60\}$.
$y$-axis shows root mean square difference in SOC vs. reference simulation with $n_z=120$.
$x$-axis shows relative step size $h = 1/ n_z$.}
\label{fig:rxn-cnvg}
\end{figure}
Figure \ref{fig:rxn-cnvg} shows the convergence.
We again fit the error using \eqref{eq:flow-convergence-fit} and find 
$\varepsilon \sim C h^{p} \cdot (1 - \alpha r^p)$ with $C = 0.408$, $p = 0.818$ and $\alpha = 0.498$.
While $p = 0.818$ suggests sublinear empirical convergence, the leading coefficient is small
and we can see that these results are in fact converging.
The reaction simulations are expensive to run, and finding consistent termination 
criteria for the different simulations is a challenge that may confound the convergence analysis.

Because of these difficulties, we were unable to replicate the full mesh convergence study
shown in Fig. \ref{fig:flow-cnvg-mesh}. 
Instead, we compared the results of a reference reaction simulation with $n_z = 128$
to an upsampled meshed simulation with $n_z = 64$ and one level of mesh refinement in three reaction models.
Table \ref{tbl:react-comp-mesh} shows the results.
All three models showed excellent agreement between the upsampled and reference simulation,
with errors in the range of $3.7 \cdot 10^{-3}$.
\begin{table}[ht]
\centering
\begin{tabular}{| l | c |}
\hline
Model & RMSE \\
\hline
Simplified BV & $3.64 \cdot 10^{-3}$ \\
Nernst & $3.50 \cdot 10^{-3}$ \\
Butler-Volmer & $3.83 \cdot 10^{-3}$ \\
\hline
\end{tabular}
\caption{Comparison of mesh refined simulations to reference.}
\label{tbl:react-comp-mesh}
\end{table}

\subsection{Consistency of Various Models for Reaction Steady State}
\label{sec:results-rxn-models}

\begin{figure}[ht]
\centering
\includegraphics[width=1.00\linewidth]{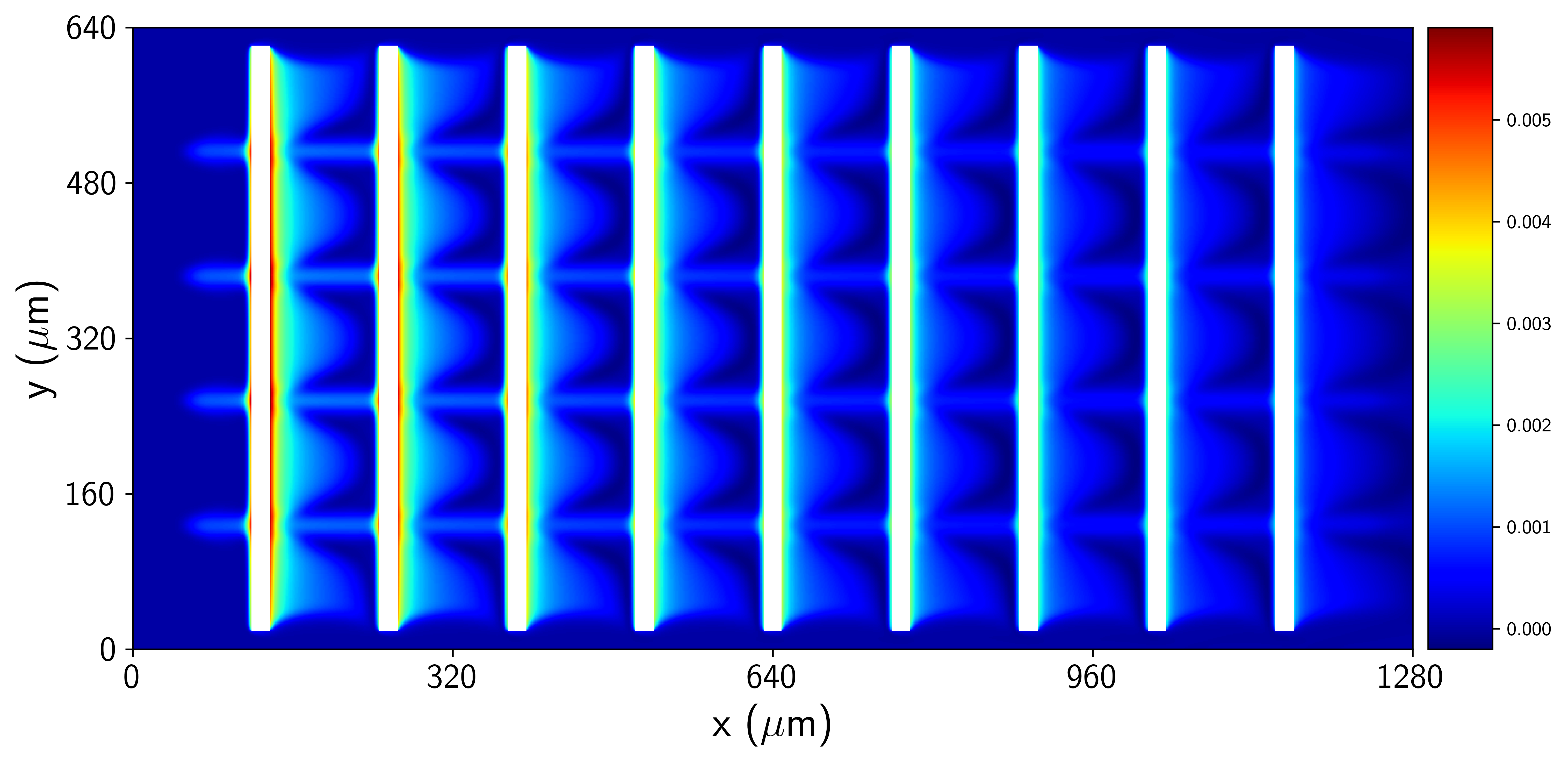}
\caption{
\textbf{SOC difference between BV and SBV models.}
Lattice geometry at $p=\qty{10}{\pa}$ and $\Var = \qty{0}{\mv}$ plotted at mid-plane $z = \qty{80}{\um}$.
Plotted values are $\abs{s_{\scriptscriptstyle\text{SBV}} - s_{\scriptscriptstyle \text{BV}}}$.}
\label{fig:soc-diff-bv}
\end{figure}

Figure \ref{fig:soc-diff-bv} plots the difference in the state of charge between the Butler-Volmer and simplified BV models.
The overall differences are small and concentrated on the y oriented filaments with good mass transport
that appear early in the flow path. The reaction proceeds fastest in these areas, and they are unusual.
As reduced product builds up and diffusion smooths out the SOC field, 
the differences between the full and simplified models gets progressively smaller.
This builds the intuition that in a larger electrode of practical interest, 
the differences between these models will be even smaller than what is shown here.
Table \ref{tbl:react-comp-model} shows the root mean square difference in SOC 
between the simplified Butler-Volmer model (as baseline) 
against the Nernst and Butler-Volmer simulation results.
Both models predict almost the same SOC field at steady state.
\begin{table}[ht]
\centering
\begin{tabular}{| l | c |}
\hline
Model & RMSE vs. SBV\\
\hline
Nernst & $2.30 \cdot 10^{-3}$ \\
Butler-Volmer & $8.61 \cdot 10^{-4}$ \\
\hline
\end{tabular}
\caption{Comparison of alternative models to SBV.}
\label{tbl:react-comp-model}
\end{table}

\begin{figure}[ht]
\centering
\begin{overpic}[permil,width=1.00\linewidth]{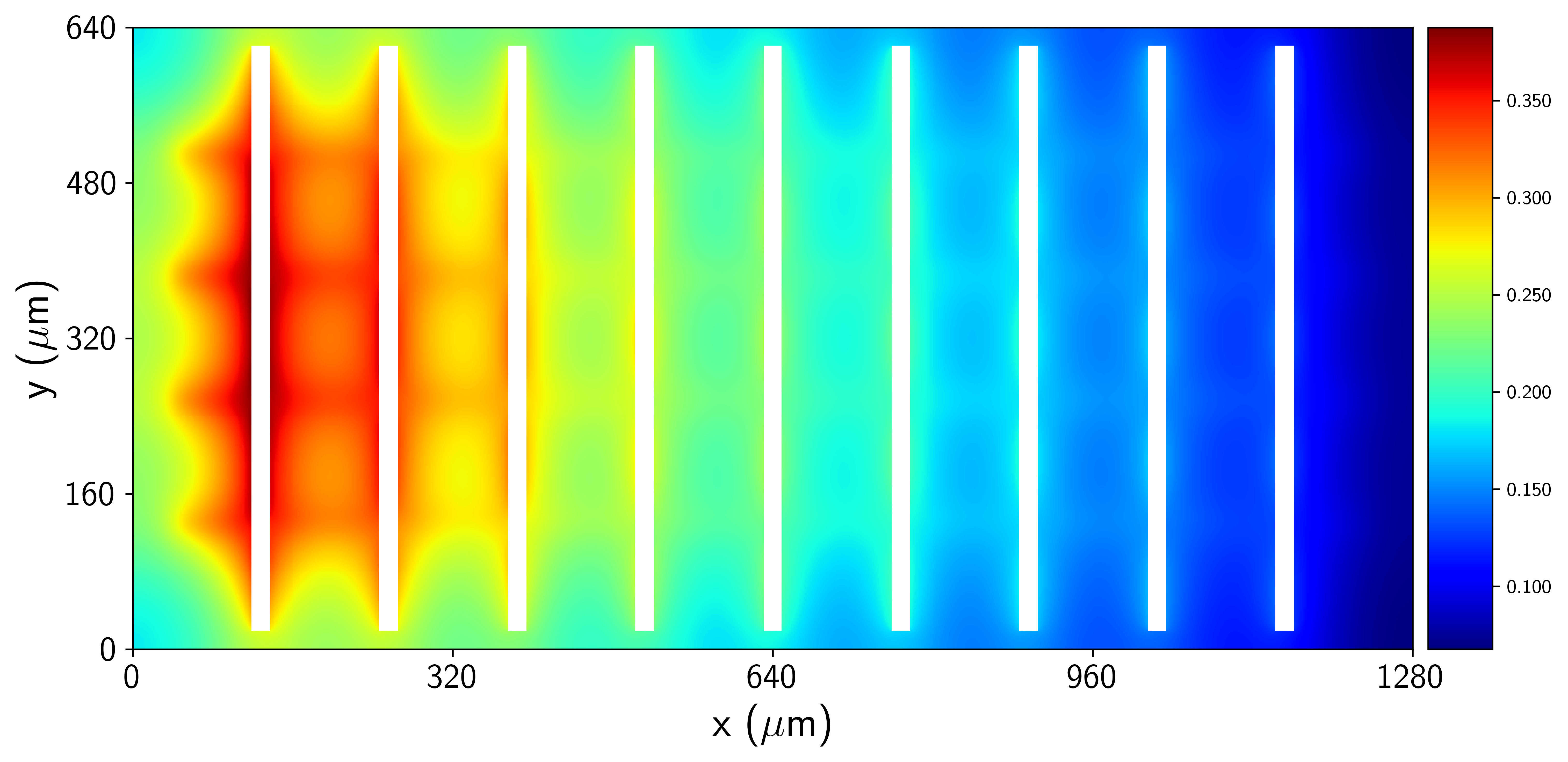}\put(-20,460){\color{black}\large{(a)}}\end{overpic}
\begin{overpic}[permil,width=1.00\linewidth]{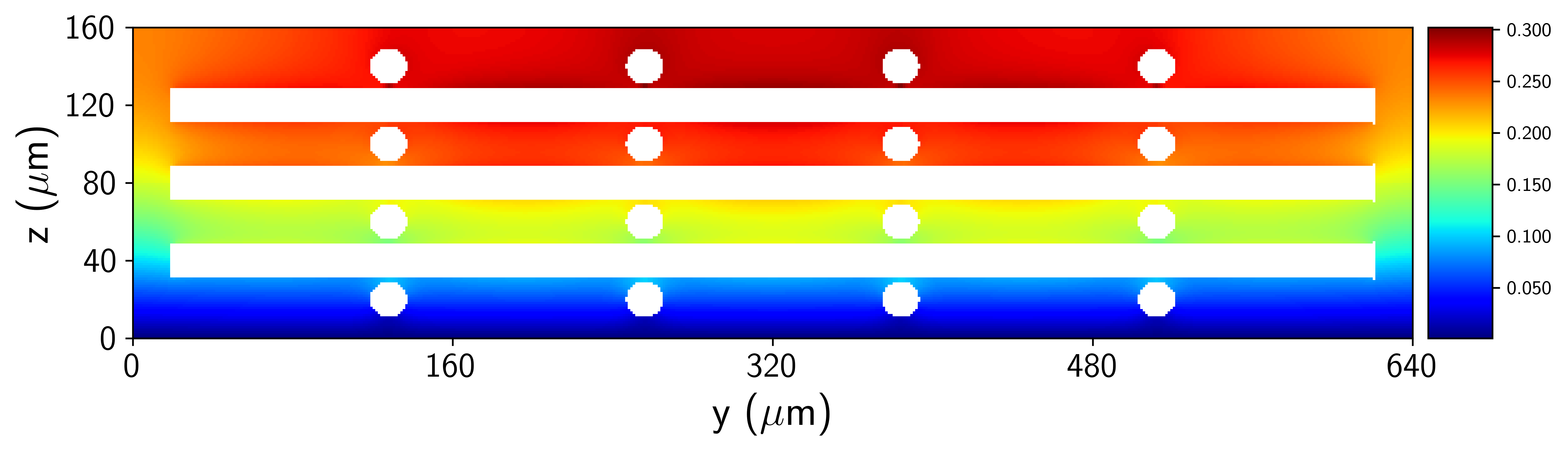}\put(-20,250){\color{black}\large{(b)}}\end{overpic}
\caption{
\textbf{Simulated potential $\phiL$ in the BV model with $\bm{p}$ = 10\,Pa and $\bm{\Var}$ = 0\,mV.}
Panel (a) shows an $xy$ projection at mid-plane $z = \qty{80}{\um}$ 
and (b) shows a $yz$ projection at mid-plane $x = \qty{512}{\um}$.
Potentials are in \unit{\mv} (scale bar at right).}
\label{fig:epot-ref}
\end{figure}

Figure \ref{fig:epot-ref} shows the simulated potential $\phiL$ in the Butler-Volmer 
model for the reaction condition discussed in the last section.
The potential in the electrolyte is minuscule. It has a mean of 0.16 mV and a max of 0.47 mV.
This is negligible in the context of flow batteries that are routinely run with overpotentials on the order of 50-150 mV.
We can see a clear trend of increasing potential moving away from the membrane
with an approximately linear gradient in $z$, but the magnitude is very small.
We emphasize that the potential in the solid does not have a direct dependence dependence on the potential in the liquid.
$\phiL$ is the solution to a Poisson equation whose right hand side depends on the movement of charged species.
The applied potential only impacts it indirectly by increasing the reaction rate.
The biggest potential buildups occur near the first few transverse filaments where 
the fastest reactions are occurring, requiring a net movement of protons towards the filament  to maintain charge balance.
The physical intuition here is simple: in the presence of a strong supporting electrolyte,
the standing electric field will be so small as to be negligible in flow battery operation.
While the precise size of the effect with vary with different electrochemical systems,
we hypothesize that in most systems of interest for energy storage,
supporting electrolytes will be sufficiently conductive to make this an insignificant effect.

\section{Conclusions}
\label{sec:conclusions}
In this work we have created a digital twin for a porous electrode in a redox flow battery,
consisting of a direct numerical solution to the governing Navier-Stokes and Nernst-Planck equations.
This numerical solution is performed with minimal compromises at a sub-fiber scale of \qty{1.25}{\um} using only open-source software that is highly performant, parallelizable, and compatible with the largest modern scientific supercomputers.
We have demonstrated that our method converges as expected on a fixed grid, and shown the efficacy of fixed mesh refinement to achieve reliable results in less time and with lower memory usage than would be required on a uniform fixed grid.
We have shown these points separately for the incompressible fluid flow and electrochemical reaction parts of the simulations.
We also briefly discuss an experimental validation of this model carried out by Barber et al \cite{BEE24}.

We have elucidated the theory of the porous electrode system and made a novel presentation of several fundamental ideas. We have shown that the total concentration of redox active species is a constant under modest assumptions that are nearly true in practice, allowing the Nernst-Planck equation to be written in terms of only the state of charge $s$. 
We have further demonstrated that in many practical systems including our experimental model, the electromigration term is negligible and can be safely disregarded, creating a simplified advection-diffusion-reaction PDE that describes the steady state condition. This will be true whenever the supporting electrolyte is a significantly better conductor than the active species, thereby minimizing standing electric fields in the electrolyte at steady state.

Our simulations are carried out using the minimal required set of materials parameters, including the electrolyte density $\rho$ and viscosity $\mu$,  the kinetic constant $k_0$, and charge transfer coefficient $\alpha$.
We do not assume a specific area $a$, but model it directly from the electrode geometry.
We similarly do not assume an empirical mass transfer coefficient $k_m$, but instead directly simulate the reaction on each cut cell, allowing $k_m$ to manifest as an emergent property of the system.

We have derived an equivalent reformulation of the steady state PDE that exchanges the Robin boundary condition on the electrode surface with a Dirichlet boundary condition, leading to a computationally efficient method (the Nernst model) for solving the steady state by iteratively constructing consistent estimates for the state of charge and overpotential.
Crucially, we have shown that the Nernst model can be implemented efficiently with an explicit algorithm once a fluid flow has been calculated, so that in practice finding the state of charge field is actually easier than solving the incompressible flow problem.
We have introduced a novel figure of merit, $\umt$, the mass-transport limiting utilization for a porous electrode geometry, and shown how it can be efficiently computed.
And we have shown the novel and broadly applicable result that under dilute-solution thermodymamics and Butler-Volmer reaction kinetics, when a battery is charged at a controlled voltage, the local reaction rate is an affine function of the local state of charge.

We have introduced a number of novel numerical techniques for speeding the computation of the electrode steady state and demonstrated their efficacy. Iterative upsampling works by simulating the system to steady state at a coarse resolution, then using the converged steady state as a starting point for a simulation at a finer resolution. It can thus be seen as a generalization of the idea behind multigrid solvers. Model refinement works by simulating the electrochemical system to steady state at a lower level of fidelity (e.g. disregarding electromigration) and then initializing the higher fidelity model. In this way, it is possible to quickly check the validity of the assumption that electromigration is small. One can then go on to complete a higher fidelity simulation if it is required and still achieve significant time savings.

Our key finding is that in an age of exascale scientific supercomputers, it is not only possible, but computationally feasible to obtain a high fidelity numerical solution of the governing equations for a porous electrode at steady state. This capability can serve many useful ends in porous electrode science. A digital twin can validate other modeling frameworks such as LBM and PNM to a higher degree of precision than is possible with experimental methods on unstructured commercial carbon electrodes. It can also be used for inverse design and to generate synthetic data to train and validate surrogate models using techniques such as machine learning. And it is a valuable adjunct to experimental characterizations of 3D printed electrodes, allowing experimental images to be compared side by side with their numerically predicted counterparts for a greater understanding of both.

\begin{acknowledgments}
The authors gratefully acknowledge support from the Department of Energy through the 
Office of Basic Energy Sciences (DE-SC0020170).

C.H.R. was supported in part by the U.S. Department of Energy, Office of Science, Office of Advanced Scientific Computing Research's Applied Mathematics Competitive Portfolios program under Contract No. AC02-05CH11231.

\end{acknowledgments}

\section*{Data Availability Statement}
\label{sec:data-availability}
All source code for this work is available at \url{github.com/memanuel/rfb-twin}.\\
All simulation outputs and images for this work are available from Zenodo
at \url{https://doi.org/10.5281/zenodo.14852325}.

\newpage
\nocite{*}
\section*{References}
\bibliography{twin}

\newpage
\appendix*
\onecolumngrid
\section{Symbols and Simulation Parameters}
\begin{longtblr}[
caption = {Symbols Used},
entry = {Symbols},
label = {tbl:symbols},
]{
colspec = {lXr}, 
width = 1.00\linewidth, 
hlines,
rowhead = 1, 
rowfoot = 0,
}
Symbol & Meaning & SI Units\\
$\rho$ & Fluid density & \unit{\kg.\m^{-3}} \\
$\uvec$ & Fluid velocity vector & \unit{\m.\sec^{-1}} \\
$(u, v, w)$ & $x$, $y$ and $z$ components of $\uvec$ & \unit{\m.\sec^{-1}} \\
$p$ & Dynamic pressure (adjusted for gravity) & \unit{\pa} =\unit{\kg.\m^{-1}.\sec^{-2}} \\
$g$ & Gravitational field & \unit{\m.\sec^{-2}} \\
$\mu$ & Dynamic viscosity & \unit{\pa.\sec} = \unit{\kg.\m^{-1}.\sec^{-1}} \\
$\kappa$ & Hydraulic permeability of porous medium & \unit{\m^2} \\
$Q$ & Volumetric flow rate through electrode & \unit{\m^{3}.\sec^{-1}} \\
$L$ & Length of channel in flow direction & \unit{\m} \\
$R_H$ & Hydraulic resistance of a porous medium & \unit{\kg.\m^{-4}.\sec^{-1}} \\
$\Delta t$ & Time step & \unit{\sec} \\
$\ccfl$ & CFL parameter& $1$\\
$C_A$ & Inverse time step - advection & \unit{\sec^{-1}} \\
$C_V$ & Inverse time step - diffusion & \unit{\sec^{-1}} \\
$C_D$ & Inverse time step - external forces & \unit{\sec^{-1}} \\
$\varepsilon_\text{flow}$ & Convergence threshold - flow steady state & $1$ \\
\ce{O} & Generic oxidized species & $1$ \\
\ce{R} & Generic reduced species & $1$ \\
$\nel$ & Number of electrons transferred in PCET reaction & $1$ \\
$\npr$ & Number of protons transferred in PCET reaction & $1$ \\
$\Estd$ & Equilibrium interfacial potential - standard conditions & \unit{\volt} \\
$\Eeq$ & Equilibrium interfacial potential - nonstandard conditions & $1$ \\
$R$ & Ideal gas constant & \unit{\joule.\mole^{-1}.\kelvin^{-1}} \\
$T$ & Temperature in Kelvin & \unit{\kelvin} \\
$\cO$ & Concentration of oxidized species & \unit{\mole.\m^{-3}} \\
$\cR$ & Concentration of reduced species & \unit{\mole.\m^{-3}} \\
$\cT$ & Total concentration of electroactive species & \unit{\mole.\m^{-3}} \\
$\cTi$ & Initial total concentration of electroactive species, a constant & \unit{\mole.\m^{-3}} \\
$\phiS$ & Potential in the solid electrode & \unit{\volt} \\
$\phiL$ & Potential in the electrolyte & \unit{\volt} \\
$\etaact$ & Activation overpotential & \unit{\volt} \\
$s$ & State of charge & $1$ \\
$F$ & Faraday's constant & \unit{\coulomb.\mole^{-1}} \\
$j_0$ & Exchange current density in Butler-Volmer model & \unit{\ampere.\m^{-2}} \\
$j$ & Current density in Butler-Volmer model & \unit{\ampere.\m^{-2}} \\
$k_0$ & Kinetic rate constant in Butler-Volmer model & \unit{\m.\sec^{-1}} \\
$A$ & Area of solid / liquid boundary in one cut cell & \unit{\m^2} \\
$V$ & Volume of electrolyte in one cell (regular or cut) & \unit{\m^3} \\
$a$ & Specific area of a cut cell & \unit{\m^{-1}} \\
$\alpha$ & Charge transfer coefficient in Butler-Volmer model & $1$ \\
$V_T$ & Thermal voltage & \unit{\volt} \\
$S$ & Chemical source term - entire reaction& \unit{\mole.\m^{-3}.\sec^{-1}} \\
$n_k$ & Stoichiometric number for $k$-th species & $1$ \\
$S_k$ & Chemical source term for $k$-th species & \unit{\mole.\m^{-3}.\sec^{-1}} \\
$\etat$ & Nondimensionalized reducing overpotential & $1$ \\
$\St$ & Nondimensionalized source term for SOC & $1$ \\
$D_j$ & Diffusivity of $j$-th species & \unit{\m^{2}.\sec^{-1}} \\
$z_j$ & Signed charge number of $j$-th species & $1$ \\
$\kappa_L$ & Ionic conductivity of the electrolyte & \unit{\kg.\m^{-3}\sec.\coulomb^{2}} \\
$S_{\phi}$ & Electron source term & \unit{\m^{-3}.\sec^{-1}.\coul}\\
$\Var$ & Applied reducing potential & \unit{\volt} \\
$\hat{n}$ & Normal vector at electrode boundary & $1$ \\
$\seq$ & Equilibrium state of charge at an applied potential & $1$ \\
$\Vart$ & Nondimensionalized applied reducing poential & $1$ \\
$\phiLt$ & Nondimensionalized potential in the electrolyte & $1$ \\
$\sigma(x)$ & Sigmoid function $\sigma(x) = (1 + e^{-x})^{-1} = \logit^{-1}(x)$ & $1$ \\
$\tau$ & Mean time for fluid to traverse the electrode & \unit{\sec} \\
$\varepsilon_s$ & Convergence threshold - reaction SOC change & $1$ \\
$\varepsilon_\phi$ & Convergence threshold - reaction potential change & $1$ \\
$\varepsilon_{I}$ & Convergence threshold - total current vs. charge flow & $1$ \\
$\norm{\Delta}$ & Root mean square $\norm{\Delta}_2 / \sqrt{N}$ & $1$ \\
$\theta_\text{mrc}$ & Maximum reactant consumption parameter & $1$ \\ 
$\mathcal{F}$ & Flow of charge in electrolyte - inlet, outlet or net & \unit{\ampere} \\
$\Itot$ & Total current on the electrode surface & \unit{\ampere} \\
\end{longtblr}

\begin{minipage}{\linewidth}
\begin{longtblr}[
caption = {Simulation Parameters},
entry = {SimulationParameters},
label = {tbl:simulation-parameters},
note{$\dag$} = {Based on unpublished experiments of open circuit voltage in a full cell and inverting the Nernst equation with \qty{200}{\mv} of potential vs. standard conditions. Results are not sensitive to this as long it's near zero; a value of exactly zero is unphysical and causes the Nernst equation to blow up.}
]{
colspec = {llll}, 
width = 1.00\linewidth, 
hlines,
rowhead = 1, 
rowfoot = 0,
}
Symbol & Value & SI Units & Source \\
$\rho$ & $997.0479$ & \unit{\kg.\m^{-3}} & \citeb{DR22,rhoH2O} \\
$\mu$ & $8.8891\times 10^{-4}$ & \unit{\kg.\m^{-1}.\sec^{-2}} & \citeb{DR22,muH2O} \\
$D_{\ce{AQDS}}$ & $4 \times 10^{-10}$ & \unit{\m^{2}.\sec^{-1}} & \citeb{DR22,HMS14} \\
$D_{\ce{H2AQDS}}$ & $4 \times 10^{-10}$ & \unit{\m^{2}.\sec^{-1}} & \citeb{DR22,HMS14} \\
$D_{\ce{H^+}}$ & $9.3 \times 10^{-9}$ & \unit{\m^{2}.\sec^{-1}} & \citeb{DR22} \\
$D_{\ce{H^+}}$ & $1.38 \times 10^{-9}$ & \unit{\m^{2}.\sec^{-1}} & \citeb{DR22} \\
$k_0$ & $7.2 \times 10^{-5} $ & \unit{\m.\sec^{-1}} & \citeb{DR22,HMS14} \\
$\alpha$ & $0.5$ & $1$ & \citeb{DR22,HMS14} \\
$\socin$ & $1.73 \times 10^{-7}$ & 1 & \TblrNote{$\dag$} \\
$c_0$ & $2.0 \times 10^{-2}$ & \unit{\mole.\m^{-3}} & NA - operating condition \\
$p_\text{in}$ & various & \unit{\kg.\m^{-1}.\sec^{-2}} & NA - operating condition \\
\end{longtblr}
\end{minipage}

\end{document}